\documentclass[12pt]{article}
\usepackage{amsmath,amssymb,amsthm}

\def\MB{\mathbf}
\def\MC{\mathcal}
\def\MR{\mathrm}
\def\BB{\mathbb}
\def\R{\BB{R}}
\def\C{\BB{C}}

\def\wh{\widehat}
\def\ol{\overline}

\def\Tr{\operatorname{\MR{Tr}}\nolimits}

\def\tsum{\textstyle\sum\limits}
\def\tprod{\textstyle\prod\limits}

\def\slacs#1{\setlength{\arraycolsep}{#1}}

\slacs{.4ex}
\begin{document}

\begin{center}
\LARGE{Nested Bethe ansatz for RTT--algebra of
$\MR{sp}(4)$ type}
\end{center}

\centerline{\v{C}. Burd\'{\i}k\,$^{1)}$, O. Navr\'atil\,$^{2)}$}

\vskip5mm

\centerline{$^{1)}$\,Faculty of Nuclear Sciences and Physical Engineering, CTU,}
\centerline{Trojanova 13, Prague, Czech Republic}

\vskip3mm

\centerline{$^{2)}$\,Faculty of Transportation Sciences, CTU,}
\centerline{Na Florenci 25, Prague, Czech Republic}

\begin{abstract}
We study the highest weight representations of the
RTT--algebras for the R--matrix $\MR{sp}(4)$ type by the nested
algebraic Bethe ansatz. These models were solved by Reshetikhin for
$\MR{sp}(2n)$ but using a very special type of representation.
The explicit construction of the Bethe vectors is carried out in the general case
and the explicit formulae for the Bethe equations are derived. In
conclusion, the direct generalization for $\MR{sp}(2n)$ is
formulated.
\end{abstract}

\section{Introduction}

The formulation of the quantum inverse scattering method, or
algebraic Bethe ansatz, by the Leningrad school \cite{FST79}
provides eigenvectors and eigenvalues of the transfer matrix.
The latter is the  generating function of the conserved quantities
of a large family of quantum integrable models. The transfer
matrix eigenvectors are constructed from the representation theory
of the RTT--algebras. In order to construct these eigenvectors, one
should first  prepare Bethe vectors, depending on a set of
complex variables. The first formulation of the Bethe vectors for
the $\MR{gl}(n)$-invariant models was given by P.P. Kulish and N.Yu.
Reshetikhin in \cite{KR83} where the nested algebraic Bethe ansatz
was introduced. These vectors are given by recursion on the rank
of the algebra. The embedding
$\MR{gl}(n-1)\subset\MR{gl}(n)$  is used  and the result for $\MR{gl}(2)$
case that is well know.
The construction was in \cite{BR08} generalized to the cases of the
RTT--(super)algebras and deformed RTT--(super)algebras with the
R--matrix of $\MR{gl}(m|\,n)$ type.
We will describe this construction for the known case of
the RTT--algebra of $\MR{gl}(3)$ type. The crucial fact for this
construction is that the RTT--algebra of $\MR{gl}(2)$ type is the
RTT--subalgebra of the RTT--algebra $\MR{gl}(3)$.  Very nice
formulas for Bethe vectors are also given in $\MR{gl}(3)$ \cite{BPRS13}.

For the R--matrices invariant under the action of the $\MR{so}(n)$
and $\MR{sp}(2n)$ groups, the problem was partially solved in
\cite{K85, K88}. The aim of this paper is to extend this calculation
to the case of $\MR{sp}(4)$ for general representations.

\medskip

The paper is structured as follows:

\smallskip

In Section 2 we give the basic definitions for the RTT--algebras
and formulate our problem.

In Section 3 we remember the known results on the algebraic Bethe ansatz
for the RTT--algebra of $\MR{gl}(2)$ type and the nested Bethe ansatz
for the RTT--algebra of $\MR{gl}(3)$ type.

The main result of the paper is the object of Section 4. In this
Section we show how it is possible to generalize the method of the
nested Bethe ansatz known for the RTT--algebra of $\MR{gl}(3)$ type
to the RTT--algebra $\MR{sp}(4)$ type. By means of this generalization
we obtain in this Section the eigenvectors and the Bethe conditions
for the RTT--algebra of $\MR{sp}(4)$ type.

In Conclusion, we mention possible generalizations of the used method
for construction of the Bethe vectors and the Bethe conditions.

The proofs of many claims are explained from the reason of transparency
of the main text in the Appendix.

\section{RTT--algebras}

We denote by $\MB{E}^k_i$ the matrix that has all elements equal to 0
with the exception of the element in the $i$--th row and $k$--th column that is 1.
Then $\MB{I}={\tsum_k}\MB{E}^k_k$ is the unit matrix and relations
$\MB{E}^k_i\MB{E}^s_r=\delta^k_r\MB{E}^s_i$ hold.

Let us consider the R--matrix
$$
\MB{R}(x,y)={\tsum_{i,k,r,s}}R^{i,r}_{k,s}(x,y)\MB{E}^k_i\otimes\MB{E}^s_r\,,
$$
which is invertible and meets the Yang--Baxter equation
\begin{equation}
 \label{YBR-0}
\MB{R}_{1,2}(x,y)\MB{R}_{1,3}(x,z)\MB{R}_{2,3}(y,z)=
\MB{R}_{2,3}(y,z)\MB{R}_{1,3}(x,z)\MB{R}_{1,2}(x,y)\,.
\end{equation}

The RTT--algebra  $\MC{A}$ is an associative algebra with the unit,
which is generated by elements $T^i_k(x)$, where for the monodromy operator
$$
\MB{T}(x)={\tsum_{i,k}}\MB{E}^k_i\otimes T^i_k(x)
$$
the RTT--equation
\begin{equation}
 \label{RTT-0}
\MB{R}_{1,2}(x,y)\MB{T}_1(x)\MB{T}_2(y)=
\MB{T}_2(y)\MB{T}_1(x)\MB{R}_{1,2}(x,y)
\end{equation}
is fulfilled.

From the invertibility of the R--matrix it follows that the element
$$
H(x)=\Tr\bigl(\MB{T}(x)\bigr)= {\tsum_i}T^i_i(x)
$$
fulfils the equation
$$
H(x)H(y)=H(y)H(x)
$$
for each $x$ and $y$.

We will deal with the RTT--algebras of types $\MR{gl}(2)$ and $\MR{gl}(3)$,
which are defined by the R--matrix of $\MR{gl}(n)$ type
$$
\MB{R}(x,y)=\frac{x-y}{x-y+1}\,\MB{I}\otimes\MB{I}+
\frac1{x-y+1}\,{\tsum_{i,k=1}^n}\MB{E}^i_k\otimes\MB{E}^k_i=
\frac1{f(x,y)}\Bigl(\MB{I}\otimes\MB{I}+
g(x,y){\tsum_{i,k=1}^n}\MB{E}^i_k\otimes\MB{E}^k_i\Bigr),
$$
where
$$
g(x,y)=\frac1{x-y}\,,\qquad f(x,y)=\frac{x-y+1}{x-y}\,,
$$
and by the RTT--algebra of $\MR{sp}(4)$ type, which is associated
with the R--matrix
$$
\MB{R}(x,y)=
\frac1{f(x,y)}\Bigl(\MB{I}\otimes\MB{I}+g(x,y){\tsum_{i,k=-2}^2}\MB{E}^i_k\otimes\MB{E}^k_i-
h(x,y){\tsum_{i,k=-2}^2}\epsilon_i\epsilon_k\MB{E}^i_k\otimes\MB{E}^{-i}_{-k}\Bigr),
$$
where the indices $i$ and $k$ take values $\pm1$ and $\pm2$ and
$$
h(x,y)=\dfrac1{x-y+3}\,,\qquad \epsilon_i=\MR{sgn}(i)\,.
$$

All of these R--matrices fulfill the Yang-Baxter equation (\ref{YBR-0}) and
the relation
$$
\MB{R}(x,y)\MB{R}(y,x)=\MB{I}\otimes\MB{I}\,.
$$

\bigskip

In the vector space $\MC{W}$ of the representation of the RTT--algebra
$\MC{A}$ the element $\omega\in\MC{W}$, vacuum vector, will be assumed to exist
for which $\MC{W}=\MC{A}\omega$ and the relations
$$
\begin{array}{l}
T^i_k(x)\omega=0\quad\MR{for}\quad i>k\,,\qquad
(\mbox{or for $i<k$})\\[4pt]
T^i_i(x)\omega=\lambda_i(x)\omega
\end{array}
$$
are valid.

In the vector space $\MC{W}=\MC{A}\omega$ we will look for eigenvectors
of $H(x)$, i.e. nonzero vectors $w\in\MC{W}$ such that
for any $x$ we have
$$
H(x)w=E(x)w\,.
$$

\section{Nested Bethe ansatz for RTT--algebras of $\MR{gl}(2)$
and $\MR{gl}(3)$ types}

We define the vacuum vector $\omega$ by the equalities
$$
T^i_k(x)\omega=0\quad \MR{for}\quad k<i\,,\hskip10mm
T^i_i(x)\omega=\lambda_i(x)\omega\,.
$$
The RTT--equation (\ref{RTT-0}) for the $\MR{gl}(n)$  RTT--algebra
is equivalent to the commutation relations
$$
\begin{array}{l}
T^i_k(x)T^r_s(y)+g(x,y)T^r_k(x)T^i_s(y)=
T^r_s(y)T^i_k(x)+g(x,y)T^r_k(y)T^i_s(x)\\[6pt]
T^i_k(x)T^r_s(y)+g(y,x)T^i_s(x)T^r_k(y)=
T^r_s(y)T^i_k(x)+g(y,x)T^i_s(y)T^r_k(x)
\end{array}
$$
It follows that for each $i$, $k$ and $x$, $y$
$T^i_k(x)T^i_k(y)=T^i_k(y)T^i_k(x)$ is fulfilled.

\subsection{Nested Bethe ansatz for the RTT--algebra of $\MR{gl}(2)$ type}

Bethe ansatz for the RTT--algebra  $\MR{gl}(2)$ is that the eigenvectors
of the operator $H(x)=T^1_1(x)+T^2_2(x)$ are looked for in the form
$$
\bigl|\ol{u}\bigr>=T^1_2(\ol{u})\omega\,,\quad\MR{where}\quad
T^1_2(\ol{u})= T^1_2(u_1)T^1_2(u_2)\ldots T^1_2(u_N)\,.
$$
In our work, we will use brand names for the sets
$$
\ol{u}=\{u_1,u_2,\ldots,u_N\}\,,\qquad
\ol{u}_k=\ol{u}\setminus\{u_k\}\,.
$$

\bigskip

 \noindent
\underbar{\textbf{Lemma 1.}}
In the RTT--algebra of $\MR{gl}(2)$ type the following relations are fulfilled
\begin{equation}
 \label{L-1}
\begin{array}{l}
T^1_1(x)T^1_2(\ol{u})= F(\ol{u},x)T^1_2(\ol{u})T^1_1(x)-
{\tsum_{u_k\in\ol{u}}}g(u_k,x)F(\ol{u}_k,u_k)T^1_2\bigl(\{\ol{u}_k,x\}\bigr)T^1_1(u_k)\,,\\[6pt]
T^2_2(x)T^1_2(\ol{u})= F(x,\ol{u})T^1_2(\ol{u})T^2_2(x)-
{\tsum_{u_k\in\ol{u}}}g(x,u_k)F(u_k,\ol{u}_k)T^1_2\bigl(\{\ol{u}_k,x\}\bigr)T^2_2(u_k)\,,
\end{array}
\end{equation}
where
$$
F(\ol{u},x)={\tprod_{u_k\in\ol{u}}}f(u_k,x)\,,\qquad
F(x,\ol{u})={\tprod_{u_k\in\ol{u}}}f(x,u_k)
$$

\bigskip

If equality (\ref{L-1}) is applied to the vacuum vector $\omega$,
we get the result

\smallskip

\underbar{\textbf{Theorem 1.}} If for any $u_k\in\ol{u}$ the
Bethe condition is fulfilled
$$
\lambda_1(u_k)F(\ol{u}_k,u_k)=\lambda_2(u_k)F(u_k,\ol{u}_k)
$$
then for any $x$ the vector $\bigl|\ol{u}\bigr>$ is the eigenvector
of operator $H(x)$ with eigenvalue
$$
E(x;\ol{u})= \lambda_1(x)F(\ol{u},x)+\lambda_2(x)F(x,\ol{u})\,.
$$

\subsection{Nested Bethe ansatz for RTT--algebra of $\MR{gl}(3)$ type}

The nested Bethe ansatz for the RTT--algebra of $\MR{gl}(3)$ type,
see \cite{KR83}, is characterized by the fact that the
eigenvectors of the operator $H(x)=T^1_1(x)+T^2_2(x)+T^3_3(x)$ are looked for in the form
$$
{\tsum_{a_1,a_2,\ldots,a_M=1}^2}T^{a_1}_3(v_1)T^{a_2}_3(v_2)\ldots
T^{a_M}_3(v_M)\Phi_{a_1,a_2,\ldots,a_M}\,,
$$
where $\Phi_{a_1,a_2,\ldots,a_M}\in\tilde{\MC{W}}=\tilde{\MC{A}}\omega$ and
$\tilde{\MC{A}}$ is the RTT--algebra generated by the elements $T^a_b(x)$,
where $a,\,b=1,\,2$, and the R--matrix
$$
\tilde{\MB{R}}(x,y)=\frac1{f(x,y)}\Bigl(\MB{I}^{(2)}\otimes\MB{I}^{(2)}+
g(x,y){\tsum_{a,b=1}^2}\MB{E}^a_b\otimes\MB{E}^b_a\Bigr),
\quad\MR{where}\quad
\MB{I}^{(2)}={\tsum_{a=1}^2}\MB{E}^a_a\,,
$$
which is of $\MR{gl}(2)$ type. Since the commutation relations
between the elements $T^a_b(x)$ are the same as in the RTT--algebra $\MC{A}$,
the RTT--algebra $\tilde{\MC{A}}$ is the RTT--subalgebra of $\MC{A}$.

It is essential for Bethe's ansatz that the subspace $\tilde{\MC{W}}$
is invariant not only at the actions of the elemets $T^1_1(x)$, $T^2_2(x)$,
but also at the action of the element $T^3_3(x)$.
This follows from the fact that for $a,\,b,\,c=1,\,2$
we obtain from the commutation relations
$$
\begin{array}{l}
T^3_3(x)T^a_b(y)=T^a_b(y)T^3_3(x)+g(x,y)T^a_3(y)T^3_b(x)-g(x,y)T^a_3(x)T^3_b(y)\,,\\[4pt]
T^3_c(x)T^a_b(y)=T^a_b(y)T^3_c(x)+g(x,y)T^a_c(y)T^3_b(x)-g(x,y)T^a_c(x)T^3_b(y)\,.
\end{array}
$$
Since, for $a=1,\,2$ is $T^3_a(x)\omega=0$, it is fulfilled for any element
$\MB{w}\in\tilde{\MC{W}}$
$$
T^3_a(x)\MB{w}=0\,,\qquad T^3_3(x)\MB{w}=\lambda_3(x)\MB{w}\,.
$$
It follows that the commutation relations between $T^a_b(x)$, $a,b=1,\,2$,
and $T^3_3(x)$ taken on the space $\tilde{\MC{W}}$ are
$$
\begin{array}{c}
T^3_3(x)T^3_3(y)=T^3_3(y)T^3_3(x)\,,\qquad
T^3_3(x)\tilde{\MB{T}}(y)=\tilde{\MB{T}}(y)T^3_3(x)\,,\\[4pt]
\tilde{\MB{R}}_{1,2}(x,y)\tilde{\MB{T}}_1(x)\tilde{\MB{T}}_2(y)=
\tilde{\MB{T}}_2(y)\tilde{\MB{T}}_1(x)\tilde{\MB{R}}_{1,2}(x,y)\,,
\end{array}
$$
where $\tilde{\MB{T}}(x)={\tsum_{a,b=1}^2}\MB{E}^b_a\otimes T^a_b(x)$.

It follows from the commutation relations that the eigenvectors of the operator
$$
\tilde{H}(x)=T^1_1(x)+T^2_2(x)\in\tilde{\MC{A}}
$$
on the space $\tilde{\MC{W}}$ are likewise eigenvectors of the operator
$H(x)\in\MC{A}$.

\bigskip

 \noindent
\underbar{\texttt{Note:}}
The original work \cite{KR83} on $\MR{gl}(n)$
uses the fact that the RTT--algebra $\tilde{\MC{A}}$ with the monodromy operator
$\tilde{\MB{T}}(x)={\tsum_{a,b=1}^{n-1}}\MB{E}^b_a\otimes T^a_b(x)$ and
R--matrix $\tilde{\MB{R}}(x,y)$ is the RTT--subalgebra of the RTT--algebra $\MC{A}$
and on the vector space $\tilde{\MC{W}}$ the element $T^n_n(x)$ acts as a multiple
of the unit operator.

However, another interpretation of commutation relations on the vector space
$\tilde{\MC{W}}$ is also possible. The commutation relations can
be written in the form of the RTT--equation
$$
\MB{R}^{(0)}_{1,2}(x,y)\MB{T}^{(0)}_1(x)\MB{T}^{(0)}_2(y)=
\MB{T}^{(0)}_2(y)\MB{T}^{(0)}_1(x)\MB{R}^{(0)}_{1,2}(x,y)
$$
where
$$
\MB{R}^{(0)}(x,y)=\MB{E}^3_3\otimes\MB{E}^3_3+\MB{E}^3_3\otimes\MB{I}^{(2)}+
\MB{I}^{(2)}\otimes\MB{E}^3_3+\tilde{\MB{R}}(x,y)\,.
$$
and $\MB{T}^{(0)}(x)=\MB{E}^3_3\otimes T^3_3(x)+\tilde{\MB{T}}(x)$.
As the R--matrix $\MB{R}^{(0)}(x,y)$ fulfills the Yang--Baxter equation and
$\MB{R}^{(0)}(x,y)\MB{R}^{(0)}(y,x)=\MB{I}\otimes\MB{I}$, we can consider
the RTT--algebra $\MC{A}^{(0)}$ generated by the elements $T^3_3(x)$
and $T^a_b(x)$ with the R--matrix $\MB{R}^{(0)}(x,y)$.
But the RTT--algebra $\MC{A}^{(0)}$ is no longer the RTT--subalgebra of $\MC{A}$.
It is actually the RTT--algebra of $\MR{gl}(2)$ type generated by the elements
$T^a_b(x)$, where $a,\,b=1,\,2$, with the R--matrix $\tilde{\MB{R}}(x,y)$
extended by the central element $T^3_3(x)$.

This is just the interpretation which we use for the nested
Bethe ansatz for the RTT--algebra of type $\MR{sp}(4)$.

\bigskip

We will now permute $T^1_1(x)$, $T^2_2(x)$ and $T^3_3(x)$ with the elements
$T^{a_1}_3(v_1)\ldots T^{a_M}_3(v_M)$.
In order not to use long expression with indices, we would introduce the markings
$$
\MB{b}(v)={\tsum_{a=1}^2}\MB{e}_a\otimes T^a_3(v)\in\MC{V}\otimes\MC{A}\,,
$$
where $\MB{e}_1$ and $\MB{e}_2$ form the basis of the two-dimensional vector space
$\MC{V}$. In general, for an ordered set of $M$ numbers
$\vec{v}=(v_1,v_2,\ldots,v_M)$ we define
$$
\begin{array}{l}
\MB{b}_{1,\ldots,M}(\vec{v})=
\MB{b}_1(v_1)\MB{b}_2(v_2)\ldots\MB{b}_M(v_M)=\\[6pt]
\hskip10mm=
{\tsum_{a_1,\ldots,a_M=1}^2}\MB{e}_{a_1}\otimes\MB{e}_{a_2}\otimes\ldots\otimes\MB{e}_{a_M}
\otimes T^{a_1}_3(v_1)T^{a_2}_3(v_2)\ldots T^{a_M}_3(v_M)\in
\underbrace{\MC{V}\otimes\ldots\otimes\MC{V}}_{M\times}\otimes\MC{A}\,.
\end{array}
$$

Let $\MB{f}^b$ be dual bases of the dual space $\MC{V}^*$, so
$\bigl<\MB{e}_a,\MB{f}^b\bigr>=\delta^a_b$ is valid.
We can write the assumed shape of the eigenvector as
$$
{\tsum_{a_1,a_2,\ldots,a_M=1}^2}T^{a_1}_3(v_1)T^{a_2}_3(v_2)\ldots
T^{a_M}_3(v_M)\Phi_{a_1,a_2,\ldots,a_M}=
\Bigl<\MB{b}_{1,\ldots,M}(\vec{v}),\MB{\Phi}\Bigr>,
$$
where
$$
\MB{\Phi}={\tsum_{b_1,\ldots,b_M=1}^2}\MB{f}^{b_1}\otimes\MB{f}^{b_2}\otimes\ldots
\otimes\MB{f}^{b_M}\otimes\Phi_{b_1,b_2,\ldots,b_M}\in
\underbrace{\MC{V}^*\otimes\ldots\otimes\MC{V}^*}_{M\times}\otimes\tilde{\MC{W}}=\wh{\MC{W}}\,.
$$

We will consider the matrix $\MB{E}^a_b$ as a linear mapping on the
space $\MC{V}$ defined by the formulae $\MB{E}^a_b\MB{e}_c=\delta^a_c\MB{e}_b$.
For adequate dual mapping we then use the abbreviation $\MB{F}^a_b$, so
$$
\bigl<\MB{E}^a_b\MB{e}_c,\MB{f}^d\bigr>=
\bigl<\MB{e}_c,\MB{F}^a_b\MB{f}^d\bigr>\,,
\quad\MR{i.e.}\quad \MB{F}^a_b\MB{f}^d=\delta^d_b\MB{f}^a
$$
is valid. Then $\MB{I}^*={\tsum_{a=1}^2}\MB{F}^a_a$ is an identical mapping
on the space $\MC{V}^*$ and contrary to similar relations for mapping $\MB{E}^a_b$
the relation $\MB{F}^a_b\MB{F}^c_d=\delta^c_b\MB{F}^a_d$ holds.

Using this notation, we can write interesting commutation relations like
$$
\begin{array}{l}
T^3_3(x)\MB{b}(v)=f(x,v)\MB{b}(v)T^3_3(x)-g(x,v)\MB{b}(x)T^3_3(v)\,,\\[6pt]
\tilde{\MB{T}}_0(x)\Bigl<\MB{b}_1(v),\MB{I}\otimes\MB{f}^b\Bigr>=
f(v,x)\Bigl<\MB{b}_1(v),\wh{\MB{T}}_{0,1^*}(x;v)\bigl(\MB{I}\otimes\MB{f}^b\bigr)\Bigr>-\\[4pt]
\hskip43mm-
g(v,x)\Bigl<\MB{b}_1(x),\wh{\BB{T}}_{0,1^*}(v)\bigl(\MB{I}\otimes\MB{f}^b\bigr)\Bigr>,
\end{array}
$$
where
$$
\begin{array}{l}
\wh{\MB{T}}_{0,1^*}(x;v)=\wh{\MB{R}}_{0,1^*}(x,v)\tilde{\MB{T}}_0(x)\,,\\[6pt]
\wh{\MB{R}}_{0,1^*}(x;v)=\dfrac1{f(v,x)}\Bigl(\MB{I}\otimes\MB{I}^*+
g(v,x){\tsum_{a,b=1}^2}\MB{E}^a_b\otimes\MB{F}^b_a\Bigr)\\[6pt]
\wh{\BB{T}}_{0,1^*}(v)=\wh{\MB{T}}_{0,1}(v;v)=
\wh{\BB{R}}_{0,1^*}\tilde{\MB{T}}_0(v)\,,\hskip10mm
\wh{\BB{R}}_{0,1^*}=\wh{\MB{R}}_{0,1^*}(v,v)={\tsum_{a,b=1}^2}\MB{E}^a_b\otimes\MB{F}^b_a\,.
\end{array}
$$
It can be shown that for any $\MB{\Phi}\in\wh{\MC{W}}$ the relations
$$
\begin{array}{l}
T^3_3(x)\Bigl<\MB{b}_{1,\ldots,M}(\vec{v}),\MB{\Phi}\Bigr>=
\lambda_3(x)F(x,\ol{v})\Bigl<\MB{b}_{1,\ldots,M}(\vec{v}),\MB{\Phi}\Bigr>-\\[6pt]
\hskip30mm-
{\tsum_{v_k\in\ol{v}}}\lambda_3(v_k)g(x,v_k)F(v_k,\ol{v}_k)
\Bigl<\MB{b}_{k;1,\ldots,M}(x,\vec{v}_k),
\tilde{\MB{R}}^{*}_{k;1,\ldots,k}(\vec{v})\MB{\Phi}\Bigr>\\[9pt]
\tilde{\MB{T}}_0(x)\Bigl<\MB{b}_{1,\ldots,M}(\vec{v}),\MB{\Phi}\Bigr>=
F(\ol{v},x)\Bigl<\MB{b}_{1,\ldots,M}(\vec{v}),\wh{\MB{T}}_{0,1,\ldots,M}(x;\vec{v})\MB{\Phi}\Bigr>-\\[4pt]
\hskip30mm- {\tsum_{v_k\in\ol{v}}}g(v_k,x)F(\ol{v}_k,v_k)
\Bigl<\MB{b}_{k;1,\ldots,M}(x,\vec{v}_k),
\tilde{\MB{R}}^{*}_{k;1,\ldots,k}(\vec{v})\wh{\BB{T}}_{k;0,1,\ldots,M}(\vec{v})\MB{\Phi}\Bigr>,
\end{array}
$$
where
$$
\begin{array}{l}
\MB{b}_{k;1,\ldots,M}(x,\vec{v}_k)=
\MB{b}_k(x)\MB{b}_1(v_1)\ldots\MB{b}_{k-1}(v_{k-1})
\MB{b}_{k+1}(v_{k+1})\ldots\MB{b}_N(v_M)\,,\\[6pt]
\tilde{\MB{R}}^*_{2,1}(y,x)=
\dfrac1{f(y,x)}\Bigl(\MB{I}^*\otimes\MB{I}^*+g(y,x){\tsum_{a,b=1}^2}\MB{F}^a_b\otimes\MB{F}^b_a\Bigr)\\[6pt]
\tilde{\MB{R}}^{*}_{k;1,\ldots,k}(\vec{v})=
\tilde{\MB{R}}^*_{k,1}(v_k,v_1)\ldots
\tilde{\MB{R}}^*_{k,k-1}(v_k,v_{k-1})\,,\hskip20mm
\tilde{\MB{R}}_{1;1\ldots,1}(\vec{v})=\MB{I}^*\,,\\[6pt]
\wh{\MB{T}}_{0,1,\ldots,M}(x;\vec{v})=
\wh{\MB{R}}_{0,1^*}(x,v_1)\wh{\MB{R}}_{0,2^*}(x,v_2)\ldots\wh{\MB{R}}_{0,M^*}(x,v_M)\tilde{\MB{T}}_0(x)\,,\\[6pt]
\wh{\BB{T}}_{k;0,1,\ldots,M}(\vec{v})=\wh{\MB{T}}_{0,1,\ldots,M}(v_k;\vec{v})\,.
\end{array}
$$
hold. If we write
$$
\wh{\MB{T}}_{0,1,\ldots,M}(x,\vec{v})=
{\tsum_{i,k=1}^2}\MB{E}^k_i\otimes\wh{T}^i_k(x;\vec{v})\,,\quad\MR{where}\quad
\wh{T}^i_k(x;\vec{v})\in\underbrace{\MC{V}^*\otimes\ldots\otimes\MC{V}^*}_{M\times}\otimes\tilde{\MC{A}}\,,
$$
we get the following

\smallskip

 \noindent
\underbar{\textbf{Proposition 1.}}
Let $\MB{\Phi}\in\wh{\MC{W}}$ be for any $x$
eigenvector of the operator
$$
\wh{H}_{1,\ldots,M}(x;\vec{v})=
\wh{T}^1_1(x;\vec{v})+\wh{T}^2_2(x;\vec{v})
$$
with eigenvalue $\wh{E}(x;\vec{v})$. If for any $v_k\in\ol{v}$
$$
\lambda_3(v_k)F(v_k,\ol{v}_k)=\wh{E}(v_k;\vec{v})F(\vec{v}_k,v_k)
$$
is fulfilled, then $\bigl<\MB{b}_{1,\ldots,M}(\vec{v}),\MB{\Phi}\bigr>$
is eigenvector of the operator
$H(x)=T^1_1(x)+T^2_2(x)+T^3_3(x)$ with eigenvalue
$$
E(x;\vec{v},\mu)=\lambda_3(x)F(x,\ol{v})+\wh{E}(x;\vec{v})F(\ol{v},x)\,.
$$

\bigskip

So we have to find our eigenvectors of the operators $\wh{H}_{1,\ldots,M}(x;\vec{v})$.
It will be easy to see that the operators
$\wh{\MB{T}}_{0,1,\ldots,M}(x;\vec{v})$ fulfill the RTT--equation
$$
\tilde{\MB{R}}_{0,0'}(x,y)\wh{\MB{T}}_{0,1,\ldots,M}(x;\vec{v})\wh{\MB{T}}_{0',1,\ldots,M}(y;\vec{v})=
\wh{\MB{T}}_{0',1,\ldots,M}(y;\vec{v})\wh{\MB{T}}_{0,1,\ldots,M}(x;\vec{v})\tilde{\MB{R}}_{0,0'}(x,y)\,.
$$
It means that they generate the RTT--algebra of type $\MR{gl}(2)$.
And so, if we find in the vector space $\wh{\MC{W}}$ the vacuum vector,
it is possible to use the results of Theorem 1 for construction of
eigenvectors of the operators $\wh{H}_{1,\ldots,M}(x;\vec{v})$.

It is possible to show that for the vector
$$
\wh{\Omega}=\underbrace{\MB{f}^2\otimes\ldots\otimes\MB{f}^2}_{M\times}\otimes\omega
$$
we have
$$
\wh{T}^2_1(x;\vec{v})\wh{\Omega}=0\,,\qquad
\wh{T}^1_1(x;\vec{v})\wh{\Omega}=\mu_1(x;\vec{v})\wh{\Omega}\,,\qquad
\wh{T}^2_2(x;\vec{v})\wh{\Omega}=\mu_2(x;\vec{v})\wh{\Omega}\,,
$$
where
$$
\mu_1(x;\vec{v})=\frac{\lambda_1(x)}{F(\ol{v},x)}\,,\qquad
\mu_2(x;\vec{v})=\lambda_2(x)\,.
$$
We get the following proposition from Theorem 1

\smallskip

 \noindent
\underbar{\textbf{Proposition 2.}}
If for any $u_k\in\ol{u}$ the equality
$$
\lambda_1(u_k)F(\ol{u}_k,u_k)=
\lambda_2(u_k)F(u_k,\ol{u}_k)F(\ol{v},u_k)\,,
$$
holds, then the vector $\MB{\Phi}(\ol{u};\vec{v})=
\wh{T}^2_1(\ol{u};\vec{v})\wh{\Omega}$
is the eigenvector of the operator
$\wh{H}_{1,\ldots,M}(x;\vec{v})$ with eigenvalue
$$
\wh{E}(x;\ol{u};\vec{v})= \lambda_1(x)
\frac{F(\ol{u},x)}{F(\ol{v},x)}+ \lambda_2(x)F(x,\ol{u})\,.
$$

\bigskip

From Propositions 1 and 2 we obtain the final result for the RTT--algebra
of type $\MR{gl}(3)$:

\smallskip

 \noindent
\underbar{\textbf{Theorem 2.}}
Let for any $u_i\in\ol{u}$ and $v_k\in\ol{v}$ the Bethe conditions
$$
\begin{array}{l}
\lambda_1(u_k)F(\ol{u}_k,u_k)=\lambda_2(u_k)F(\ol{v},u_k)F(u_k,\ol{u}_k)\,,\\[4pt]
\lambda_3(v_k)F(v_k,\ol{v}_k)=\lambda_2(v_k)F(\ol{v}_k,v_k)F(v_k,\ol{u})
\end{array}
$$
be fulfilled. Then the vector
$\bigl|\vec{v},\ol{u}\bigr>=\Bigl<\MB{b}_{1,\ldots,M}(\vec{v});
\MB{\Phi}(\ol{u};\vec{v})\Bigr>$ is eigenvector of the operator $H(x)$ with
eigenvalue
$$
E(x;\ol{u};\vec{v})= \lambda_1(x)F(\ol{u},x)+
\lambda_2(x)F(x,\ol{u})F(\ol{v},x)+\lambda_3(x)F(x,\ol{v})
$$

\section{Nested Bethe ansatz for RTT--algebra of $\MR{sp}(4)$ type}

The vacuum vector $\omega$ for the RTT--algebra $\MC{A}$ of $\MR{sp}(4)$ type
is defined by
$$
T^i_k(x)\omega=0\quad\MR{for}\quad i<k\,,\hskip15mm
T^i_i(x)\omega=\lambda_i(x)\omega\quad\MR{for}\quad
i=\pm1,\,\pm2\,.
$$
From the RTT--equation we obtain the commutation relations
\begin{equation}
 \label{KR-sp4}
\begin{array}{l}
T^i_k(x)T^r_s(y)+g(x,y)T^r_k(x)T^i_s(y)+
\delta^{i,-r}h(x,y){\tsum_{p=-2}^2}\epsilon_p\epsilon_rT^p_k(x)T^{-p}_s(y)=\\[4pt]
\hskip20mm= T^r_s(y)T^i_k(x)+g(x,y)T^r_k(y)T^i_s(x)+
\delta_{k,-s}h(x,y){\tsum_{p=-2}^2}\epsilon_k\epsilon_pT^r_p(y)T^i_{-p}(x)\\[6pt]
T^i_k(x)T^r_s(y)+g(y,x)T^i_s(x)T^r_k(y)+
\delta_{k,-s}h(y,x){\tsum_{p=-2}^2}\epsilon_s\epsilon_pT^i_p(x)T^r_{-p}(y)=\\[4pt]
\hskip20mm= T^r_s(y)T^i_k(x)+g(y,x)T^i_s(y)T^r_k(x)+
h(y,x)\delta^{i,-r}{\tsum_{p=-2}^2}\epsilon_i\epsilon_pT^p_s(y)T^{-p}_k(x).
\end{array}
\end{equation}
It follows that for each $i,\,k$ and $x,\,y$
$T^i_k(x)T^i_k(y)=T^i_k(y)T^i_k(x)$ is valid.

\subsection{RTT--algebra $\tilde{\MC{A}}$}

In the RTT--algebra $\MC{A}$ there are two RTT--subalgebras $\MC{A}^{(+)}$
and $\MC{A}^{(-)}$, respectively, which are generated by the elements $T^i_k(x)$
and $T^{-i}_{-k}(x)$, where $i,\,k=1,\,2$.

First, we deal with the vector space
$$
\MC{W}_0=\MC{A}^{(+)}\MB{A}^{(-)}\omega\subset\MC{W}=\MC{A}\omega\,.
$$

\medskip

 \noindent
\underbar{\textbf{Lemma 2.}}
For any $i,\,k=1,\,2$ and any $\MB{w}\in\MC{W}_0$ is
$T^{-i}_k(x)\MB{w}=0$.

\bigskip

 \noindent
\underbar{\textbf{Lemma 3.}}
We denote
$$
\MB{T}^{(+)}(x)={\tsum_{i,k=1}^2}\MB{E}^k_i\otimes T^i_k(x)\,,
\hskip10mm
\MB{T}^{(-)}(x)={\tsum_{i,k=1}^2}\MB{E}^{-k}_{-i}\otimes
T^{-i}_{-k}(x)\,,
$$
and
$$
\begin{array}{ll}
\MB{R}^{(+,+)}(x,y)=
\dfrac1{f(x,y)}\,\Bigl(\MB{I}_+\otimes\MB{I}_++
g(x,y){\tsum_{i,k=1}^2}\MB{E}^i_k\otimes\MB{E}^k_i\Bigr),
\qquad&
\MB{I}_+={\tsum_{i=1}^2}\MB{E}^i_i\,,\\[6pt]
\MB{R}^{(-,-)}(x,y)=
\dfrac1{f(x,y)}\,\Bigr(\MB{I}_-\otimes\MB{I}_-+
g(x,y){\tsum_{i,k=1}^2}\MB{E}^{-i}_{-k}\otimes\MB{E}^{-k}_{-i}\Bigr),
\qquad&
\MB{I}_-={\tsum_{i=1}^2}\MB{E}^{-i}_{-i}\,,\\[6pt]
\MB{R}^{(+,-)}(x,y)=
\MB{I}_+\otimes\MB{I}_--k(x,y){\tsum_{i,k=1}^2}\MB{E}^i_k\otimes\MB{E}^{-i}_{-k}\,,
\qquad&
k(x,y)=\dfrac1{x-y-1}\,,\\[6pt]
\MB{R}^{(-,+)}(x,y)=
\MB{I}_-\otimes\MB{I}_+-h(x,y){\tsum_{i,k=1}^2}\MB{E}^{-i}_{-k}\otimes\MB{E}^i_k\,.
\end{array}
$$
On the vector space $\MC{W}_0$ the relations
$$
\MB{R}^{(\epsilon_1,\epsilon_2)}_{1,2}(x,y)
\MB{T}^{(\epsilon_1)}_1(x)\MB{T}^{(\epsilon_2)}_2(y)=
\MB{T}^{(\epsilon_2)}_2(y)\MB{T}^{(\epsilon_1)}_1(x)
\MB{R}^{(\epsilon_1,\epsilon_2)}_{1,2}(x,y)
$$
are valid for any $\epsilon_1,\,\epsilon_2=\pm$.

\bigskip

 \noindent
\underbar{\textbf{Proposition 3.}}
If we define
\begin{eqnarray*}
\tilde{\MB{R}}(x,y)&=&\MB{R}^{(+,+)}(x,y)+\MB{R}^{(+,-)}(x,y)+
\MB{R}^{(-,+)}(x,y)+\MB{R}^{(-,-)}(x,y)\\
\tilde{\MB{T}}(x)&=&\MB{T}^{(+)}(x)+\MB{T}^{(-)}(x)
\end{eqnarray*}
the RTT--equation
$$
\tilde{\MB{R}}_{1,2}(x,y)\tilde{\MB{T}}_1(x)\tilde{\MB{T}}_2(y)=
\tilde{\MB{T}}_2(y)\tilde{\MB{T}}_1(x)\tilde{\MB{R}}_{1,2}(x,y)
$$
is fulfilled on the space $\MC{W}_0$.
Further, the R--matrix $\tilde{\MB{R}}(x,y)$ meets the Yang--Baxter equation
$$
\tilde{\MB{R}}_{1,2}(x,y)\tilde{\MB{R}}_{1,3}(x,z)\tilde{\MB{R}}_{2,3}(y,z)=
\tilde{\MB{R}}_{2,3}(y,z)\tilde{\MB{R}}_{1,3}(x,z)\tilde{\MB{R}}_{1,2}(x,y)
$$
and
$\tilde{\MB{R}}_{1,2}(x,y)\tilde{\MB{R}}_{2,1}(y,x)=\MB{I}\otimes\MB{I}$
is hold.\footnote{Here is
\begin{eqnarray*}
\tilde{\MB{R}}_{1,2}(x,y)&=&
\MB{R}^{(+,+)}_{1,2}(x,y)+\MB{R}^{(+,-)}_{1,2}(x,y)+
\MB{R}^{(-,+)}_{1,2}(x,y)+\MB{R}^{(-,-)}_{1,2}(x,y)\,,\\
\tilde{\MB{R}}_{2,1}(y,x)&=&
\MB{R}^{(+,+)}_{2,1}(y,x)+\MB{R}^{(+,-)}_{2,1}(y,x)+
\MB{R}^{(-,+)}_{2,1}(y,x)+\MB{R}^{(-,-)}_{2,1}(y,x)\,,
\end{eqnarray*}
where
\begin{eqnarray*}
\MB{R}^{(+,+)}_{1,2}(x,y)&=&
\frac1{f(x,y)}\,\Bigl(\MB{I}_+\otimes\MB{I}_++g(x,y){\tsum_{i,k=1}^2}\MB{E}^i_k\otimes\MB{E}^k_i\Bigr),\\
\MB{R}^{(-,-)}_{1,2}(x,y)&=&
\frac1{f(x,y)}\,\Bigr(\MB{I}_-\otimes\MB{I}_-+g(x,y){\tsum_{i,k=1}^2}\MB{E}^{-i}_{-k}\otimes\MB{E}^{-k}_{-i}\Bigr),\\
\MB{R}^{(+,-)}_{1,2}(x,y)&=&
\MB{I}_+\otimes\MB{I}_--k(x,y){\tsum_{i,k=1}^2}\MB{E}^i_k\otimes\MB{E}^{-i}_{-k}\,,\\
\MB{R}^{(-,+)}_{1,2}(x,y)&=&
\MB{I}_-\otimes\MB{I}_+-h(x,y){\tsum_{i,k=1}^2}\MB{E}^{-i}_{-k}\otimes\MB{E}^i_k\,,\\
\MB{R}^{(+,+)}_{2,1}(y,x)&=&
\frac1{f(y,x)}\,\Bigl(\MB{I}_+\otimes\MB{I}_++g(y,x){\tsum_{i,k=1}^2}\MB{E}^i_k\otimes\MB{E}^k_i\Bigr),\\
\MB{R}^{(-,-)}_{2,1}(y,x)&=&
\frac1{f(y,x)}\,\Bigr(\MB{I}_-\otimes\MB{I}_-+g(y,x){\tsum_{i,k=1}^2}\MB{E}^{-i}_{-k}\otimes\MB{E}^{-k}_{-i}\Bigr),\\
\MB{R}^{(+,-)}_{2,1}(y,x)&=&
\MB{I}_-\otimes\MB{I}_+-k(y,x){\tsum_{i,k=1}^2}\MB{E}^{-i}_{-k}\otimes\MB{E}^i_k\,,\\
\MB{R}^{(-,+)}_{2,1}(y,x)&=&
\MB{I}_+\otimes\MB{I}_--h(y,x){\tsum_{i,k=1}^2}\MB{E}^i_k\otimes\MB{E}^{-i}_{-k}\,.
\end{eqnarray*}
}

\smallskip

 \noindent
\underbar{\textsc{Proof:}}
These claims can be proved by direct calculation.

\bigskip

 \noindent
\underbar{\textbf{Definition.}}
We denote by $\tilde{\MC{A}}$ the RTT--algebra, which is defined by
the R--matrix $\tilde{\MB{R}}(x,y)$.

\bigskip

For generators of the RTT--algebra $\tilde{\MC{A}}$ we use the designation
$\tilde{T}^i_k(x)$ and $\tilde{T}^{-i}_{-k}(x)$. Specifically, we label
these operators $T^i_k(x)$ and $T^{-i}_{-k}(x)$ on
the vector space $\MC{W}_0$.

\bigskip

 \noindent
\underbar{\textbf{Lemma 4.}}
In the RTT--algebra $\tilde{\MC{A}}$ the following relations are
fulfilled for any $x$ and $y$
\begin{enumerate}
\itemsep=0pt
\item
for any $i$ and $k$
$$
\tilde{T}^i_k(x)\tilde{T}^i_k(y)=
\tilde{T}^i_k(y)\tilde{T}^i_k(x)\,,\qquad
\tilde{T}^{-i}_{-k}(x)\tilde{T}^{-i}_{-k}(y)=
\tilde{T}^{-i}_{-k}(y)\tilde{T}^{-i}_{-k}(x)\,,
$$
\item
for any $i\neq k$
$$
\tilde{T}^i_k(x)\tilde{T}^{-k}_{-i}(y)=
\tilde{T}^{-k}_{-i}(y)\tilde{T}^i_k(x)\,,
$$
\item
for the operators $\tilde{H}^{(+)}(x)=\tilde{T}^1_1(x)+\tilde{T}^2_2(x)$ and
$\tilde{H}^{(-)}(x)=\tilde{T}^{-1}_{-1}(x)+\tilde{T}^{-2}_{-2}(x)$ we have
$$
\begin{array}{l}
\tilde{H}^{(+)}(x)\tilde{H}^{(+)}(y)=\tilde{H}^{(+)}(y)\tilde{H}^{(+)}(x)\,,\\[4pt]
\tilde{H}^{(-)}(x)\tilde{H}^{(-)}(y)=\tilde{H}^{(-)}(y)\tilde{H}^{(-)}(x)\,,\\[4pt]
\tilde{H}^{(+)}(x)\tilde{H}^{(-)}(y)=\tilde{H}^{(-)}(y)\tilde{H}^{(+)}(x)\,.
\end{array}
$$
\end{enumerate}

\smallskip

 \noindent
\underbar{\textsc{Proof:}}
These relations are an easy consequence of the commutation relations
in the RTT--algebra $\tilde{\MC{A}}$.

\subsection{The general shape of eigenvectors}

We denote by $\vec{u}$ an ordered set of mutually different numbers,
i.e. $\vec{u}=(u_1,u_2,\ldots,u_N)$. The eigenvectors of the RTT--algebra
is looked for in the shape
$$
{\tsum_{i_1,\ldots,i_N,k_1,\ldots,k_N=1}^2}T^{i_1}_{-k_1}(u_1)
T^{i_2}_{-k_2}(u_2)\ldots T^{i_N}_{-k_N}(u_N)
\Phi^{-k_1,-k_2,\ldots,-k_N}_{i_1,i_2,\ldots,i_N}
$$
where
$\Phi^{-k_1,-k_2,\ldots,-k_N}_{i_1,i_2,\ldots,i_N}\in\MC{W}_0$.

Let $\MB{e}_1$, $\MB{e}_2$ be the basis of the two-dimensional space
$\MC{V}_+\sim\C^2$ and $\MB{f}^1$, $\MB{f}^2$ its dual basis in
the space $\MC{V}^*_+$. Similarly, $\MB{e}_{-1}$, $\MB{e}_{-2}$ is the basis
of the vector space $\MC{V}_-\sim\C^2$ and $\MB{f}^{-1}$, $\MB{f}^{-2}$ its dual
basis in the space $\MC{V}_-^*$.

We denote
$$
\MB{B}_1(u)={\tsum_{i,k=1}^2}\MB{e}_i\otimes\MB{f}^{-k}\otimes
T^i_{-k}(u)\in\MC{V}_{1_+}\otimes\MC{V}^*_{1_-}\otimes\MC{A}
$$
and define
$$
\MB{B}_{1,\ldots,N}(\vec{u})=\MB{B}_1(u_1)\MB{B}_2(u_2)\ldots\MB{B}_N(u_N)\in
\MC{V}_+\otimes\MC{V}^*_-\otimes\MC{A}\,,
$$
where
$$
\MC{V}_+=\MC{V}_{1_+}\otimes\MC{V}_{2_+}\otimes\ldots\otimes\MC{V}_{N_+}\,,\qquad
\MC{V}^*_-=\MC{V}^*_{1_-}\otimes\MC{V}^*_{2_-}\otimes\ldots\otimes\MC{V}^*_{N_-}\,.
$$
Specifically,
$$
\MB{B}_{1,\ldots,N}(\vec{u})=
{\tsum_{\vec{i},\vec{k}}}\MB{e}_{\vec{i}}\otimes\MB{f}^{-\vec{k}}\otimes
T^{\vec{i}}_{-\vec{k}}(\vec{u})
$$
where
$$
\begin{array}{l}
\MB{e}_{\vec{i}}=\MB{e}_{i_1}\otimes\ldots\otimes\MB{e}_{i_N}\in
\MC{V}_{1_+}\otimes\ldots\otimes\MC{V}_{N_+}=\MC{V}_+\\[4pt]
\MB{f}^{-\vec{k}}=\MB{f}^{-k_1}\otimes\ldots\otimes\MB{f}^{-k_N}\in
\MC{V}^*_{1_-}\otimes\ldots\otimes\MC{V}^*_{N_-}=\MC{V}^*_-\\[4pt]
T^{\vec{i}}_{-\vec{k}}(\vec{u})= T^{i_1}_{-k_1}(u_1)\ldots
T^{i_N}_{-k_N}(u_N)\in\MC{A}\,.
\end{array}
$$
The general shape of the eigenvector can be written as
$\Bigl<\MB{B}_{1,\ldots,N}(\vec{u}),\MB{\Phi}\Bigr>$, where
$$
\begin{array}{l}
\MB{\Phi}= \MB{f}^{(\vec{s})}\otimes\MB{e}_{-\vec{r}}\otimes
\Phi^{-\vec{r}}_{\vec{s}}\in\MC{V}^*_+\otimes\MC{V}_-\otimes\MC{W}_0=
\wh{\MC{W}}_0\,,\\[6pt]
\MB{f}^{(\vec{s})}=\MB{f}^{s_1}\otimes\ldots\otimes\MB{f}^{s_N}\in
\MC{V}^*_{1_+}\otimes\ldots\otimes\MC{V}^*_{N_+}=\MC{V}^*_+\,,\\[4pt]
\MB{e}_{-\vec{r}}=\MB{e}_{-r_1}\otimes\ldots\otimes\MB{e}_{-r_N}\in
\MC{V}_{1_-}\otimes\ldots\otimes\MC{V}_{N_-}=\MC{V}_-\,,\\[4pt]
\Phi^{-\vec{r}}_{\vec{s}}=
\Phi^{-r_1,\ldots,-r_N}_{s_1,\ldots,s_N}\in\MC{W}_0\,.
\end{array}
$$

\subsection{Some commutation relations and their consequences}

We will now look for the operator action of $\MB{T}^{(+)}(x)$ and $\MB{T}^{(-)}(x)$
on the assumed shape of eigenvectors
$\Bigl<\MB{B}_{1,\ldots,N}(\vec{u}),\MB{\Phi}\Bigr>$.

\medskip

 \noindent
\underbar{\textbf{Lemma 5.}}
In the RTT--algebra of $\MR{sp}(4)$ type the following relations
\begin{eqnarray*}
\MB{T}^{(+)}_0(x)\Bigl<\MB{B}_1(u),\MB{f}^r\otimes\MB{e}_{-s}\Bigr>&=&
f(u,x)\Bigl<\MB{B}_1(u),\wh{\MB{T}}^{(+)}_{0,1}(x;u)
\bigl(\MB{I}\otimes\MB{f}^r\otimes\MB{e}_{-s}\bigr)\Bigr>-\\
&&- g(u,x)\Bigl<\MB{B}_1(x),\wh{\BB{T}}^{(+)}_{0,1}(u)
\bigl(\MB{I}\otimes\MB{f}^r\otimes\MB{e}_{-s}\bigr)\Bigr>\\[6pt]
\MB{T}^{(-)}_0(x)\Bigl<\MB{B}_1(u),\MB{f}^r\otimes\MB{e}_{-s}\Bigr>&=&
f(x,u)\Bigl<\MB{B}_1(u),\wh{\MB{T}}^{(-)}_{0,1}(x;u)
\bigl(\MB{I}\otimes\MB{f}^r\otimes\MB{e}_{-s}\bigr)\Bigr>-\\
&&- g(x,u)\Bigl<\MB{B}_1(x),\wh{\BB{T}}^{(-)}_{0,1}(u)
\bigl(\MB{I}\otimes\MB{f}^r\otimes\MB{e}_{-s}\bigr)\Bigr>
\end{eqnarray*}
are fulfilled, where
$$
\begin{array}{l}
\wh{\MB{T}}^{(\epsilon)}_{0,1}(x;u)=
\wh{\MB{R}}^{(\epsilon,+)}_{0,1_+^*}(x,u)
\MB{T}^{(\epsilon)}_0(x)
\MB{R}^{(\epsilon,-)}_{0,1_-}(x,u)\,,\\[6pt]
\wh{\MB{R}}^{(+,+)}_{0,1_+^*}(x,u)=
\dfrac1{f(u,x)}\Bigl(\MB{I}_{0_+}\otimes\MB{I}^*_{1_+}+
g(u,x){\tsum_{r,s=1}^2}\MB{E}^r_s\otimes\MB{F}^s_r\Bigr)\,,\\[4pt]
\wh{\MB{R}}^{(-,+)}_{0,1_+^*}(x,u)=
\MB{I}_{0_-}\otimes\MB{I}^*_{1_+}-
k(u,x){\tsum_{r,s=1}^2}\MB{E}^{-r}_{-s}\otimes\MB{F}^r_s\,,\\[4pt]
\wh{\BB{T}}^{(\epsilon)}_{0,1}(u)=
\wh{\R}^{(\epsilon,+)}_{0,1^*_+}
\MB{T}^{(\epsilon)}_0(u)
\R^{(\epsilon,-)}_{0,1_-}\\[4pt]
\wh{\R}^{(+,+)}_{0,1^*_+}=\wh{\MB{R}}^{(+,+)}_{0,1_+^*}(u,u)=
{\tsum_{r,s=1}^2}\MB{E}^r_s\otimes\MB{F}^s_r\,,
\end{array}
$$
$$
\begin{array}{l}
\wh{\R}^{(-,+)}_{0,1^*_+}= \wh{\MB{R}}^{(-,+)}_{0,1^*_+}(u,u)=
\MB{I}_{0_-}\otimes\MB{I}^*_{1_+}+
{\tsum_{r,s=1}^2}\MB{E}^{-r}_{-s}\otimes\MB{F}^r_s\,,\\[4pt]
\R^{(+,-)}_{0,1_-}= \MB{R}^{(+,-)}_{0,1_-}(u,u)=
\MB{I}_{0_+}\otimes\MB{I}_{1_-}+{\tsum_{r,s=1}^2}\MB{E}^r_s\otimes\MB{E}^{-r}_{-s}\,,\\[4pt]
\R^{(-,-)}_{0,1_-}= \MB{R}^{(-,-)}_{0,1_-}(u,u)=
{\tsum_{r,s=1}^2}\MB{E}^{-r}_{-s}\otimes\MB{E}^{-s}_{-r}
\end{array}
$$

\bigskip

For the formulation and proof of the following lemma, it is preferable
to define linear mapping
$\bigl(\MB{R}^*\bigr)^{(+,+)}(x,y)$ by the relation
$$
\Bigl<\MB{R}^{(+,+)}(x,y)\bigl(\MB{e}_{s_1}\otimes\MB{e}_{s_2}\bigr),
\MB{f}^{r_1}\otimes\MB{f}^{r_2}\Bigr>=
\Bigl<\MB{e}_{s_1}\otimes\MB{e}_{s_2},
\bigl(\MB{R}^*\bigr)^{(+,+)}(x,y)\bigl(\MB{f}^{r_1}\otimes\MB{f}^{r_2}\bigr)\Bigr>,
$$
i.e.
$$
\bigl(\MB{R}^*\bigr)^{(+,+)}(x,y)=
\frac1{f(x,y)}\Bigl(\MB{I}^*_+\otimes\MB{I}^*_++
g(x,y){\tsum_{i,k=1}^2}\MB{F}^i_k\otimes\MB{F}^k_i\Bigr).
$$

 \noindent
\underbar{\textbf{Lemma 6.}}
If we denote by $\ol{u}$ the set of elements
$\vec{u}$, i.e. $\ol{u}=\{u_1,u_2,\ldots,u_N\}$, the following relations
$$
\begin{array}{l}
\MB{T}^{(+)}_0(x)\Bigl<\MB{B}_{1,\ldots,N}(\vec{u}),
\MB{f}^{\vec{r}}\otimes\MB{e}_{-\vec{s}}\Bigr>=
F(\ol{u},x)\Bigl<\MB{B}_{1,\ldots,N}(\vec{u}),\wh{\MB{T}}^{(+)}_{0;1,\ldots,N}(x;\vec{u})
\bigl(\MB{I}\otimes\MB{f}^{\vec{r}}\otimes\MB{e}_{-\vec{s}}\bigr)\Bigr>-\\[6pt]
\hskip3mm- {\tsum_{u_k\in\ol{u}}}g(u_k,x)F(\ol{u}_k,u_k)
\Bigl<\MB{B}_{k;1,\ldots,N}(x,\vec{u}_k),
\bigl(\MB{R}^*\bigr)^{(+,+)}_{1,\ldots,k}(\vec{u})
\MB{R}^{(-,-)}_{1,\ldots,k}(\vec{u})
\wh{\BB{T}}^{(+)}_{k;0,1,\ldots,N}(\vec{u})
\bigl(\MB{I}\otimes\MB{f}^{\vec{r}}\otimes\MB{e}_{-\vec{s}}\bigr)\Bigr>\\[9pt]
\MB{T}^{(-)}_0(x)\Bigl<\MB{B}_{1,\ldots,N}(\vec{u}),
\MB{f}^{\vec{r}}\otimes\MB{e}_{-\vec{s}}\Bigr>=
F(x,\ol{u})\Bigl<\MB{B}_{1,\ldots,N}(\vec{u}),\wh{\MB{T}}^{(-)}_{0;1,\ldots,N}(x;\vec{u})
\bigl(\MB{I}\otimes\MB{f}^{\vec{r}}\otimes\MB{e}_{-\vec{s}}\bigr)\Bigr>-\\[6pt]
\hskip3mm- {\tsum_{u_k\in\ol{u}}}g(x,u_k)F(u_k,\ol{u}_k)
\Bigl<\MB{B}_{k;1,\ldots,N}(x,\vec{u}_k),
\bigl(\MB{R}^*\bigr)^{(+,+)}_{1,\ldots,k}(\vec{u})
\MB{R}^{(-,-)}_{1,\ldots,k}(\vec{u})
\wh{\BB{T}}^{(-)}_{k;0,1,\ldots,N}(\vec{u})
\bigl(\MB{I}\otimes\MB{f}^{\vec{r}}\otimes\MB{e}_{-\vec{s}}\bigr)\Bigr>
\end{array}
$$
are valid, where
$$
\begin{array}{l}
\wh{\MB{T}}^{(\epsilon)}_{0;1,\ldots,N}(x;\vec{u})=
\wh{\MB{R}}^{(\epsilon,+)}_{0;1^*_+,\ldots,N^*_+}(x;\vec{u})
\MB{T}^{(\epsilon)}_0(x)
\MB{R}^{(\epsilon,-)}_{0;1_-,\ldots,N_-}(x;\vec{u})\\[6pt]
\wh{\MB{R}}^{(\epsilon,+)}_{0;1^*_+,\ldots,N^*_+}(x;\vec{u})=
\wh{\MB{R}}^{(\epsilon,+)}_{0,1^*_+}(x,u_1)
\wh{\MB{R}}^{(\epsilon,+)}_{0,2^*_+}(x,u_2)\ldots
\wh{\MB{R}}^{(\epsilon,+)}_{0,N^*_+}(x,u_N)\\[6pt]
\MB{R}^{(\epsilon,-)}_{0;1_-,\ldots,N_-}(x;\vec{u})=
\MB{R}^{(\epsilon,-)}_{0,N_-}(x,u_N)\ldots
\MB{R}^{(\epsilon,-)}_{0,2_-}(x,u_2)
\MB{R}^{(\epsilon,-)}_{0,1_-}(x,u_1)\\[6pt]
\MB{B}_{k;1,\ldots,N}(x,\vec{u}_k)=
\MB{B}_k(x)\MB{B}_1(u_1)\ldots\MB{B}_{k-1}(u_{k-1})
\MB{B}_{k+1}(u_{k+1})\ldots\MB{B}_N(u_N)\\[6pt]
\bigl(\MB{R}^*\bigr)^{(+,+)}_{1,\ldots,k}(\vec{u})=
\bigl(\MB{R}^*\bigr)^{(+,+)}_{k^*_+,1^*_+}(u_k,u_1)
\bigl(\MB{R}^*\bigr)^{(+,+)}_{k^*_+,2^*_+}(u_k,u_2)\ldots
\bigl(\MB{R}^*\bigr)^{(+,+)}_{k^*_+,(k-1)^*_+}(u_k,u_{k-1})\\[6pt]
\MB{R}^{(-,-)}_{1,\ldots,k}(\vec{u})=
\MB{R}^{(-,-)}_{1_-,k_-}(u_1,u_k)
\MB{R}^{(-,-)}_{2_-,k_-}(u_2,u_k)\ldots
\MB{R}^{(-,-)}_{(k-1)_-,k_-}(u_{k-1},u_k)\\[6pt]
\wh{\BB{T}}^{(\epsilon)}_{k;0,1,\ldots,N}(\vec{u})=
\wh{\MB{T}}^{(\epsilon)}_{0;1,\ldots,N}(u_k;\vec{u})=
\wh{\MB{R}}^{(\epsilon,+)}_{0;1^*_+,\ldots,N^*_+}(u_k;\vec{u})
\MB{T}^{(\epsilon)}_0(u_k)
\MB{R}^{(\epsilon,-)}_{0;1_-,\ldots,N_-}(u_k;\vec{u})
\end{array}
$$
and empty products $\bigl(\MB{R}^*\bigr)^{(+,+)}_{1,\ldots,1}(\vec{u})$ and
$\MB{R}^{(-,-)}_{1,\ldots,1}(\vec{u})$ are understood as identical maps.

\bigskip

 \noindent
\underbar{\textbf{Proposition 4.}}
Let $\MB{\Phi}\in\wh{\MC{W}}_0$ be the eigenvectors of
the operators
$$
\begin{array}{l}
\wh{H}^{(+)}_{1,\ldots,N}(x;\vec{u})=
\bigl(\wh{\MB{T}}^{(+)}_{0,1,\ldots,N}(x;\vec{u})\bigr)^1_1+
\bigl(\wh{\MB{T}}^{(+)}_{0,1,\ldots,N}(x;\vec{u})\bigr)^2_2\\[6pt]
\wh{H}^{(-)}_{1,\ldots,N}(x;\vec{u})=
\bigl(\wh{\MB{T}}^{(-)}_{0,1,\ldots,N}(x;\vec{u})\bigr)^{-1}_{-1}+
\bigl(\wh{\MB{T}}^{(-)}_{0,1,\ldots,N}(x;\vec{u})\bigr)^{-2}_{-2}
\end{array}
$$
with eigenvalues $E^{(+)}(x;\vec{u})$ and
$E^{(-)}(x;\vec{u})$. If for any $u_k\in\ol{u}$
$$
F(\ol{u}_k,u_k)E^{(+)}(u_k;\vec{u}_k)=
F(u_k,\ol{u}_k)E^{(-)}(u_k;\vec{u}_k)
$$
is valid, then
$\Bigl<\MB{B}_{1,\ldots,N}(\vec{u}),\MB{\Phi}\Bigr>$
are the eigenvectors of the operator $H(x)$ with eigenvalue
$$
E(x;\vec{u})=F(\ol{u},x)E^{(+)}(x;\vec{u})+F(x,\ol{u})E^{(-)}(x;\vec{u})\,.
$$

\smallskip

 \noindent
\underbar{\textsc{Proof.}}
The claim comes from Lemma 6.

\subsection{The operators $\wh{\MB{T}}^{(\pm)}_{1,\ldots,N}(x;\vec{u})$
and RTT--algebra $\tilde{\MC{A}}$}

The Proposition 4 translates the original problem into the task of finding
eigenvectors of the operators
$\wh{H}^{(\pm)}(x;\vec{u})=\wh{H}^{(\pm)}_{1,\ldots,N}(x;\vec{u})$
on the vector space $\wh{\MC{W}}_0$. We show that the operators
$\wh{T}^i_k(x;\vec{u})$ and $\wh{T}^{-i}_{-k}(x,\vec{u})$, where
$i,\,k=1,\,2$, and
$$
\wh{\MB{T}}^{(+)}_{0;1,\ldots,N}(x;\vec{u})=
{\tsum_{i,k=1}^2}\MB{E}^k_i\otimes\wh{T}^i_k(x;\vec{u})\,,
\qquad
\wh{\MB{T}}^{(-)}_{0;1,\ldots,N}(x;\vec{u})=
{\tsum_{i,k=1}^2}\MB{E}^{-k}_{-i}\otimes\wh{T}^{-i}_{-k}(x;\vec{u})
$$
are on the space $\wh{\MC{W}}_0$ the generators of the RTT--algebra
$\tilde{\MC{A}}$ and that in the representation there is a vacuum vector.

\bigskip

 \noindent
\underbar{\textbf{Lemma 7.}}
For any $\MB{\Phi}\in\wh{\MC{W}}_0$ and
$\epsilon_0,\,\epsilon_{0'}=\pm$, the RTT--relation
$$
\MB{R}^{(\epsilon_0,\epsilon_{0'})}_{0,0'}(x,y)
\wh{\MB{T}}^{(\epsilon_0)}_{0,1,\ldots,N}(x,\vec{u})
\wh{\MB{T}}^{(\epsilon_{0'})}_{0',1,\ldots,N}(y;\vec{u})\MB{\Phi}=
\wh{\MB{T}}^{(\epsilon_{0'})}_{0',1,\ldots,N}(y;\vec{u})
\wh{\MB{T}}^{(\epsilon_0)}_{0,1,\ldots,N}(x,\vec{u})
\MB{R}^{(\epsilon_0,\epsilon_{0'})}_{0,0'}(x,y)\MB{\Phi}
$$
is fulfilled.

\bigskip

 \noindent
\underbar{\textbf{Lemma 8.}} For the vector
$$
\wh{\Omega}=\underbrace{\MB{f}^1\otimes\ldots\otimes\MB{f}^1}_{N\times}\otimes
\underbrace{\MB{e}_{-1}\otimes\ldots\otimes\MB{e}_{-1}}_{N\times}\otimes\,\omega\in\wh{\MC{W}}_0
$$
we have
$$
\begin{array}{lll}
\wh{T}^1_2(x;\vec{u})\wh{\Omega}=0\,,\qquad&
\wh{T}^1_1(x;\vec{u})\wh{\Omega}=\mu_1(x,\ol{u})\wh{\Omega}\,,\qquad&
\wh{T}^2_2(x;\vec{u})\wh{\Omega}=\mu_2(x,\ol{u})\wh{\Omega}\,,\\[6pt]
\wh{T}^{-2}_{-1}(x;\vec{u})\wh{\Omega}=0\,,\qquad&
\wh{T}^{-1}_{-1}(x;\vec{u})\wh{\Omega}=\mu_{-1}(x,\ol{u})\wh{\Omega}\,,\qquad&
\wh{T}^{-2}_{-2}(x;\vec{u})\wh{\Omega}=\mu_{-2}(x,\ol{u})\wh{\Omega}
\end{array}
$$
where
$$
\begin{array}{ll}
\mu_1(x;\ol{u})=\lambda_1(x)F(\ol{u},x-1)\,,\qquad&
\mu_2(x;\ol{u})=\dfrac{\lambda_2(x)}{F(\ol{u},x)}\,,\\[6pt]
\mu_{-1}(x;\ol{u})=
\lambda_{-1}(x)F(x+1,\ol{u})\,,\qquad&
\mu_{-2}(x;\ol{u})=\dfrac{\lambda_{-2}(x)}{F(x,\ol{u})}.
\end{array}
$$

\subsection{Nested Bethe ansatz for RTT--algebra $\tilde{\MC{A}}$}

We want to find in the space $\wh{\MC{W}}_0$ the common eigenvectors of operators
$$
\wh{H}^{(+)}_{1,\ldots,N}(x;\vec{u})=\wh{T}^1_1(x;\vec{u})+\wh{T}^2_2(x;\vec{u})\,,
\qquad
\wh{H}^{(-)}_{1,\ldots,N}(x;\vec{u})=\wh{T}^{-1}_{-1}(x;\vec{u})+\wh{T}^{-2}_{-2}(x;\vec{u})\,,
$$
for any $\vec{u}$. We will study the representation of the RTT--algebra
$\tilde{\MC{A}}$ with the generators $\tilde{T}^i_k(x)$ and $\tilde{T}^{-i}_{-k}(x)$
on the space $\tilde{\MC{W}}_0=\tilde{\MC{A}}\tilde{\omega}$, where $\tilde{\omega}$ is
the vacuum vector for which
$$
\tilde{T}^1_2(x)\tilde{\omega}=\tilde{T}^{-2}_{-1}(x)\tilde{\omega}=0\,,
\qquad
\tilde{T}^i_i(x)\tilde{\omega}=\mu_i(x)\tilde{\omega}\,,\quad
\tilde{T}^{-i}_{-i}(x)\tilde{\omega}=\mu_{-i}(x)\tilde{\omega}
$$
is valid. Common eigenvectors of the operators
$$
\tilde{H}^{(+)}(x)=\tilde{T}^1_1(x)+\tilde{T}^2_2(x)\,,\qquad
\tilde{H}^{(-)}(x)=\tilde{T}^{-1}_{-1}(x)+\tilde{T}^{-2}_{-2}(x)
$$
will be looked for in the shape
$$
\bigl|\ol{v},\ol{w}\bigr>=
\tilde{T}^2_1(\ol{v})\tilde{T}^{-1}_{-2}(\ol{w})\tilde{\omega}=
\tilde{T}^2_1(w_1)\ldots\tilde{T}^2_1(v_P)
\tilde{T}^{-1}_{-2}(w_1)\ldots\tilde{T}^{-1}_{-2}(w_Q)\tilde{\omega}\,,
$$
where $\ol{v}=\bigl\{v_1,\ldots,w_P\bigr\}$ and
$\ol{w}=\bigl\{w_1,\ldots,w_Q\bigr\}$.

\bigskip

 \noindent
\underbar{\textbf{Lemma 9.}}
In the RTT--algebra $\tilde{\MC{A}}$ the following relations are valid:
$$
\begin{array}{l}
\tilde{T}^1_1(x)\tilde{T}^2_1(\ol{v})= F(x,\ol{v})\tilde{T}^2_1(\ol{v})\tilde{T}^1_1(x)-
{\tsum_{v_r\in\ol{v}}}g(x,v_r)F(v_r,\ol{v}_r)\tilde{T}^2_1(\ol{v}_r,x)\tilde{T}^1_1(v_r)\\[6pt]
\tilde{T}^2_2(x)\tilde{T}^2_1(\ol{v})= F(\ol{v},x)\tilde{T}^2_1(\ol{v})\tilde{T}^2_2(x)-
{\tsum_{v_r\in\ol{v}}}g(v_r,x)F(\ol{v}_r,v_r)\tilde{T}^2_1(\ol{v}_r,x)\tilde{T}^2_2(v_r)\\[6pt]
\tilde{T}^{-1}_{-1}(x)\tilde{T}^{-1}_{-2}(\ol{w})=
F(\ol{w},x)\tilde{T}^{-1}_{-2}(\ol{w})\tilde{T}^{-1}_{-1}(x)-\\[4pt]
\hskip60mm-
{\tsum_{w_s\in\ol{w}}}g(w_s,x)F(\ol{w}_s,w_s)
\tilde{T}^{-1}_{-2}(\ol{w}_s,x)\tilde{T}^{-1}_{-1}(w_s)\\[6pt]
\tilde{T}^{-2}_{-2}(x)\tilde{T}^{-1}_{-2}(\ol{w})=
F(x,\ol{w})\tilde{T}^{-1}_{-2}(\ol{w})\tilde{T}^{-2}_{-2}(x)-\\[4pt]
\hskip60mm-
{\tsum_{w_s\in\ol{w}}}g(x,w_s)F(w_s,\ol{w}_s)
\tilde{T}^{-1}_{-2}(\ol{w}_s,x)\tilde{T}^{-2}_{-2}(w_s)\\[9pt]
%
%%%
%
\tilde{T}^1_1(x)\tilde{T}^{-1}_{-2}(\ol{w})=
F(x-2,\ol{w})\tilde{T}^{-1}_{-2}(\ol{w})\tilde{T}^1_1(x)+\\[4pt]
\hskip60mm+
{\tsum_{w_s\in\ol{w}}}g(x-2,w_s)F(w_s,\ol{w}_s)
\tilde{T}^2_1(x)\tilde{T}^{-1}_{-2}(\ol{w}_s)\tilde{T}^{-2}_{-2}(w_s)\\[6pt]
\tilde{T}^2_2(x)\tilde{T}^{-1}_{-2}(\ol{w})=
F(\ol{w},x-2)\tilde{T}^{-1}_{-2}(\ol{w})\tilde{T}^2_2(x)+\\[4pt]
\hskip60mm+
{\tsum_{w_s\in\ol{w}}}g(w_s,x-2)F(\ol{w}_s,w_s)
\tilde{T}^2_1(x)\tilde{T}^{-1}_{-2}(\ol{w}_s)\tilde{T}^{-1}_{-1}(w_s)\\[6pt]
\tilde{T}^{-1}_{-1}(x)\tilde{T}^2_1(\ol{v})=
F(\ol{v},x+2)\tilde{T}^2_1(\ol{v})\tilde{T}^{-1}_{-1}(x)+
{\tsum_{v_r\in\ol{v}}}g(v_r,x+2)F(\ol{v}_r,v_r)
\tilde{T}^2_1(\ol{v}_r)\tilde{T}^{-1}_{-2}(x)\tilde{T}^2_2(v_r)\\[6pt]
\tilde{T}^{-2}_{-2}(x)\tilde{T}^2_1(\ol{v})=
F(x+2,\ol{v})\tilde{T}^2_1(\ol{v})\tilde{T}^{-2}_{-2}(x)+
{\tsum_{v_r\in\ol{v}}}g(x+2,v_r)F(v_r,\ol{v}_r)
\tilde{T}^2_1(\ol{v}_r)\tilde{T}^{-1}_{-2}(x)\tilde{T}^1_1(v_r)
\end{array}
$$

\bigskip

Applying the results of Lemma 9 to the assumed shape
eigenvector, we get the following statement

\medskip

 \noindent
\underbar{\textbf{Lemma 10.}}
For the representation of the RTT--algebra $\tilde{\MC{A}}$
with the highest weight $\mu_i(x)$ we obtain
$$
\begin{array}{rcl}
\tilde{T}^1_1(x)\bigl|\ol{v};\ol{w}\bigr>&=&
\mu_1(x)F(x,\ol{v})F(x-2,\ol{w})\bigl|\ol{v};\ol{w}\bigr>-\\[4pt]
&&-
{\tsum_{v_r\in\ol{v}}}\mu_1(v_r)g(x,v_r)F(v_r,\ol{v}_r)F(v_r-2,\ol{w})
\bigl|\{\ol{v}_r,x\};\ol{w}\bigr>+\\[4pt]
&&+
{\tsum_{w_s\in\ol{w}}}\mu_{-2}(w_s)g(x-2,w_s)F(w_s+2,\ol{v})F(w_s,\ol{w}_s)
\bigl|\{\ol{v},x\};\ol{w}_s\bigr>\\[6pt]
\tilde{T}^2_2(x)\bigl|\ol{v};\ol{w}\bigr>&=&
\mu_2(x)F(\ol{v},x)F(\ol{w},x-2)\bigl|\ol{v};\ol{w}\bigr>-\\[4pt]
&&-
{\tsum_{v_r\in\ol{v}}}\mu_2(v_r)g(v_r,x)F(\ol{v}_r,v_r)F(\ol{w},v_r-2)
\bigl|\{\ol{v}_r,x\};\ol{w}\bigr>+\\[4pt]
&&+
{\tsum_{w_s\in\ol{w}}}\mu_{-1}(w_s)g(w_s,x-2)F(\ol{v},w_s+2)F(\ol{w}_s,w_s)
\bigl|\{\ol{v},x\};\ol{w}_s\bigr>\\[6pt]
\tilde{T}^{-1}_{-1}(x)\bigl|\ol{v};\ol{w}\bigr>&=&
\mu_{-1}(x)F(\ol{v},x+2)F(\ol{w},x)\bigl|\ol{v};\ol{w}\bigr>+\\[4pt]
&&+
{\tsum_{v_r\in\ol{v}}}\mu_2(v_r)g(v_r,x+2)F(\ol{v}_r,v_r)F(\ol{w},v_r-2)
\bigl|\ol{v}_r;\{\ol{w},x\}\bigr>-\\[4pt]
&&-
{\tsum_{w_s\in\ol{w}}}\mu_{-1}(w_s)g(w_s,x)F(\ol{v},w_s+2)F(\ol{w}_s,w_s)
\bigl|\ol{v};\{\ol{w}_s,x\}\bigr>\\[6pt]
\tilde{T}^{-2}_{-2}(x)\bigl|\ol{v};\ol{w}\bigr>&=&
\mu_{-2}(x)F(x+2,\ol{v})F(x,\ol{w})\bigl|\ol{v};\ol{w}\bigr>+\\[4pt]
&&+
{\tsum_{v_r\in\ol{v}}}\mu_1(v_r)g(x+2,v_r)F(v_r,\ol{v}_r)F(v_r-2,\ol{w})
\bigl|\ol{v}_r;\{\ol{w},x\}\bigr>-\\[4pt]
&&-
{\tsum_{w_s\in\ol{w}}}\mu_{-2}(w_s)g(x,w_s)F(w_s+2,\ol{v})F(w_s,\ol{w}_s)
\bigl|\ol{v};\{\ol{w}_s,x\}\bigr>\,,
\end{array}
$$

\bigskip

From Lemma 10 we almost immediately have the following sentence, which
gives Bethe's conditions for eigenvectors in the RTT--algebra $\tilde{\MC{A}}$.

\bigskip

 \noindent
\underbar{\textbf{Theorem 3.}} If for any
$v_r\in\ol{v}$ and $w_s\in\ol{w}$ the conditions
\begin{equation}
 \label{BP-TA}
\begin{array}{l}
\mu_1(v_r)F(v_r,\ol{v}_r)F(v_r-2,\ol{w})=
\mu_2(v_r)F(\ol{v}_r,v_r)F(\ol{w},v_r-2)\\[6pt]
\mu_{-1}(w_s)F(\ol{v},w_s+2)F(\ol{w}_s,w_s)=
\mu_{-2}(w_s)F(w_s+2,\ol{v})F(w_s,\ol{w}_s)
\end{array}
\end{equation}
are fulfilled, the vector
$\bigl|\ol{v};\ol{w}\bigr>=\tilde{T}^2_1(\ol{v})\tilde{T}^{-1}_{-2}(\ol{w})\omega$
is the eigenvector $\tilde{H}^{(+)}(x)$ a $\tilde{H}^{(-)}(x)$
with eigenvalue
$$
\begin{array}{l}
\mu^{(+)}(x;\ol{v},\ol{w})=
\mu_1(x)F(x,\ol{v})F(x-2,\ol{w})+
\mu_2(x)F(\ol{v},x)F(\ol{w},x-2)\\[6pt]
\mu^{(-)}(x;\ol{v},\ol{w})=
\mu_{-1}(x)F(\ol{v},x+2)F(\ol{w},x)+
\mu_{-2}(x)F(x+2,\ol{v})F(x,\ol{w})\,.
\end{array}
$$

\subsection{The eigenvectors for RTT--algebra of $\MR{sp}(4)$ type}

The consequence of the Theorem 3 is the following statement

\medskip

 \noindent
\underbar{\textbf{Proposition 5.}}
If for any  $v_r\in\ol{v}$ and $w_s\in\ol{w}$ the following conditions
$$
\begin{array}{l}
\lambda_1(v_r)F(\ol{u},v_r)F(\ol{u},v_r-1)F(v_r,\ol{v}_r)F(v_r-2,\ol{w})=
\lambda_2(v_r)F(\ol{v}_r,v_r)F(\ol{w},v_r-2)\\[6pt]
\lambda_{-1}(w_s)F(w_s,\ol{u})F(w_s+1,\ol{u})F(\ol{v},w_s+2)F(\ol{w}_s,w_s)=
\lambda_{-2}(w_s)F(w_s+2,\ol{v})F(w_s,\ol{w}_s)
\end{array}
$$
are fulfilled, then the vectors
$$
\MB{\Phi}(\vec{u};\ol{v};\ol{w})=
\wh{T}^2_1(\ol{v};\vec{u})\wh{T}^{-1}_{-2}(\ol{w};\vec{u})\wh{\Omega}\in\wh{\MC{W}}_0
$$
are common eigenvectors of the operators
$\wh{H}^{(+)}_{1,\ldots,N}(x;\vec{u})$ and
$\wh{H}^{(-)}_{1,\ldots,N}(x;\vec{u})$ with eigenvalues
\begin{eqnarray*}
E^{(+)}(x;\vec{u},\ol{v},\ol{w})&=&
\lambda_1(x)F(\ol{u},x-1)F(x,\ol{v})F(x-2,\ol{w})+
\lambda_2(x)\,\dfrac{F(\ol{v},x)F(\ol{w},x-2)}{F(\ol{u},x)}\\
E^{(-)}(x;\vec{u},\ol{v},\ol{w})&=&
\lambda_{-1}(x)F(x+1,\ol{u})F(\ol{v},x+2)F(\ol{w},x)+
\lambda_{-2}(x)\dfrac{F(x+2,\ol{v})F(x,\ol{w})}{F(x,\ol{u})}
\end{eqnarray*}

\bigskip

Since the relations
$$
\frac{F(\ol{u}_k,u_k)}{F(\ol{u},u_k)}=
\frac{F(u_k,\ol{u}_k)}{F(u_k,\ol{u})}=
\frac1{f(u_k,u_k)}=0\,,
$$
are true, we immediately derive from Propositions 4 and 5 our main result
that gives Bethe vectors and Bethe conditions for the RTT--algebra
of $\MR{sp}(4)$ type.

\bigskip

 \noindent
\underbar{\textbf{Theorem 4.}} Let for any $u_k\in\ol{u}$,
$v_r\in\ol{v}$ and $w_s\in\ol{w}$ the conditions
$$
\begin{array}{l}
\lambda_1(u_k)F(\ol{u}_k,u_k-1)F(\ol{u}_k,u_k)F(u_k,\ol{v})F(u_k-2,\ol{w})=\\[4pt]
\hskip40mm=
\lambda_{-1}(u_k)F(u_k+1,\ol{u}_k)F(u_k,\ol{u}_k)F(\ol{v},u_k+2)F(\ol{w},u_k)\\[6pt]
\lambda_1(v_r)F(\ol{u},v_r-1)F(\ol{u},v_r)F(v_r,\ol{v}_r)F(v_r-2,\ol{w})=
\lambda_2(v_r)F(\ol{v}_r,v_r)F(\ol{w},v_r-2)\\[6pt]
\lambda_{-1}(w_s)F(w_s+1,\ol{u})F(w_s,\ol{u})F(\ol{v},w_s+2)F(\ol{w}_s,w_s)=
\lambda_{-2}(w_s)F(w_s+2,\ol{v})F(w_s,\ol{w}_s)\,.
\end{array}
$$
be fulfilled. Then the vector
$$
\bigl|\vec{u};\ol{v};\ol{w}\bigr>=
\Bigl<\MB{B}_{1,\ldots,N}(\vec{u}),\MB{\Phi}(\vec{u};\ol{v};\ol{w})\Bigr>
$$
is an eigenvector of $H(x)$ with eigenvalue
$$
\begin{array}{l}
E(x;\vec{u},\ol{v},\ol{w})=
\lambda_1(x)F(\ol{u},x)F(\ol{u},x-1)F(x,\ol{v})F(x-2,\ol{w})+
\lambda_2(x)F(\ol{v},x)F(\ol{w},x-2)+\\[4pt]
\hskip25mm+
\lambda_{-1}(x)F(x,\ol{u})F(x+1,\ol{u})F(\ol{v},x+2)F(\ol{w},x)+
\lambda_{-2}(x)F(x+2,\ol{v})F(x,\ol{w})
\end{array}
$$

\section{Conclusion}

In the paper we formulated the nested Bethe ansatz for the RTT--algebra
of $\MR{sp}(4)$ type. By means of this ansatz we found the Bethe vectors
and the Bethe conditions for this RTT--algebra in the general case.
We suppose that a similar ansatz can be used to find the Bethe vectors
and the Bethe conditions for all RTT--algebras of $sp(2n)$ and $o(n)$ type.
Some of these results will be published in the following paper.

We believe that generalizing our method to the RTT--superalgebras of these
types is also possible.

\section*{Appendix}

The Appendix provides evidence of some of the propositions in the main text.

For many calculations in the article, we use the following proposition:

\medskip

\underbar{\textbf{Lemma 0.}}
For any $x,\,y\notin\ol{u}$, where $x\neq y$ we have
$$
\begin{array}{l}
{\tsum_{u_k\in\ol{u}}}g(x,u_k)g(u_k,y)F(u_k,\ol{u}_k)=
g(x,y)\bigl(F(x,\ol{u})-F(y,\ol{u})\bigr)\,,\\[6pt]
{\tsum_{u_k\in\ol{u}}}g(x,u_k)g(u_k,y)F(\ol{u}_k,u_k)=
g(x,y)\bigl(F(\ol{u},y)-F(\ol{u},x)\bigr)\,.
\end{array}
$$

\smallskip

\underbar{\textsc{Proof:}}
When we multiply the first proven relationship by expression
$(x-y){\tprod_{u_s\in\ol{u}}}(x-u_s)$, we obtain
$$
\begin{array}{l}
{\tsum_{u_k\in\ol{u}}}\dfrac{x-y}{u_k-y}\,F(u_k,\ol{u}_k){\tprod_{u_s\in\ol{u}_k}}(x-u_s)=
{\tprod_{u_s\in\ol{u}}}(x-u_s+1)-{\tprod_{u_s\in\ol{u}}}\dfrac{x-u_s}{y-u_s}\,(y-u_s+1)\,.
\end{array}
$$
The left and right side of this relation are polynomials of the order $N$ in the
variable $x$. Since they are equal at $N+1$ points $x=y$ and $x=u_r$ for any
$u_r\in\ol{u}$, they are equal for any $x$.

The second equality can be shown similarly.
\qed

\subsection*{A.1. Proof of Lemma 1}

The lemma assertion can be induced through a
number of elements $N$ of the set $\ol{u}$.

The assertions for $N=1$ are the commutation relations.

If these relations are valid for $N$, for $N+1$ we have
$$
\begin{array}{l}
T^1_1(x)T^1_2(\{\ol{u},u_{N+1}\})=
F(\ol{u},x)T^1_2(\ol{u})T^1_1(x)T^1_2(u_{N+1})-\\[4pt]
\hskip40mm-
{\tsum_{u_k\in\ol{u}}}g(u_k,x)F(\ol{u}_k,u_k)T^1_2\bigl(\{\ol{u}_k,x\}\bigr)T^1_1(u_k)T^1_2(u_{N+1})\,,\\[6pt]
T^2_2(x)T^1_2(\{\ol{u},u_{N+1}\})=
F(x,\ol{u})T^1_2(\ol{u})T^2_2(x)T^1_2(u_{N+1})-\\[4pt]
\hskip40mm-
{\tsum_{u_k\in\ol{u}}}g(x,u_k)F(u_k,\ol{u}_k)T^1_2\bigl(\{\ol{u}_k,x\}\bigr)T^2_2(u_k)T^1_2(u_{N+1})\,,
\end{array}
$$
By using the commutation relations we have
$$
\begin{array}{l}
T^1_1(x)T^1_2(\{\ol{u},u_{N+1}\})=
F(\{\ol{u},u_{N+1}\},x)T^1_2(\{\ol{u},u_{N+1}\})T^1_1(x)-\\[4pt]
\hskip15mm-
{\tsum_{u_k\in\ol{u}}}g(u_k,x)F(\{\ol{u}_k,u_{N+1}\},u_k)
T^1_2\bigl(\{\ol{u}_k,u_{N+1},x\}\bigr)T^1_1(u_k)-\\[4pt]
\hskip15mm-
\Bigl(g(u_{N+1},x)F(\ol{u},x)-{\tsum_{u_k\in\ol{u}}}g(u_{N+1},u_k)g(u_k,x)F(\ol{u}_k,u_k)\Bigr)
T^1_2\bigl(\{\ol{u},x\}\bigr)T^1_1(u_{N+1})\,,\\[9pt]
T^2_2(x)T^1_2(\{\ol{u},u_{N+1})=
F(x,\{\ol{u},u_{N+1}\})T^1_2(\{\ol{u},u_{N+1}\})T^2_2(x)-\\[4pt]
\hskip15mm-
{\tsum_{u_k\in\ol{u}}}g(x,u_k)F(u_k,\{\ol{u}_k,u_{N+1}\})
T^1_2\bigl(\{\ol{u}_k,,u_{N+1},x\}\bigr)T^2_2(u_k)-\\[4pt]
\hskip15mm-
\Bigl(g(x,u_{N+1})F(x,\ol{u})-{\tsum_{u_k\in\ol{u}}}g(x,u_k)g(u_k,u_{N+1})F(u_k,\ol{u}_k)\Bigr)
T^1_2\bigl(\{\ol{u},x\}\bigr)T^2_2(u_{N+1})\,.
\end{array}
$$
And the proposition will be obtained from Lemma  0.
\qed

\subsection*{A.2. Proof of Lemma 2}

For $i,\,k,\,r,\,s>0$ the first commutation relation in (\ref{KR-sp4}) give
$$
\begin{array}{l}
T^{-i}_k(x)T^r_s(y)+g(x,y)T^r_k(x)T^{-i}_s(y)+
\delta^{i,r}h(x,y){\tsum_{p=-2}^2}\varepsilon_pT^p_k(x)T^{-p}_s(y)=\\[4pt]
\hskip20mm= T^r_s(y)T^{-i}_k(x)+g(x,y)T^r_k(y)T^{-i}_s(x)
\end{array}
$$
i.e.
$$
\begin{array}{l}
{\tsum_{p=1}^2}\bigl(\delta^i_p\delta^r_q-\delta^{i,r}\delta_{p,q}h(x,y)\bigr)T^{-p}_k(x)T^q_s(y)=
T^r_s(y)T^{-i}_k(x)+g(x,y)T^r_k(y)T^{-i}_s(x)-\\[4pt]
\hskip40mm-
g(x,y)T^r_k(x)T^{-i}_s(y)-\delta^{i,r}h(x,y){\tsum_{p=1}^2}T^p_k(x)T^{-p}_s(y)\,.
\end{array}
$$
If we multiply this relation by expression
$$
\bigl(\delta^a_i\delta^b_r-k(y,x)\delta^{a,b}\delta_{i,r}\bigr)\,,
\quad\MR{kde}\quad k(y,x)=\frac1{y-x-1}\,,
$$
and sum by $i$ and $r$ and rename the indices, we have
$$
\begin{array}{l}
T^{-i}_k(x)T^r_s(y)=
T^r_s(y)T^{-i}_k(x)+g(x,y)T^r_k(y)T^{-i}_s(x)-g(x,y)T^r_k(x)T^{-i}_s(y)-\\[6pt]
\hskip7mm-
\delta^{i,r}{\tsum_{p=1}^2}\Bigl(k(y,x)T^p_s(y)T^{-p}_k(x)+
g(x,y)k(y,x)T^p_k(y)T^{-p}_s(x)+g(x,y)T^p_k(x)T^{-p}_s(y)\Bigr).
\end{array}
$$
Similarly, we obtain from the second commutation relation
in (\ref{KR-sp4}) the equality
$$
\begin{array}{l}
T^{-i}_k(x)T^{-r}_{-s}(y)=
T^{-r}_{-s}(y)T^{-i}_k(x)+g(y,x)T^{-i}_{-s}(y)T^{-r}_k(x)-g(y,x)T^{-i}_{-s}(x)T^{-r}_k(y)-\\[6pt]
\hskip7mm-
\delta_{k,s}{\tsum_{p=1}^2}\Bigl(k(x,y)T^{-r}_{-p}(y)T^{-i}_p(x)+
g(y,x)k(x,y)T^{-i}_{-p}(y)T^{-r}_p(x)+g(y,x)T^{-i}_{-p}(x)T^{-r}_p(y)\Bigr).
\end{array}
$$
This means that for $i,\,k>0$, we can commute $T^{-i}_k(x)$ to the
vacuum vector, and we have $T^{-i}_k(x)\MB{w}=0$ for any $\MB{w}\in\MC{W}_0$.
\qed

\subsection*{A.3. Proof of Lemma 3}

It follows from the commutation relations that for $i,\,k,\,r,\,s>0$
\begin{eqnarray*}
T^i_k(x)T^r_s(y)+g(x,y)T^r_k(x)T^i_s(y)&=&
T^r_s(y)T^i_k(x)+g(x,y)T^r_k(y)T^i_s(x)\\
T^{-i}_{-k}(x)T^{-r}_{-s}(y)+g(x,y)T^{-r}_{-k}(x)T^{-i}_{-s}(y)&=&
T^{-r}_{-s}(y)T^{-i}_{-k}(x)+g(x,y)T^{-r}_{-k}(y)T^{-i}_{-s}(x)
\end{eqnarray*}
hold.

If we use the commutation relations for $T^i_k(x)T^{-r}_{-s}(y)$ and
for $T^{-i}_{-k}(x)T^r_s(y)$, we obtain, according to Lemma 2,
that on the vector space $\MC{W}_0$
\begin{eqnarray*}
T^i_k(x)T^{-r}_{-s}(y)-
\delta_{k,s}h(y,x){\tsum_{p=1}^2}T^i_p(x)T^{-r}_{-p}(y)&=&
T^{-r}_{-s}(y)T^i_k(x)-
\delta^{i,r}h(y,x){\tsum_{p=1}^2}T^{-p}_{-s}(y)T^p_k(x)\\
T^{-i}_{-k}(x)T^r_s(y)-\delta^{i,r}h(x,y){\tsum_{p=1}^2}T^{-p}_{-k}(x)T^p_s(y)&=&
T^r_s(y)T^{-i}_{-k}(x)-\delta_{k,s}h(x,y){\tsum_{p=1}^2}T^r_p(y)T^{-i}_{-p}(x)
\end{eqnarray*}
are valid.

Equation from Lemma 3 is only matrix expression of these relations.
\qed

\subsection*{A.4. Proof of Lemma 5}

For the commutation $T^i_k(x)T^r_{-s}(u)$ and
$T^{-i}_{-k}(x)T^r_{-s}(u)$, where $i,k,r,s=1,\,2$, in the RTT--algebra
of $\MR{sp}(4)$ type  we use the commutation relations
$$
\begin{array}{l}
T^i_k(x)T^r_{-s}(u)=
T^r_{-s}(u)T^i_k(x)+ g(u,x)T^i_{-s}(u)T^r_k(x)-\\[6pt]
\hskip15mm- k(x,u)\delta_{k,s}{\tsum_{p=1}^2}T^r_{-p}(u)T^i_p(x)-
g(u,x)k(x,u)\delta_{k,s}{\tsum_{p=1}^2}T^i_{-p}(u)T^r_p(x)-\\[4pt]
\hskip20mm- g(u,x)T^i_{-s}(x)T^r_k(u)-
g(u,x)\delta_{k,s}{\tsum_{p=1}^2}T^i_{-p}(x)T^r_p(u)\\[9pt]
T^{-i}_{-k}(x)T^r_{-s}(u)= T^r_{-s}(u)T^{-i}_{-k}(x)+
g(x,u)T^r_{-k}(u)T^{-i}_{-s}(x)-\\[4pt]
\hskip15mm-
k(u,x)\delta^{i,r}{\tsum_{p=1}^2}T^p_{-s}(u)T^{-p}_{-k}(x)-
k(u,x)g(x,u)\delta^{i,r}{\tsum_{p=1}^2}T^p_{-k}(u)T^{-p}_{-s}(x)-\\[4pt]
\hskip25mm- g(x,u)T^r_{-k}(x)T^{-i}_{-s}(u)-
g(x,u)\delta^{i,r}{\tsum_{p=1}^2}T^p_{-k}(x)T^{-p}_{-s}(u)
\end{array}
$$
The equations listed in Lemma 5 are only another entry of these
commutation relations.
\qed

\subsection*{A.5. Proof of Lemma 6}

The assertion of the Lemma can be proved by induction with respect
to numbers $N$ of the set $\ol{u}$.

For $N=1$ the relations are the same as in Lemma 5.

Let the assertion be true for $N$. If we write
$$
\begin{array}{ll}
\vec{u}=(u_1,\ldots,u_{N+1})\,,\qquad&
\vec{u}_1=(u_2,\ldots,u_{N+1})\,,\\[4pt]
\vec{r}=(r_1,\ldots,r_{N+1})\,,\qquad&
\vec{r}_1=(r_2,\ldots,r_{N+1})\,,\\[4pt]
\vec{s}=(s_1,\ldots,s_{N+1})\,,\qquad&
\vec{s}_1=(s_2,\ldots,s_{N+1})\,,
\end{array}
$$
we have
$$
\begin{array}{l}
\MB{T}^{(\epsilon)}_0(x)\Bigl<\MB{B}_{1,\ldots,N+1}(\vec{u}),
\MB{I}\otimes\MB{f}^{\vec{r}}\otimes\MB{e}_{-\vec{s}}\Bigr>=\\[4pt]
\hskip20mm=
\MB{T}^{(\epsilon)}_0(x)\Bigl<\MB{B}_1(u_1),
\MB{f}^{r_1}\otimes\MB{e}_{-s_1}\Bigr>
\Bigl<\MB{B}_{2,\ldots,N+1}(\vec{u}_1),
\MB{I}\otimes\MB{f}^{\vec{r}_1}\otimes\MB{e}_{-\vec{s}_1}\Bigr>.
\end{array}
$$
Lemma 5 then leads to the relations
$$
\begin{array}{l}
\MB{T}^{(+)}_0(x)\Bigl<\MB{B}_{1,\ldots,N+1}(\vec{u}),
\MB{I}\otimes\MB{f}^{\vec{r}}\otimes\MB{e}_{-\vec{s}}\Bigr>=\\[4pt]
\hskip10mm=
f(u_1,x)\Bigl<\MB{B}_1(u_1),\wh{\MB{T}}^{(+)}_{0,1}(x;u_1)
\bigl(\MB{I}\otimes\MB{f}^{r_1}\otimes\MB{e}_{-s_1}\bigr)\Bigr>
\Bigl<\MB{B}_{2,\ldots,N+1}(\vec{u}_1),
\MB{I}\otimes\MB{f}^{\vec{r}_1}\otimes\MB{e}_{-\vec{s}_1}\Bigr>-\\[4pt]
\hskip20mm- g(u_1,x)\Bigl<\MB{B}_1(x),\wh{\BB{T}}^{(+)}_{0,1}(u_1)
\bigl(\MB{I}\otimes\MB{f}^{r_1}\otimes\MB{e}_{-s_1}\bigr)\Bigr>
\Bigl<\MB{B}_{2,\ldots,N+1}(\vec{u}_1),
\MB{I}\otimes\MB{f}^{\vec{r}_1}\otimes\MB{e}_{-\vec{s}_1}\Bigr>=\\[6pt]
\hskip10mm=
f(u_1,x)\Bigl<\MB{B}_1(u_1),
\wh{\MB{R}}^{(+,+)}_{0,1_+^*}(x,u_1)
\MB{T}^{(+)}_0(x)
\MB{R}^{(+,-)}_{0,1_-}(x,u_1)
\bigl(\MB{I}\otimes\MB{f}^{r_1}\otimes\MB{e}_{-s_1}\bigr)\Bigr>\\[4pt]
\hskip40mm
\Bigl<\MB{B}_{2,\ldots,N+1}(\vec{u}_1),
\MB{I}\otimes\MB{f}^{\vec{r}_1}\otimes\MB{e}_{-\vec{s}_1}\Bigr>-\\[4pt]
\hskip15mm- g(u_1,x)\Bigl<\MB{B}_1(x),
\wh{\R}^{(+,+)}_{0,1^*_+}
\MB{T}^{(+)}_0(u_1)
\R^{(+,-)}_{0,1_-}
\bigl(\MB{I}\otimes\MB{f}^{r_1}\otimes\MB{e}_{-s_1}\bigr)\Bigr>\\[4pt]
\hskip40mm
\Bigl<\MB{B}_{2,\ldots,N+1}(\vec{u}_1),
\MB{I}\otimes\MB{f}^{\vec{r}_1}\otimes\MB{e}_{-\vec{s}_1}\Bigr>,
\end{array}
$$
and so
$$
\begin{array}{l}
\MB{T}^{(+)}_0(x)\Bigl<\MB{B}_{1,\ldots,N+1}(\vec{u}),
\MB{I}\otimes\MB{f}^{\vec{r}}\otimes\MB{e}_{-\vec{s}}\Bigr>=\\[4pt]
\hskip10mm=
f(u_1,x)\Bigl<\MB{B}_1(u_1),
\wh{\MB{R}}^{(+,+)}_{0,1_+^*}(x,u_1)
\MB{T}^{(+)}_0(x)\Bigl<\MB{B}_{2,\ldots,N+1}(\vec{u}_1),
\MB{I}\otimes\MB{f}^{\vec{r}_1}\otimes\MB{e}_{-\vec{s}_1}\Bigr>\\[4pt]
\hskip60mm
\MB{R}^{(+,-)}_{0,1_-}(x,u_1)
\bigl(\MB{I}\otimes\MB{f}^{r_1}\otimes\MB{e}_{-s_1}\bigr)\Bigr>-\\[4pt]
\hskip15mm- g(u_1,x)\Bigl<\MB{B}_1(x),
\wh{\R}^{(+,+)}_{0,1^*_+}
\MB{T}^{(+)}_0(u_1)\Bigl<\MB{B}_{2,\ldots,N+1}(\vec{u}_1),
\MB{I}\otimes\MB{f}^{\vec{r}_1}\otimes\MB{e}_{-\vec{s}_1}\Bigr>\\[4pt]
\hskip60mm
\R^{(+,-)}_{0,1_-}
\bigl(\MB{I}\otimes\MB{f}^{r_1}\otimes\MB{e}_{-s_1}\bigr)\Bigr>.
\end{array}
$$
is valid.

Similarly, we have
$$
\begin{array}{l}
\MB{T}^{(-)}_0(x)\Bigl<\MB{B}_{1,\ldots,N+1}(\vec{u}),
\MB{I}\otimes\MB{f}^{\vec{r}}\otimes\MB{e}_{-\vec{s}}\Bigr>=\\[4pt]
\hskip10mm=
f(x,u_1)\Bigl<\MB{B}_1(u_1),
\wh{\MB{R}}^{(-,+)}_{0,1_+^*}(x,u_1)
\MB{T}^{(-)}_0(x)
\Bigl<\MB{B}_{2,\ldots,N+1}(\vec{u}_1),
\MB{I}\otimes\MB{f}^{\vec{r}_1}\otimes\MB{e}_{-\vec{s}_1}\Bigr>\\[4pt]
\hskip60mm
\MB{R}^{(-,-)}_{0,1_-}(x,u_1)
\bigl(\MB{I}\otimes\MB{f}^{r_1}\otimes\MB{e}_{-s_1}\bigr)\Bigr>-\\[4pt]
\hskip15mm-
g(x,u_1)\Bigl<\MB{B}_1(x),
\wh{\R}^{(-,+)}_{0,1^*_+}
\MB{T}^{(-)}_0(u_1)
\Bigl<\MB{B}_{2,\ldots,N+1}(\vec{u}_1),
\MB{I}\otimes\MB{f}^{\vec{r}_1}\otimes\MB{e}_{-\vec{s}_1}\Bigr>\\[4pt]
\hskip60mm
\R^{(-,-)}_{0,1_-}
\bigl(\MB{I}\otimes\MB{f}^{r_1}\otimes\MB{e}_{-s_1}\bigr)\Bigr>.
\end{array}
$$

The inductive assumption now gives
$$
\begin{array}{l}
\MB{T}^{(+)}_0(x)\Bigl<\MB{B}_{1,\ldots,N+1}(\vec{u}),
\MB{I}\otimes\MB{f}^{\vec{r}}\otimes\MB{e}_{-\vec{s}}\Bigr>=\\[4pt]
\hskip5mm=
F(\ol{u},x)\Bigl<\MB{B}_{1,\ldots,N+1}(\vec{u}),
\wh{\MB{T}}^{(+)}_{0;1,\ldots,N+1}(x;\vec{u})
\bigl(\MB{I}\otimes\MB{f}^{\vec{r}}\otimes\MB{e}_{-\vec{s}}\bigr)\Bigr>-\\[4pt]
\hskip10mm-
g(u_1,x)F(\ol{u}_1,u_1)\Bigl<\MB{B}_{1;1,\ldots,N+1}(x,\vec{u}_1),
\wh{\BB{T}}^{(+)}_{1;0,1,\ldots,N+1}(\vec{u})
\bigl(\MB{I}\otimes\MB{f}^{\vec{r}}\otimes\MB{e}_{-\vec{s}}\bigr)\Bigr>-\\[4pt]
\hskip10mm-
{\tsum_{u_k\in\ol{u}_1}}g(u_k,x)f(u_1,x)F(\ol{u}_{1,k},u_k)
\Bigl<\MB{B}_1(u_1)\MB{B}_{k;2,\ldots,N+1}(x,\vec{u}_{1,k}),\\[4pt]
\hskip15mm
\bigl(\MB{R}^*\bigr)^{(+,+)}_{2,\ldots,k}(\vec{u}_1)
\MB{R}^{(-,-)}_{2,\ldots,k}(\vec{u}_1)
\wh{\MB{R}}^{(+,+)}_{0,1_+^*}(x,u_1)
\wh{\BB{T}}^{(+)}_{k;0,2,\ldots,N}(\vec{u})
\MB{R}^{(+,-)}_{0,1_-}(x,u_1)
\bigl(\MB{I}\otimes\MB{f}^{\vec{r}}\otimes\MB{e}_{-\vec{s}}\bigr)\Bigr>+\\[4pt]
\hskip10mm+
{\tsum_{u_k\in\ol{u}_1}}g(u_1,x)g(u_k,u_1)F(\ol{u}_{1,k},u_k)
\Bigl<\MB{B}_1(x)\MB{B}_{k;2,\ldots,N+1}(u_1,\vec{u}_{1,k}),\\[4pt]
\hskip15mm
\bigl(\MB{R}^*\bigr)^{(+,+)}_{2,\ldots,k}(\vec{u}_1)
\MB{R}^{(-,-)}_{2,\ldots,k}(\vec{u}_1)
\wh{\R}^{(+,+)}_{0,1^*_+}
\wh{\BB{T}}^{(+)}_{k;0,2,\ldots,N}(\vec{u})
\R^{(+,-)}_{0,1_-}
\bigl(\MB{I}\otimes\MB{f}^{\vec{r}}\otimes\MB{e}_{-\vec{s}}\bigr)\Bigr>\\[12pt]
\MB{T}^{(-)}_0(x)\Bigl<\MB{B}_{1,\ldots,N+1}(\vec{u}),
\MB{I}\otimes\MB{f}^{\vec{r}}\otimes\MB{e}_{-\vec{s}}\Bigr>=\\[4pt]
\hskip5mm=
F(x,\ol{u})\Bigl<\MB{B}_{1,\ldots,N+1}(\vec{u}),
\wh{\MB{T}}^{(-)}_{0;1,\ldots,N+1}(x;\vec{u})
\bigl(\MB{I}\otimes\MB{f}^{\vec{r}}\otimes\MB{e}_{-\vec{s}}\bigr)\Bigr>-\\[4pt]
\hskip10mm-
g(x,u_1)F(u_1,\ol{u}_1)\Bigl<\MB{B}_{1;1,\ldots,N+1}(x,\vec{u}_1),
\wh{\BB{T}}^{(-)}_{1;0,1,\ldots,N+1}(\vec{u})
\bigl(\MB{I}\otimes\MB{f}^{\vec{r}}\otimes\MB{e}_{-\vec{s}}\bigr)\Bigr>-\\[4pt]
\hskip10mm-
{\tsum_{u_k\in\ol{u}_1}}g(x,u_k)f(x,u_1)F(u_k,\ol{u}_{1,k})
\Bigl<\MB{B}_1(u_1)\MB{B}_{k;2,\ldots,N+1}(x,\vec{u}_{1,k}),\\[4pt]
\hskip15mm
\bigl(\MB{R}^*\bigr)^{(+,+)}_{2,\ldots,k}(\vec{u}_1)
\MB{R}^{(-,-)}_{2,\ldots,k}(\vec{u}_1)
\wh{\MB{R}}^{(-,+)}_{0,1_+^*}(x,u_1)
\wh{\BB{T}}^{(-)}_{k;0,2,\ldots,N+1}(\vec{u}_1)
\MB{R}^{(-,-)}_{0,1_-}(x,u_1)
\bigl(\MB{I}\otimes\MB{f}^{\vec{r}}\otimes\MB{e}_{-\vec{s}}\bigr)\Bigr>+\\[4pt]
\hskip10mm+
{\tsum_{u_k\in\ol{u}_1}}g(x,u_1)g(u_1,u_k)F(u_k,\ol{u}_{1,k})
\Bigl<\MB{B}_1(x)\MB{B}_{k;2,\ldots,N+1}(u_1,\vec{u}_{1,k}),\\[4pt]
\hskip15mm
\bigl(\MB{R}^*\bigr)^{(+,+)}_{2,\ldots,k}(\vec{u}_1)
\MB{R}^{(-,-)}_{2,\ldots,k}(\vec{u}_1)
\wh{\R}^{(-,+)}_{0,1^*_+}
\wh{\BB{T}}^{(-)}_{k;0,2,\ldots,N+1}(\vec{u}_1)
\R^{(-,-)}_{0,1_-}
\bigl(\MB{I}\otimes\MB{f}^{\vec{r}}\otimes\MB{e}_{-\vec{s}}\bigr)\Bigr>.
\end{array}
$$
To prove our claim, we have to show that for any $u_k\in\ol{u}_1$
the equations
$$
\begin{array}{l}
g(u_k,x)f(u_1,x)F(\ol{u}_{1,k},u_k)
\Bigl<\MB{B}_1(u_1)\MB{B}_{k;2,\ldots,N+1}(x,\vec{u}_{1,k}),\\[4pt]
\hskip15mm
\bigl(\MB{R}^*\bigr)^{(+,+)}_{2,\ldots,k}(\vec{u}_1)
\MB{R}^{(-,-)}_{2,\ldots,k}(\vec{u}_1)
\wh{\MB{R}}^{(+,+)}_{0,1_+^*}(x,u_1)
\wh{\BB{T}}^{(+)}_{k;0,2,\ldots,N}(\vec{u})
\MB{R}^{(+,-)}_{0,1_-}(x,u_1)
\bigl(\MB{I}\otimes\MB{f}^{\vec{r}}\otimes\MB{e}_{-\vec{s}}\bigr)\Bigr>-\\[6pt]
\hskip10mm-
g(u_1,x)g(u_k,u_1)F(\ol{u}_{1,k},u_k)
\Bigl<\MB{B}_1(x)\MB{B}_{k;2,\ldots,N+1}(u_1,\vec{u}_{1,k}),\\[4pt]
\hskip15mm
\bigl(\MB{R}^*\bigr)^{(+,+)}_{2,\ldots,k}(\vec{u}_1)
\MB{R}^{(-,-)}_{2,\ldots,k}(\vec{u}_1)
\wh{\R}^{(+,+)}_{0,1^*_+}
\wh{\BB{T}}^{(+)}_{k;0,2,\ldots,N}(\vec{u})
\R^{(+,-)}_{0,1_-}
\bigl(\MB{I}\otimes\MB{f}^{\vec{r}}\otimes\MB{e}_{-\vec{s}}\bigr)\Bigr>=\\[6pt]
\hskip5mm=
g(u_k,x)F(\ol{u}_k,u_k)
\Bigl<\MB{B}_{k;1,\ldots,N}(x,\vec{u}_k),
\bigl(\MB{R}^*\bigr)^{(+,+)}_{1,\ldots,k}(\vec{u})
\MB{R}^{(-,-)}_{1,\ldots,k}(\vec{u})
\wh{\BB{T}}^{(+)}_{k;0,1,\ldots,N}(\vec{u})
\bigl(\MB{I}\otimes\MB{f}^{\vec{r}}\otimes\MB{e}_{-\vec{s}}\bigr)\Bigr>,\\[12pt]
g(x,u_k)f(x,u_1)F(u_k,\ol{u}_{1,k})
\Bigl<\MB{B}_1(u_1)\MB{B}_{k;2,\ldots,N+1}(x,\vec{u}_{1,k}),\\[4pt]
\hskip15mm
\bigl(\MB{R}^*\bigr)^{(+,+)}_{2,\ldots,k}(\vec{u}_1)
\MB{R}^{(-,-)}_{2,\ldots,k}(\vec{u}_1)
\wh{\MB{R}}^{(-,+)}_{0,1_+^*}(x,u_1)
\wh{\BB{T}}^{(-)}_{k;0,2,\ldots,N+1}(\vec{u}_1)
\MB{R}^{(-,-)}_{0,1_-}(x,u_1)
\bigl(\MB{I}\otimes\MB{f}^{\vec{r}}\otimes\MB{e}_{-\vec{s}}\bigr)\Bigr>-\\[6pt]
\hskip10mm-
g(x,u_1)g(u_1,u_k)F(u_k,\ol{u}_{1,k})
\Bigl<\MB{B}_1(x)\MB{B}_{k;2,\ldots,N+1}(u_1,\vec{u}_{1,k}),\\[4pt]
\hskip15mm
\bigl(\MB{R}^*\bigr)^{(+,+)}_{2,\ldots,k}(\vec{u}_1)
\MB{R}^{(-,-)}_{2,\ldots,k}(\vec{u}_1)
\wh{\R}^{(-,+)}_{0,1^*_+}
\wh{\BB{T}}^{(-)}_{k;0,2,\ldots,N+1}(\vec{u}_1)
\R^{(-,-)}_{0,1_-}
\bigl(\MB{I}\otimes\MB{f}^{\vec{r}}\otimes\MB{e}_{-\vec{s}}\bigr)\Bigr>=\\[6pt]
\hskip5mm=
g(x,u_k)F(u_k,\ol{u}_k)
\Bigl<\MB{B}_{k;1,\ldots,N}(x,\vec{u}_k),
\bigl(\MB{R}^*\bigr)^{(+,+)}_{1,\ldots,k}(\vec{u})
\MB{R}^{(-,-)}_{1,\ldots,k}(\vec{u})
\wh{\BB{T}}^{(-)}_{k;0,1,\ldots,N}(\vec{u})
\bigl(\MB{I}\otimes\MB{f}^{\vec{r}}\otimes\MB{e}_{-\vec{s}}\bigr)\Bigr>.
\end{array}
$$
are valid.
\bigskip

From the commutation relations
$$
T^i_{-k}(x)T^r_{-s}(y)+g(x,y)T^r_{-k}(x)T^i_{-s}(y)=
T^r_{-s}(y)T^i_{-k}(x)+g(x,y)T^r_{-k}(y)T^i_{-s}(x)
$$
we obtain
$$
\begin{array}{l}
T^i_{-k}(x)T^r_{-s}(y)=
\dfrac1{f(x,y)f(y,x)}\Bigl(T^r_{-s}(y)T^i_{-k}(x)+
g(x,y)T^r_{-k}(y)T^i_{-s}(x)+\\[6pt]
\hskip55mm+
g(y,x)T^i_{-s}(y)T^r_{-k}(x)+
g(x,y)g(y,x)T^i_{-k}(y)T^r_{-s}(x)\Bigr).
\end{array}
$$
The last relations can be expressed in the form
$$
\begin{array}{l}
\Bigl<\MB{B}_1(x)\MB{B}_2(y),
\MB{f}^{r_1}\otimes\MB{f}^{r_2}\otimes\MB{e}_{-s_1}\otimes\MB{e}_{-s_2}\Bigr>=\\[4pt]
\hskip30mm= \Bigl<\MB{B}_2(y)\MB{B}_1(x),
\bigl(\MB{R}^*\bigr)^{(+,+)}_{2_+,1_+}(y,x)\MB{R}^{(-,-)}_{1_-,2_-}(x,y)
\bigl(\MB{f}^{r_1}\otimes\MB{f}^{r_2}\otimes\MB{e}_{-s_1}\otimes\MB{e}_{-s_2}\bigr)\Bigr>
\end{array}
$$
It is easy to verify that
$$
\begin{array}{l}
\Bigl<\MB{B}_1(x)\MB{B}_2(y),
\MB{f}^{r_1}\otimes\MB{f}^{r_2}\otimes\MB{e}_{-s_1}\otimes\MB{e}_{-s_2}\Bigr>=\\[4pt]
\hskip30mm=
\Bigl<\MB{B}_2(x)\MB{B}_1(y),\bigl(\R^*\bigr)^{(+,+)}_{2_+,1_+}\R^{(-,-)}_{1_-,2_-}
\bigl(\MB{f}^{r_1}\otimes\MB{f}^{r_2}\otimes\MB{e}_{-s_1}\otimes\MB{e}_{-s_2}\bigr)\Bigr>,
\end{array}
$$
where $\bigl(\R^*\bigr)^{(+,+)}_{2_+,1_+}=
\bigl(\MB{R}^*\bigr)^{(+,+)}_{2_+,1_+}(x,x)$,
is valid.

It follows from these relationships that it is enough to prove equalities
$$
\begin{array}{l}
g(u_k,x)f(u_1,x)
\bigl(\MB{R}^*\bigr)^{(+,+)}_{k^*_+,1^*_+}(x,u_1)
\bigl(\MB{R}^*\bigr)^{(+,+)}_{2,\ldots,k}(\vec{u}_1)
\MB{R}^{(-,-)}_{1_-,k_-}(u_1,x)
\MB{R}^{(-,-)}_{2,\ldots,k}(\vec{u}_1)\\[4pt]
\hskip60mm
\wh{\MB{R}}^{(+,+)}_{0,1_+^*}(x,u_1)
\wh{\BB{T}}^{(+)}_{k;0,2,\ldots,N+1}(\vec{u}_1)
\MB{R}^{(+,-)}_{0,1_-}(x,u_1)-\\[4pt]
\hskip25mm-
g(u_1,x)g(u_k,u_1)
\bigl(\R^*\bigr)^{(+,+)}_{k^*_+,1^*_+}
\bigl(\MB{R}^*\bigr)^{(+,+)}_{2,\ldots,k}(\vec{u}_1)
\R^{(-,-)}_{1_-,k_-}
\MB{R}^{(-,-)}_{2,\ldots,k}(\vec{u}_1)\\[4pt]
\hskip60mm
\wh{\R}^{(+,+)}_{0,1^*_+}
\wh{\BB{T}}^{(+)}_{k;0,2,\ldots,N+1}(\vec{u}_1)
\R^{(+,-)}_{0,1_-}=\\[6pt]
\hskip10mm=
g(u_k,x)f(u_1,u_k)
\bigl(\MB{R}^*\bigr)^{(+,+)}_{1,\ldots,k}(\vec{u})
\MB{R}^{(-,-)}_{1,\ldots,k}(\vec{u})
\wh{\BB{T}}^{(+)}_{k;0,1,\ldots,N+1}(\vec{u})\\[12pt]
g(x,u_k)f(x,u_1)
\bigl(\MB{R}^*\bigr)^{(+,+)}_{k^*_+,1^*_+}(x,u_1)
\bigl(\MB{R}^*\bigr)^{(+,+)}_{2,\ldots,k}(\vec{u}_1)
\MB{R}^{(-,-)}_{1_-,k_-}(u_1,x)
\MB{R}^{(-,-)}_{2,\ldots,k}(\vec{u}_1)\\[4pt]
\hskip60mm
\wh{\MB{R}}^{(-,+)}_{0,1_+^*}(x,u_1)
\wh{\BB{T}}^{(-)}_{k;0,2,\ldots,N+1}(\vec{u}_1)
\MB{R}^{(-,-)}_{0,1_-}(x,u_1)-\\[4pt]
\hskip25mm-
g(x,u_1)g(u_1,u_k)
\bigl(\R^*\bigr)^{(+,+)}_{k^*_+,1^*_+}
\bigl(\MB{R}^*\bigr)^{(+,+)}_{2,\ldots,k}(\vec{u}_1)
\R^{(-,-)}_{1_-,k_-}
\MB{R}^{(-,-)}_{2,\ldots,k}(\vec{u}_1)\\[4pt]
\hskip60mm
\wh{\R}^{(-,+)}_{0,1^*_+}
\wh{\BB{T}}^{(-)}_{k;0,2,\ldots,N+1}(\vec{u}_1)
\R^{(-,-)}_{0,1_-}=\\[6pt]
\hskip10mm=
g(x,u_k)f(u_k,u_1)
\bigl(\MB{R}^*\bigr)^{(+,+)}_{1,\ldots,k}(\vec{u})
\MB{R}^{(-,-)}_{1,\ldots,k}(\vec{u})
\wh{\BB{T}}^{(-)}_{k;0,1,\ldots,N+1}(\vec{u})\,,
\end{array}
$$

If we use the definitions of these operators, we get after a suitable arrangement
$$
\begin{array}{l}
\bigl(\MB{R}^*\bigr)^{(+,+)}_{k^*_+,1^*_+}(x,u_1)
\bigl(\MB{R}^*\bigr)^{(+,+)}_{2,\ldots,k}(\vec{u}_1)
\MB{R}^{(-,-)}_{1_-,k_-}(u_1,x)
\MB{R}^{(-,-)}_{2,\ldots,k}(\vec{u}_1)\\[4pt]
\hskip60mm
\wh{\MB{R}}^{(+,+)}_{0,1_+^*}(x,u_1)
\wh{\BB{T}}^{(+)}_{k;0,2,\ldots,N+1}(\vec{u}_1)
\MB{R}^{(+,-)}_{0,1_-}(x,u_1)=\\[6pt]
\hskip10mm=
\Bigl(\bigl(\MB{R}^*\bigr)^{(+,+)}_{k^*_+,1^*_+}(x,u_1)
\wh{\MB{R}}^{(+,+)}_{0,1_+^*}(x,u_1)\Bigr)
\Bigl(\bigl(\MB{R}^*\bigr)^{(+,+)}_{k^*_+,2^*_+}(u_k,u_2)
\wh{\MB{R}}^{(+,+)}_{0,2^*_+}(u_k,u_2)\Bigr)
\ldots\\[4pt]
\hskip30mm
\Bigl(\bigl(\MB{R}^*\bigr)^{(+,+)}_{k^*_+,(k-2)^*_+}(u_k,u_{k-2})
\wh{\MB{R}}^{(+,+)}_{0,(k-2)^*_+}(u_k,u_{k-2})\Bigr)\\[4pt]
\hskip40mm
\Bigl(\bigl(\MB{R}^*\bigr)^{(+,+)}_{k^*_+,(k-1)^*_+}(u_k,u_{k-1})
\wh{\MB{R}}^{(+,+)}_{0,(k-1)^*_+}(u_k,u_{k-1})
\wh{\R}^{(+,+)}_{0,k^*_+}\Bigr)\\[4pt]
\hskip30mm
\Bigl(\wh{\MB{R}}^{(+,+)}_{0,(k+1)^*_+}(u_k,u_{k+1})
\ldots
\wh{\MB{R}}^{(+,+)}_{0,(N+1)^*_+}(u_k,u_{N+1})\Bigr)
\MB{T}^{(+)}_0(u_k)\\[4pt]
\hskip30mm
\Bigl(\MB{R}^{(+,-)}_{0,(N+1)_-}(u_k,u_{N+1})
\ldots
\MB{R}^{(+,-)}_{0,(k+1)_-}(u_k,u_{k+1})\Bigr)\\[4pt]
\hskip30mm
\MB{R}^{(-,-)}_{1_-,k_-}(u_1,x)
\Bigl(\MB{R}^{(-,-)}_{2_-,k_-}(u_2,u_k)\ldots
\MB{R}^{(-,-)}_{(k-2)_-,k_-}(u_{k-2},u_k)\Bigr)\\[4pt]
\hskip40mm
\Bigl(\MB{R}^{(-,-)}_{(k-1)_-,k_-}(u_{k-1},u_k)
\R^{(+,-)}_{0,k_-}
\MB{R}^{(+,-)}_{0,(k-1)_-}(u_k,u_{k-1})\Bigr)\\[4pt]
\hskip30mm
\Bigl(\MB{R}^{(+,-)}_{0,(k-2)_-}(u_k,u_{k-2})
\ldots
\MB{R}^{(+,-)}_{0,2_-}(u_k,u_2)\Bigr)
\MB{R}^{(+,-)}_{0,1_-}(x,u_1)
\end{array}
$$
$$
\begin{array}{l}
\bigl(\R^*\bigr)^{(+,+)}_{k^*_+,1^*_+}
\bigl(\MB{R}^*\bigr)^{(+,+)}_{2,\ldots,k}(\vec{u}_1)
\R^{(-,-)}_{1_-,k_-}
\MB{R}^{(-,-)}_{2,\ldots,k}(\vec{u}_1)
\wh{\R}^{(+,+)}_{0,1^*_+}
\wh{\BB{T}}^{(+)}_{k;0,2,\ldots,N+1}(\vec{u}_1)
\R^{(+,-)}_{0,1_-}=\\[6pt]
\hskip10mm=
\Bigl(\bigl(\R^*\bigr)^{(+,+)}_{k^*_+,1^*_+}
\wh{\R}^{(+,+)}_{0,1^*_+}\Bigr)
\Bigl(\bigl(\MB{R}^*\bigr)^{(+,+)}_{k^*_+,2^*_+}(u_k,u_2)
\wh{\MB{R}}^{(+,+)}_{0,2^*_+}(u_k,u_2)\Bigr)\ldots\\[4pt]
\hskip30mm
\Bigl(\bigl(\MB{R}^*\bigr)^{(+,+)}_{k^*_+,(k-2)^*_+}(u_k,u_{k-2})
\wh{\MB{R}}^{(+,+)}_{0,(k-2)^*_+}(u_k,u_{k-2})\Bigr)\\[4pt]
\hskip40mm
\Bigl(\bigl(\MB{R}^*\bigr)^{(+,+)}_{k^*_+,(k-1)^*_+}(u_k,u_{k-1})
\wh{\MB{R}}^{(+,+)}_{0,(k-1)^*_+}(u_k,u_{k-1})
\wh{\R}^{(+,+)}_{0,k^*_+}\Bigr)\\[4pt]
\hskip30mm
\Bigl(\wh{\MB{R}}^{(+,+)}_{0,(k+1)^*_+}(u_k,u_{k+1})\ldots
\wh{\MB{R}}^{(+,+)}_{0,(N+1)^*_+}(u_k,u_{N+1})\Bigr)
\MB{T}^{(+)}_0(u_k)\\[4pt]
\hskip30mm
\Bigl(\MB{R}^{(+,-)}_{0,(N+1)_-}(u_k,u_{N+1})\ldots
\MB{R}^{(+,-)}_{0,(k+1)_-}(u_k,u_{k+1})\Bigr)\\[4pt]
\hskip30mm
\R^{(-,-)}_{1_-,k_-}
\Bigl(\MB{R}^{(-,-)}_{2_-,k_-}(u_2,u_k)\ldots
\MB{R}^{(-,-)}_{(k-2)_-,k_-}(u_{k-2},u_k)\Bigr)\\[4pt]
\hskip40mm
\Bigl(\MB{R}^{(-,-)}_{(k-1)_-,k_-}(u_{k-1},u_k)\R^{(+,-)}_{0,k_-}
\MB{R}^{(+,-)}_{0,(k-1)_-}(u_k,u_{k-1})\Bigr)\\[4pt]
\hskip30mm
\Bigl(\MB{R}^{(+,-)}_{0,(k-2)_-}(u_k,u_{k-2})
\ldots
\MB{R}^{(+,-)}_{0,2_-}(u_k,u_2)\Bigr)
\R^{(+,-)}_{0,1_-}
\end{array}
$$
$$
\begin{array}{l}
\bigl(\MB{R}^*\bigr)^{(+,+)}_{1,\ldots,k}(\vec{u})
\MB{R}^{(-,-)}_{1,\ldots,k}(\vec{u})
\wh{\BB{T}}^{(+)}_{k;0,1,\ldots,N+1}(\vec{u})=\\[6pt]
\hskip10mm=
\Bigl(\bigl(\MB{R}^*\bigr)^{(+,+)}_{k^*_+,1^*_+}(u_k,u_1)
\wh{\MB{R}}^{(+,+)}_{0,1^*_+}(u_k,u_1)\Bigr)
\ldots
\Bigl(\bigl(\MB{R}^*\bigr)^{(+,+)}_{k^*_+,(k-2)^*_+}(u_k,u_{k-2})
\wh{\MB{R}}^{(+,+)}_{0,(k-2)^*_+}(u_k,u_{k-2})\Bigr)\\[4pt]
\hskip40mm
\Bigl(\bigl(\MB{R}^*\bigr)^{(+,+)}_{k^*_+,(k-1)^*_+}(u_k,u_{k-1})
\wh{\MB{R}}^{(+,+)}_{0,(k-1)^*_+}(u_k,u_{k-1})
\wh{\R}^{(+,+)}_{0,k^*_+}\Bigr)\\[4pt]
\hskip30mm
\Bigl(\wh{\MB{R}}^{(+,+)}_{0,(k+1)^*_+}(u_k,u_{k+1})
\ldots
\wh{\MB{R}}^{(+,+)}_{0,(N+1)^*_+}(u_k,u_{N+1})\Bigr)
\MB{T}^{(+)}_0(u_k)\\[4pt]
\hskip30mm
\Bigl(\MB{R}^{(+,-)}_{0,(N+1)_-}(u_k,u_{N+1})
\ldots
\MB{R}^{(+,-)}_{0,(k+1)_-}(u_k,u_{k+1})\Bigr)\\[4pt]
\hskip30mm
\Bigl(\MB{R}^{(-,-)}_{1_-,k_-}(u_1,u_k)
\ldots
\MB{R}^{(-,-)}_{(k-2)_-,k_-}(u_{k-2},u_k)\Bigr)\\[4pt]
\hskip40mm
\Bigl(\MB{R}^{(-,-)}_{(k-1)_-,k_-}(u_{k-1},u_k)
\R^{(+,-)}_{0,k_-}
\MB{R}^{(+,-)}_{0,(k-1)_-}(u_k,u_{k-1})\Bigr)\\[4pt]
\hskip30mm
\Bigl(\MB{R}^{(+,-)}_{0,(k-2)_-}(u_k,u_{k-2})\ldots
\MB{R}^{(+,-)}_{0,1_-}(u_k,u_1)\Bigr)
\end{array}
$$
$$
\begin{array}{l}
\bigl(\MB{R}^*\bigr)^{(+,+)}_{k^*_+,1^*_+}(x,u_1)
\bigl(\MB{R}^*\bigr)^{(+,+)}_{2,\ldots,k}(\vec{u}_1)
\MB{R}^{(-,-)}_{1_-,k_-}(u_1,x)
\MB{R}^{(-,-)}_{2,\ldots,k}(\vec{u}_1)\\[4pt]
\hskip60mm
\wh{\MB{R}}^{(-,+)}_{0,1_+^*}(x,u_1)
\wh{\BB{T}}^{(-)}_{k;0,2,\ldots,N+1}(\vec{u}_1)
\MB{R}^{(-,-)}_{0,1_-}(x,u_1)=\\[6pt]
\hskip10mm=
\Bigl(\bigl(\MB{R}^*\bigr)^{(+,+)}_{k^*_+,1^*_+}(x,u_1)
\wh{\MB{R}}^{(-,+)}_{0,1_+^*}(x,u_1)\Bigr)
\Bigl(\bigl(\MB{R}^*\bigr)^{(+,+)}_{k^*_+,2^*_+}(u_k,u_2)
\wh{\MB{R}}^{(-,+)}_{0,2^*_+}(u_k,u_2)\Bigr)
\ldots\\[4pt]
\hskip30mm
\Bigl(\bigl(\MB{R}^*\bigr)^{(+,+)}_{k^*_+,(k-2)^*_+}(u_k,u_{k-2})
\wh{\MB{R}}^{(-,+)}_{0,(k-2)^*_+}(u_k,u_{k-2})\Bigr)\\[4pt]
\hskip40mm
\Bigl(\bigl(\MB{R}^*\bigr)^{(+,+)}_{k^*_+,(k-1)^*_+}(u_k,u_{k-1})
\wh{\MB{R}}^{(-,+)}_{0,(k-1)^*_+}(u_k,u_{k-1})
\wh{\R}^{(-,+)}_{0,k^*_+}\Bigr)\\[4pt]
\hskip30mm
\Bigl(\wh{\MB{R}}^{(-,+)}_{0,(k+1)^*_+}(u_k,u_{k+1})\ldots
\wh{\MB{R}}^{(-,+)}_{0,(N+1)^*_+}(u_k,u_{N+1})\Bigr)
\MB{T}^{(-)}_0(u_k)\\[4pt]
\hskip30mm
\Bigl(\MB{R}^{(-,-)}_{0,(N+1)_-}(u_k,u_{N+1})\ldots
\MB{R}^{(-,-)}_{0,(k+1)_-}(u_k,u_{k+1})\Bigr)\\[4pt]
\hskip30mm
\MB{R}^{(-,-)}_{1_-,k_-}(u_1,x)
\Bigl(\MB{R}^{(-,-)}_{2_-,k_-}(u_2,u_k)
\ldots
\MB{R}^{(-,-)}_{(k-2)_-,k_-}(u_{k-2},u_k)\Bigr)\\[4pt]
\hskip40mm
\Bigl(\MB{R}^{(-,-)}_{(k-1)_-,k_-}(u_{k-1},u_k)
\R^{(-,-)}_{0,k_-}
\MB{R}^{(-,-)}_{0,(k-1)_-}(u_k,u_{k-1})\Bigr)\\[4pt]
\hskip30mm
\Bigl(\MB{R}^{(-,-)}_{0,(k-2)_-}(u_k,u_{k-2})
\ldots
\MB{R}^{(-,-)}_{0,2_-}(u_k,u_2)\Bigr)
\MB{R}^{(-,-)}_{0,1_-}(x,u_1)
\end{array}
$$
$$
\begin{array}{l}
\bigl(\R^*\bigr)^{(+,+)}_{k^*_+,1^*_+}
\bigl(\MB{R}^*\bigr)^{(+,+)}_{2,\ldots,k}(\vec{u}_1)
\R^{(-,-)}_{1_-,k_-}
\MB{R}^{(-,-)}_{2,\ldots,k}(\vec{u}_1)
\wh{\R}^{(-,+)}_{0,1^*_+}
\wh{\BB{T}}^{(-)}_{k;0,2,\ldots,N+1}(\vec{u}_1)
\R^{(-,-)}_{0,1_-}=\\[6pt]
\hskip10mm=
\Bigl(\bigl(\R^*\bigr)^{(+,+)}_{k^*_+,1^*_+}
\wh{\R}^{(-,+)}_{0,1^*_+}\Bigr)
\Bigl(\bigl(\MB{R}^*\bigr)^{(+,+)}_{k^*_+,2^*_+}(u_k,u_2)
\wh{\MB{R}}^{(-,+)}_{0,2^*_+}(u_k,u_2)\Bigr)
\ldots\\[4pt]
\hskip30mm
\Bigl(\bigl(\MB{R}^*\bigr)^{(+,+)}_{k^*_+,(k-2)^*_+}(u_k,u_{k-2})
\wh{\MB{R}}^{(-,+)}_{0,(k-2)^*_+}(u_k,u_{k-2})\Bigr)\\[4pt]
\hskip40mm
\Bigl(\bigl(\MB{R}^*\bigr)^{(+,+)}_{k^*_+,(k-1)^*_+}(u_k,u_{k-1})
\wh{\MB{R}}^{(-,+)}_{0,(k-1)^*_+}(u_k,u_{k-1})
\wh{\R}^{(-,+)}_{0,k^*_+}\Bigr)\\[4pt]
\hskip30mm
\Bigl(\wh{\MB{R}}^{(-,+)}_{0,(k+1)^*_+}(u_k,u_{k+1})
\ldots
\wh{\MB{R}}^{(-,+)}_{0,(N+1)^*_+}(u_k,u_{N+1})\Bigr)
\MB{T}^{(-)}_0(u_k)\\[4pt]
\hskip30mm
\Bigl(\MB{R}^{(-,-)}_{0,(N+1)_-}(u_k,u_{N+1})
\ldots
\MB{R}^{(-,-)}_{0,(k+1)_-}(u_k,u_{k+1})\Bigr)\\[4pt]
\hskip30mm
\R^{(-,-)}_{1_-,k_-}
\Bigl(\MB{R}^{(-,-)}_{2_-,k_-}(u_2,u_k)
\ldots
\MB{R}^{(-,-)}_{(k-2)_-,k_-}(u_{k-2},u_k)\Bigr)\\[4pt]
\hskip40mm
\Bigl(\MB{R}^{(-,-)}_{(k-1)_-,k_-}(u_{k-1},u_k)
\R^{(-,-)}_{0,k_-}
\MB{R}^{(-,-)}_{0,(k-1)_-}(u_k,u_{k-1})\Bigr)\\[4pt]
\hskip30mm
\Bigl(\MB{R}^{(-,-)}_{0,(k-2)_-}(u_k,u_{k-2})
\ldots
\MB{R}^{(-,-)}_{0,2_-}(u_k,u_2)\Bigr)
\R^{(-,-)}_{0,1_-}
\end{array}
$$
$$
\begin{array}{l}
\bigl(\MB{R}^*\bigr)^{(+,+)}_{1,\ldots,k}(\vec{u})
\MB{R}^{(-,-)}_{1,\ldots,k}(\vec{u})
\wh{\BB{T}}^{(-)}_{k;0,1,\ldots,N+1}(\vec{u})=\\[6pt]
\hskip10mm=
\Bigl(\bigl(\MB{R}^*\bigr)^{(+,+)}_{k^*_+,1^*_+}(u_k,u_1)
\wh{\MB{R}}^{(-,+)}_{0,1^*_+}(u_k,u_1)\Bigr)
\ldots
\Bigl(\bigl(\MB{R}^*\bigr)^{(+,+)}_{k^*_+,(k-2)^*_+}(u_k,u_{k-2})
\wh{\MB{R}}^{(-,+)}_{0,(k-2)^*_+}(u_k,u_{k-2})\Bigr)\\[4pt]
\hskip40mm
\Bigl(\bigl(\MB{R}^*\bigr)^{(+,+)}_{k^*_+,(k-1)^*_+}(u_k,u_{k-1})
\wh{\MB{R}}^{(-,+)}_{0,(k-1)^*_+}(u_k,u_{k-1})
\wh{\R}^{(-,+)}_{0,k^*_+}\Bigr)\\[4pt]
\hskip30mm
\Bigl(\wh{\MB{R}}^{(-,+)}_{0,(k+1)^*_+}(u_k,u_{k+1})
\ldots
\wh{\MB{R}}^{(-,+)}_{0,(N+1)^*_+}(u_k,u_{N+1})\Bigr)
\MB{T}^{(-)}_0(u_k)\\[4pt]
\hskip30mm
\Bigl(\MB{R}^{(-,-)}_{0,(N+1)_-}(u_k,u_{N+1})
\ldots
\MB{R}^{(-,-)}_{0,(k+1)_-}(u_k,u_{k+1})\Bigr)\\[4pt]
\hskip30mm
\Bigl(\MB{R}^{(-,-)}_{1_-,k_-}(u_1,u_k)
\ldots
\MB{R}^{(-,-)}_{(k-2)_-,k_-}(u_{k-2},u_k)\Bigr)\\[4pt]
\hskip40mm
\Bigl(\MB{R}^{(-,-)}_{(k-1)_-,k_-}(u_{k-1},u_k)
\R^{(-,-)}_{0,k_-}
\MB{R}^{(-,-)}_{0,(k-1)_-}(u_k,u_{k-1})\Bigr)\\[4pt]
\hskip30mm
\Bigl(\MB{R}^{(-,-)}_{0,(k-2)_-}(u_k,u_{k-2})
\ldots
\MB{R}^{(-,-)}_{0,1_-}(u_k,u_1)\Bigr)
\end{array}
$$

By direct calculation it is possible to show that the relations
$$
\begin{array}{l}
\bigl(\MB{R}^*\bigr)^{(+,+)}_{2^*_+,1^*_+}(u_2,u_1)
\wh{\MB{R}}^{(+,+)}_{0,1^*_+}(u_2,u_1)
\wh{\R}^{(+,+)}_{0,2^*_+}=
\wh{\R}^{(+,+)}_{0,2^*_+}
\wh{\MB{R}}^{(+,+)}_{0,1^*_+}(u_2,u_1)
\bigl(\MB{R}^*\bigr)^{(+,+)}_{2^*_+,1^*_+}(u_2,u_1)=
\wh{\R}^{(+,+)}_{0,2^*_+}\\[6pt]
\MB{R}^{(-,-)}_{1_-,2_-}(u_1,u_2)
\R^{(+,-)}_{0,2_-}
\MB{R}^{(+,-)}_{0,1_-}(u_2,u_1)=
\MB{R}^{(+,-)}_{0,1_-}(u_2,u_1)
\R^{(+,-)}_{0,2_-}
\MB{R}^{(-,-)}_{1_-,2_-}(u_1,u_2)\\[6pt]
\bigl(\MB{R}^*\bigr)^{(+,+)}_{2^*_+,1^*_+}(u_2,u_1)
\wh{\MB{R}}^{(-,+)}_{0,1^*_+}(u_2,u_1)
\wh{\R}^{(-,+)}_{0,2^*_+}=
\wh{\R}^{(-,+)}_{0,2^*_+}
\wh{\MB{R}}^{(-,+)}_{0,1^*_+}(u_2,u_1)
\bigl(\MB{R}^*\bigr)^{(+,+)}_{2^*_+,1^*_+}(u_2,u_1)\\[6pt]
\MB{R}^{(-,-)}_{1_-,2_-}(u_1,u_2)
\R^{(-,-)}_{0,2_-}
\MB{R}^{(-,-)}_{0,1_-}(u_2,u_1)=
\MB{R}^{(-,-)}_{0,1_-}(u_2,u_1)
\R^{(-,-)}_{0,2_-}
\MB{R}^{(-,-)}_{1_-,2_-}(u_1,u_2)=\R^{(-,-)}_{0,2_-}\,,
\end{array}
$$
are valid. By using these relations we can write
$$
\begin{array}{l}
\bigl(\MB{R}^*\bigr)^{(+,+)}_{k^*_+,1^*_+}(x,u_1)
\bigl(\MB{R}^*\bigr)^{(+,+)}_{2,\ldots,k}(\vec{u}_1)
\MB{R}^{(-,-)}_{1_-,k_-}(u_1,x)
\MB{R}^{(-,-)}_{2,\ldots,k}(\vec{u}_1)\\[4pt]
\hskip60mm
\wh{\MB{R}}^{(+,+)}_{0,1_+^*}(x,u_1)
\wh{\BB{T}}^{(+)}_{k;0,2,\ldots,N+1}(\vec{u}_1)
\MB{R}^{(+,-)}_{0,1_-}(x,u_1)=\\[6pt]
\hskip10mm=
\wh{\R}^{(+,+)}_{0,k^*_+}
\Bigl(\wh{\MB{R}}^{(+,+)}_{0,(k+1)^*_+}(u_k,u_{k+1})
\ldots
\wh{\MB{R}}^{(+,+)}_{0,(N+1)^*_+}(u_k,u_{N+1})\Bigr)
\MB{T}^{(+)}_0(u_k)\\[4pt]
\hskip40mm
\Bigl(\MB{R}^{(+,-)}_{0,(N+1)_-}(u_k,u_{N+1})
\ldots
\MB{R}^{(+,-)}_{0,(k+1)_-}(u_k,u_{k+1})\Bigr)\\[4pt]
\hskip40mm
\Bigl(\MB{R}^{(+,-)}_{0,(k-1)_-}(u_k,u_{k-1})
\ldots
\MB{R}^{(+,-)}_{0,(2)_-}(u_k,u_2)\Bigr)\\[4pt]
\hskip30mm
\Bigl(\MB{R}^{(-,-)}_{1_-,k_-}(u_1,x)
\R^{(+,-)}_{0,k_-}
\MB{R}^{(+,-)}_{0,1_-}(x,u_1)\Bigr)\\[4pt]
\hskip40mm
\Bigl(\MB{R}^{(-,-)}_{2_-,k_-}(u_2,u_k)
\ldots
\MB{R}^{(-,-)}_{(k-1)_-,k_-}(u_{k-1},u_k)\Bigr)
\end{array}
$$
$$
\begin{array}{l}
\bigl(\R^*\bigr)^{(+,+)}_{k^*_+,1^*_+}
\bigl(\MB{R}^*\bigr)^{(+,+)}_{2,\ldots,k}(\vec{u}_1)
\R^{(-,-)}_{1_-,k_-}
\MB{R}^{(-,-)}_{2,\ldots,k}(\vec{u}_1)
\wh{\R}^{(+,+)}_{0,1^*_+}
\wh{\BB{T}}^{(+)}_{k;0,2,\ldots,N+1}(\vec{u}_1)
\R^{(+,-)}_{0,1_-}=\\[6pt]
\hskip10mm=
\wh{\R}^{(+,+)}_{0,k^*_+}
\Bigl(\wh{\MB{R}}^{(+,+)}_{0,(k+1)^*_+}(u_k,u_{k+1})\ldots
\wh{\MB{R}}^{(+,+)}_{0,(N+1)^*_+}(u_k,u_{N+1})\Bigr)
\MB{T}^{(+)}_0(u_k)\\[4pt]
\hskip40mm
\Bigl(\MB{R}^{(+,-)}_{0,(N+1)_-}(u_k,u_{N+1})\ldots
\MB{R}^{(+,-)}_{0,(k+1)_-}(u_k,u_{k+1})\Bigr)\\[4pt]
\hskip40mm
\Bigl(\MB{R}^{(+,-)}_{0,(k-1)_-}(u_k,u_{k-1})
\ldots
\MB{R}^{(+,-)}_{0,2_-}(u_k,u_2)\Bigr)\\[4pt]
\hskip30mm
\Bigl(\R^{(-,-)}_{1_-,k_-}
\R^{(+,-)}_{0,k_-}
\R^{(+,-)}_{0,1_-}\Bigr)\\[4pt]
\hskip40mm
\Bigl(\MB{R}^{(-,-)}_{2_-,k_-}(u_2,u_k)
\ldots
\MB{R}^{(-,-)}_{(k-1)_-,k_-}(u_{k-1},u_k)\Bigr)
\end{array}
$$
$$
\begin{array}{l}
\bigl(\MB{R}^*\bigr)^{(+,+)}_{1,\ldots,k}(\vec{u})
\MB{R}^{(-,-)}_{1,\ldots,k}(\vec{u})
\wh{\BB{T}}^{(+)}_{k;0,1,\ldots,N+1}(\vec{u})=\\[6pt]
\hskip10mm=
\wh{\R}^{(+,+)}_{0,k^*_+}
\Bigl(\wh{\MB{R}}^{(+,+)}_{0,(k+1)^*_+}(u_k,u_{k+1})
\ldots
\wh{\MB{R}}^{(+,+)}_{0,(N+1)^*_+}(u_k,u_{N+1})\Bigr)
\MB{T}^{(+)}_0(u_k)\\[4pt]
\hskip40mm
\Bigl(\MB{R}^{(+,-)}_{0,(N+1)_-}(u_k,u_{N+1})
\ldots
\MB{R}^{(+,-)}_{0,(k+1)_-}(u_k,u_{k+1})\Bigr)\\[4pt]
\hskip40mm
\Bigl(\MB{R}^{(+,-)}_{0,(k-1)_-}(u_k,u_{k-1})
\ldots
\MB{R}^{(+,-)}_{0,2_-}(u_k,u_2)\Bigr)\\[4pt]
\hskip30mm
\Bigl(\MB{R}^{(-,-)}_{1_-,k_-}(u_1,u_k)
\R^{(+,-)}_{0,k_-}
\MB{R}^{(+,-)}_{0,1_-}(u_k,u_1)\Bigr)\\[4pt]
\hskip40mm
\Bigl(\MB{R}^{(-,-)}_{2_-,k_-}(u_2,u_k)
\ldots
\MB{R}^{(-,-)}_{(k-1)_-,k_-}(u_{k-1},u_k)\Bigr)
\end{array}
$$

$$
\begin{array}{l}
\bigl(\MB{R}^*\bigr)^{(+,+)}_{k^*_+,1^*_+}(x,u_1)
\bigl(\MB{R}^*\bigr)^{(+,+)}_{2,\ldots,k}(\vec{u}_1)
\MB{R}^{(-,-)}_{1_-,k_-}(u_1,x)
\MB{R}^{(-,-)}_{2,\ldots,k}(\vec{u}_1)\\[4pt]
\hskip60mm
\wh{\MB{R}}^{(-,+)}_{0,1_+^*}(x,u_1)
\wh{\BB{T}}^{(-)}_{k;0,2,\ldots,N+1}(\vec{u}_1)
\MB{R}^{(-,-)}_{0,1_-}(x,u_1)=\\[6pt]
\hskip10mm=
\Bigl(\bigl(\MB{R}^*\bigr)^{(+,+)}_{k^*_+,1^*_+}(x,u_1)
\wh{\MB{R}}^{(-,+)}_{0,1_+^*}(x,u_1)
\wh{\R}^{(-,+)}_{0,k^*_+}\Bigr)
\Bigl(\wh{\MB{R}}^{(-,+)}_{0,2^*_+}(u_k,u_2)
\ldots
\wh{\MB{R}}^{(-,+)}_{0,(k-1)^*_+}(u_k,u_{k-1})\Bigr)\\[4pt]
\hskip40mm
\Bigl(\wh{\MB{R}}^{(-,+)}_{0,(k+1)^*_+}(u_k,u_{k+1})\ldots
\wh{\MB{R}}^{(-,+)}_{0,(N+1)^*_+}(u_k,u_{N+1})\Bigr)
\MB{T}^{(-)}_0(u_k)\\[4pt]
\hskip40mm
\Bigl(\MB{R}^{(-,-)}_{0,(N+1)_-}(u_k,u_{N+1})\ldots
\MB{R}^{(-,-)}_{0,(k+1)_-}(u_k,u_{k+1})\Bigr)\R^{(-,-)}_{0,k_-}\\[4pt]
\hskip40mm
\Bigl(\bigl(\MB{R}^*\bigr)^{(+,+)}_{k^*_+,2^*_+}(u_k,u_2)
\ldots
\bigl(\MB{R}^*\bigr)^{(+,+)}_{k^*_+,(k-1)^*_+}(u_k,u_{k-1})\Bigr)\\[6pt]
%
%%%
%
\bigl(\R^*\bigr)^{(+,+)}_{k^*_+,1^*_+}
\bigl(\MB{R}^*\bigr)^{(+,+)}_{2,\ldots,k}(\vec{u}_1)
\R^{(-,-)}_{1_-,k_-}
\MB{R}^{(-,-)}_{2,\ldots,k}(\vec{u}_1)
\wh{\R}^{(-,+)}_{0,1^*_+}
\wh{\BB{T}}^{(-)}_{k;0,2,\ldots,N+1}(\vec{u}_1)
\R^{(-,-)}_{0,1_-}=\\[6pt]
\hskip10mm=
\Bigl(\bigl(\R^*\bigr)^{(+,+)}_{k^*_+,1^*_+}
\wh{\R}^{(-,+)}_{0,1^*_+}
\wh{\R}^{(-,+)}_{0,k^*_+}\Bigr)
\Bigl(\wh{\MB{R}}^{(-,+)}_{0,2^*_+}(u_k,u_2)
\ldots
\wh{\MB{R}}^{(-,+)}_{0,(k-1)^*_+}(u_k,u_{k-1})\Bigr)\\[4pt]
\hskip40mm
\Bigl(\wh{\MB{R}}^{(-,+)}_{0,(k+1)^*_+}(u_k,u_{k+1})
\ldots
\wh{\MB{R}}^{(-,+)}_{0,(N+1)^*_+}(u_k,u_{N+1})\Bigr)
\MB{T}^{(-)}_0(u_k)\\[4pt]
\hskip40mm
\Bigl(\MB{R}^{(-,-)}_{0,(N+1)_-}(u_k,u_{N+1})
\ldots
\MB{R}^{(-,-)}_{0,(k+1)_-}(u_k,u_{k+1})\Bigr)\R^{(-,-)}_{0,k_-}\\[4pt]
\hskip40mm
\Bigl(\bigl(\MB{R}^*\bigr)^{(+,+)}_{k^*_+,2^*_+}(u_k,u_2)
\ldots
\bigl(\MB{R}^*\bigr)^{(+,+)}_{k^*_+,(k-1)^*_+}(u_k,u_{k-1})\Bigr)\\[6pt]
%
%%%
%
\bigl(\MB{R}^*\bigr)^{(+,+)}_{1,\ldots,k}(\vec{u})
\MB{R}^{(-,-)}_{1,\ldots,k}(\vec{u})
\wh{\BB{T}}^{(-)}_{k;0,1,\ldots,N+1}(\vec{u})=\\[6pt]
\hskip10mm=
\Bigl(\bigl(\MB{R}^*\bigr)^{(+,+)}_{k^*_+,1^*_+}(u_k,u_1)
\wh{\MB{R}}^{(-,+)}_{0,1^*_+}(u_k,u_1)
\wh{\R}^{(-,+)}_{0,k^*_+}\Bigr)
\Bigl(\wh{\MB{R}}^{(-,+)}_{0,2^*_+}(u_k,u_2)
\ldots
\wh{\MB{R}}^{(-,+)}_{0,(k-1)^*_+}(u_k,u_{k-1})\Bigr)\\[4pt]
\hskip40mm
\Bigl(\wh{\MB{R}}^{(-,+)}_{0,(k+1)^*_+}(u_k,u_{k+1})
\ldots
\wh{\MB{R}}^{(-,+)}_{0,(N+1)^*_+}(u_k,u_{N+1})\Bigr)
\MB{T}^{(-)}_0(u_k)\\[4pt]
\hskip40mm
\Bigl(\MB{R}^{(-,-)}_{0,(N+1)_-}(u_k,u_{N+1})
\ldots
\MB{R}^{(-,-)}_{0,(k+1)_-}(u_k,u_{k+1})\Bigr)
\R^{(-,-)}_{0,k_-}\\[4pt]
\hskip40mm
\Bigl(\bigl(\MB{R}^*\bigr)^{(+,+)}_{k^*_+,2^*_+}(u_k,u_2)
\ldots
\bigl(\MB{R}^*\bigr)^{(+,+)}_{k^*_+,(k-1)^*_+}(u_k,u_{k-1})\Bigr)
\end{array}
$$
From these expressions it follows that it is enough to prove
that for any $u_k\in\ol{u}_1$
$$
\begin{array}{l}
g(u_k,x)f(u_1,x)
\MB{R}^{(-,-)}_{1_-,k_-}(u_1,x)\R^{(+,-)}_{0,k_-}\MB{R}^{(+,-)}_{0,1_-}(x,u_1)-\\[4pt]
\hskip40mm-
g(u_1,x)g(u_k,u_1)
\R^{(-,-)}_{1_-,k_-}\R^{(+,-)}_{0,k_-}\R^{(+,-)}_{0,1_-}=\\[4pt]
\hskip55mm=
g(u_k,x)f(u_1,u_k)
\MB{R}^{(-,-)}_{1_-,k_-}(u_1,u_k)\R^{(+,-)}_{0,k_-}\MB{R}^{(+,-)}_{0,1_-}(u_k,u_1)\\[9pt]
g(x,u_k)f(x,u_1)
\bigl(\MB{R}^*\bigr)^{(+,+)}_{k^*_+,1^*_+}(x,u_1)
\wh{\MB{R}}^{(-,+)}_{0,1_+^*}(x,u_1)\wh{\R}^{(-,+)}_{0,k^*_+}-\\[4pt]
\hskip40mm-
g(x,u_1)g(u_1,u_k)
\bigl(\R^*\bigr)^{(+,+)}_{k^*_+,1^*_+}\wh{\R}^{(-,+)}_{0,1^*_+}\wh{\R}^{(-,+)}_{0,k^*_+}=\\[4pt]
\hskip55mm=
g(x,u_k)f(u_k,u_1)
\bigl(\MB{R}^*\bigr)^{(+,+)}_{k^*_+,1^*_+}(u_k,u_1)
\wh{\MB{R}}^{(-,+)}_{0,1^*_+}(u_k,u_1)
\wh{\R}^{(-,+)}_{0,k^*_+}\,,
\end{array}
$$
are valid. However, the last statements can be easily verified by direct calculation.
\qed

\subsection*{A.6. The proof of Lemma 7}

Since we have
$$
\begin{array}{l}
\wh{\MB{T}}^{(\epsilon_0)}_{0,1,\ldots,N}(x,\vec{u})
\wh{\MB{T}}^{(\epsilon_{0'})}_{0',1,\ldots,N}(y;\vec{u})=\\[4pt]
\hskip5mm=
\wh{\MB{R}}^{(\epsilon_0,+)}_{0;1^*_+,\ldots,N^*_+}(x;\vec{u})
\MB{T}^{(\epsilon_0)}_0(x)
\MB{R}^{(\epsilon_0,-)}_{0';1_-,\ldots,N_-}(x;\vec{u})
\wh{\MB{R}}^{(\epsilon_{0'},+)}_{0;1^*_+,\ldots,N^*_+}(y;\vec{u})
\MB{T}^{(\epsilon_{0'})}_{0'}(y)
\MB{R}^{(\epsilon_{0'},-)}_{0';1_-,\ldots,N_-}(y;\vec{u})=\\[4pt]
\hskip5mm=
\wh{\MB{R}}^{(\epsilon_0,+)}_{0;1^*_+,\ldots,N^*_+}(x;\vec{u})
\wh{\MB{R}}^{(\epsilon_{0'},+)}_{0;1^*_+,\ldots,N^*_+}(y;\vec{u})
\MB{T}^{(\epsilon_0)}_0(x)\MB{T}^{(\epsilon_{0'})}_{0'}(y)
\MB{R}^{(\epsilon_0,-)}_{0';1_-,\ldots,N_-}(x;\vec{u})
\MB{R}^{(\epsilon_{0'},-)}_{0';1_-,\ldots,N_-}(y;\vec{u})
\end{array}
$$
it is enough to show that the relations
$$
\begin{array}{l}
\MB{R}^{(\epsilon_0,\epsilon_{0'})}_{0,0'}(x,y)
\wh{\MB{R}}^{(\epsilon_0,+)}_{0;1^*_+,\ldots,N^*_+}(x;\vec{u})
\wh{\MB{R}}^{(\epsilon_{0'},+)}_{0;1^*_+,\ldots,N^*_+}(y;\vec{u})=\\[4pt]
\hskip50mm=
\wh{\MB{R}}^{(\epsilon_{0'},+)}_{0;1^*_+,\ldots,N^*_+}(y;\vec{u})
\wh{\MB{R}}^{(\epsilon_0,+)}_{0;1^*_+,\ldots,N^*_+}(x;\vec{u})
\MB{R}^{(\epsilon_0,\epsilon_{0'})}_{0,0'}(x,y)\\[6pt]
\MB{R}^{(\epsilon_0,\epsilon_{0'})}_{0,0'}(x,y)
\tilde{\MB{T}}^{(\epsilon_0)}_0(x)\tilde{\MB{T}}^{(\epsilon_{0'})}_{0'}(y)=
\tilde{\MB{T}}^{(\epsilon_{0'})}_{0'}(y)\tilde{\MB{T}}^{(\epsilon_0)}_0(x)
\MB{R}^{(\epsilon_0,\epsilon_{0'})}_{0,0'}(x,y)\\[6pt]
\MB{R}^{(\epsilon_0,\epsilon_{0'})}_{0,0'}(x,y)
\MB{R}^{(\epsilon_0,-)}_{0';1_-,\ldots,N_-}(x;\vec{u})
\MB{R}^{(\epsilon_{0'},-)}_{0';1_-,\ldots,N_-}(y;\vec{u})=\\[4pt]
\hskip50mm=
\MB{R}^{(\epsilon_{0'},-)}_{0';1_-,\ldots,N_-}(y;\vec{u})
\MB{R}^{(\epsilon_0,-)}_{0';1_-,\ldots,N_-}(x;\vec{u})
\MB{R}^{(\epsilon_0,\epsilon_{0'})}_{0,0'}(x,y)
\end{array}
$$
are true.

The second relation is the RTT--equation for $\MB{T}^{(\epsilon)}(x)$
restricted to the subspace $\MC{W}_0$.
Therefore, according to Lemma 3, it is valid.

If we write
$$
\begin{array}{l}
\MB{R}^{(\epsilon_0,-)}_{0';1_-,\ldots,N_-}(x;\vec{u})
\MB{R}^{(\epsilon_{0'},-)}_{0';1_-,\ldots,N_-}(y;\vec{u})=\\[4pt]
\hskip10mm=
\MB{R}^{(\epsilon_0,-)}_{0,N_-}(x,u_N)\ldots
\MB{R}^{(\epsilon_0,-)}_{0,1_-}(x,u_1)
\MB{R}^{(\epsilon_{0'},-)}_{0',N_-}(y,u_N)\ldots
\MB{R}^{(\epsilon_{0'},-)}_{0',1_-}(y,u_1)=\\[6pt]
\hskip10mm=
\Bigl(\MB{R}^{(\epsilon_0,-)}_{0,N_-}(x,u_N)
\MB{R}^{(\epsilon_{0'},-)}_{0',N_-}(y,u_N)\Bigr)
\ldots
\Bigl(\MB{R}^{(\epsilon_0,-)}_{0,1_-}(x,u_1)
\MB{R}^{(\epsilon_{0'},-)}_{0',1_-}(y,u_1)\Bigr),
\end{array}
$$
the last relation from the Yang--Baxter equation for the
R--matrix $\tilde{\MB{R}}(x,y)$ follows.

Similarly, we obtain
$$
\begin{array}{l}
\wh{\MB{R}}^{(\epsilon_0,+)}_{0;1^*_+,\ldots,N^*_+}(x;\vec{u})
\wh{\MB{R}}^{(\epsilon_{0'},+)}_{0;1^*_+,\ldots,N^*_+}(y;\vec{u})=\\[6pt]
\hskip10mm=
\Bigl(\wh{\MB{R}}^{(\epsilon_0,+)}_{0,1_+^*}(x,u_1)
\wh{\MB{R}}^{(\epsilon_{0'},+)}_{0',1_+^*}(y,u_1)\Bigr)\ldots
\Bigl(\wh{\MB{R}}^{(\epsilon_0,+)}_{0,N_+^*}(x,u_N)
\wh{\MB{R}}^{(\epsilon_{0'},+)}_{0',N_+^*}(y,u_N)\Bigr),
\end{array}
$$
and it is enough to show that the relation
$$
\MB{R}^{(\epsilon_1,\epsilon_2)}_{1,2}(x,y)
\wh{\MB{R}}^{(\epsilon_1,+)}_{1,3_+^*}(x,z)
\wh{\MB{R}}^{(\epsilon_2,+)}_{2,1_+^*}(y,z)=
\wh{\MB{R}}^{(\epsilon_2,+)}_{2,1_+^*}(y,z)
\wh{\MB{R}}^{(\epsilon_1,+)}_{1,3_+^*}(x,z)
\MB{R}^{(\epsilon_1,\epsilon_2)}_{1,2}(x,y)\,.
$$
is true.
This equality\footnote{which is actually the consequence
of the Yang--Baxter equation for an R--matrix
$\tilde{\MB{R}}(x,y)$}
can be proved by direct calculation.
\qed

\subsection*{A.7. Proof of Lemma 8}

If we write
$$
\begin{array}{ll}
\wh{\MB{R}}^{(+,+)}_{0,1^*}(x,u)=\wh{R}^{r,a}_{s,b}(x,u)\MB{E}^s_r\otimes\MB{F}^b_a\otimes\MB{I}\,,
\qquad&
\wh{R}^{r,a}_{s,b}(x,u)=\dfrac1{f(u,x)}\Bigl(\delta^r_s\delta^a_b+g(u,x)\delta^r_b\delta^a_s\Bigr)\\[6pt]
\MB{R}^{(+,-)}_{0,1}(x,u)=R^{p,-c}_{q,-d}(x,u)\MB{E}^q_p\otimes\MB{I}^*\otimes\MB{E}^{-d}_{-c}\,,
\qquad&
R^{p,-c}_{q,-d}(x,u)=\delta^p_q\delta^c_d-k(x,u)\delta^{p,c}\delta_{q,d}\\[6pt]
\wh{\MB{R}}^{(-,+)}_{0,1^*}(x,u)=\wh{R}^{-r,a}_{-s,b}(x,u)\MB{E}^{-s}_{-r}\otimes\MB{F}^b_a\otimes\MB{I}\,,
\qquad&
\wh{R}^{-r,a}_{-s,b}(x,u)=\delta^r_s\delta^a_b-k(u,x)\delta^{r,a}\delta_{s,b}\\[6pt]
\MB{R}^{(-,-)}_{0,1}(x,u)=R^{-p,-c}_{-q,-d}(x,u)\MB{E}^{-q}_{-p}\otimes\MB{I}^*\MB{E}^{-d}_{-c}\,,
\qquad&
R^{-p,-c}_{-q,-d}(x,u)=\dfrac1{f(x,u)}\Bigl(\delta^p_q\delta^c_d+g(x,u)\delta^p_d\delta^c_q\Bigr),
\end{array}
$$
we get the expression
$$
\begin{array}{l}
\wh{\MB{T}}^{(+)}_{0,1,\ldots,N}(x;\vec{u})=
\wh{R}^{r_1,a_1}_{s_1,b_1}(x,u_1)\wh{R}^{s_1,a_2}_{s_2,b_2}(x,u_2)\ldots
\wh{R}^{s_{N-1},a_N}_{s_N,b_N}(x,u_N)\\[4pt]
\hskip35mm
R^{p_N,-c_N}_{q_N,-d_N}(x,u_N)R^{q_N,-c_{N-1}}_{q_{N-1},-d_{N-1}}(x,u_{N-1})\ldots
R^{q_2,-c_1}_{q_1,-d_1}(x,u_1)\\[4pt]
\hskip40mm
\MB{E}^{q_1}_{r_1}\otimes\Bigl(\MB{F}^{b_1}_{a_1}\otimes\ldots\otimes\MB{F}^{b_N}_{a_N}\Bigr)\otimes
\Bigl(\MB{E}^{-d_1}_{-c_1}\otimes\ldots\otimes\MB{E}^{-d_N}_{-c_N}\Bigr)\otimes
T^{s_N}_{p_N}(x)\\[9pt]
\wh{\MB{T}}^{(-)}_{0,1,\ldots,N}(x;\vec{u})=
\wh{R}^{-r_1,a_1}_{-s_1,b_1}(x,u_1)\wh{R}^{-s_1,a_2}_{-s_2,b_2}(x,u_2)\ldots
\wh{R}^{-s_{N-1},a_N}_{-s_N,b_N}(x,u_N)\\[4pt]
\hskip35mm
R^{-p_N,-c_N}_{-q_N,-d_N}(x,u_N)R^{-q_N,-c_{N-1}}_{-q_{N-1},-d_{N-1}}(x,u_{N-1})\ldots
R^{-q_2,-c_1}_{-q_1,-d_1}(x,u_1)\\[4pt]
\hskip40mm
\MB{E}^{-q_1}_{-r_1}\otimes\Bigl(\MB{F}^{b_1}_{a_1}\otimes\ldots\otimes\MB{F}^{b_N}_{a_N}\Bigr)\otimes
\Bigl(\MB{E}^{-d_1}_{-c_1}\otimes\ldots\otimes\MB{E}^{-d_N}_{-c_N}\Bigr)\otimes
T^{-s_N}_{-p_N}(x)
\end{array}
$$
or, in the coordinates
$$
\begin{array}{l}
\wh{T}^i_k(x;\vec{u})=\wh{R}^{i,a_1}_{s_1,b_1}(x,u_1)\wh{R}^{s_1,a_2}_{s_2,b_2}(x,u_2)\ldots
\wh{R}^{s_{N-1},a_N}_{s_N,b_N}(x,u_N)\\[4pt]
\hskip30mm
R^{p_N,-c_N}_{p_{N-1},-d_N}(x,u_N)R^{p_{N-1},-c_{N-1}}_{p_{N-2},-d_{N-1}}(x,u_{N-1})\ldots
R^{p_1,-c_1}_{k,-d_1}(x,u_1)\\[4pt]
\hskip40mm
\Bigl(\MB{F}^{b_1}_{a_1}\otimes\ldots\otimes\MB{F}^{b_N}_{a_N}\Bigr)\otimes
\Bigl(\MB{E}^{-d_1}_{-c_1}\otimes\ldots\otimes\MB{E}^{-d_N}_{-c_N}\Bigr)\otimes
T^{s_N}_{p_N}(x)\\[9pt]
\wh{T}^{-i}_{-k}(x;\vec{u})=
\wh{R}^{-i,a_1}_{-s_1,b_1}(x,u_1)\wh{R}^{-s_1,a_2}_{-s_2,b_2}(x,u_2)\ldots
\wh{R}^{-s_{N-1},a_N}_{-s_N,b_N}(x,u_N)\\[4pt]
\hskip30mm
R^{-p_N,-c_N}_{-p_{N-1},-d_N}(x,u_N)R^{-p_{N-1},-c_{N-1}}_{-p_{N-2},-d_{N-1}}(x,u_{N-1})\ldots
R^{-p_1,-c_1}_{-k,-d_1}(x,u_1)\\[4pt]
\hskip40mm
\Bigl(\MB{F}^{b_1}_{a_1}\otimes\ldots\otimes\MB{F}^{b_N}_{a_N}\Bigr)\otimes
\Bigl(\MB{E}^{-d_1}_{-c_1}\otimes\ldots\otimes\MB{E}^{-d_N}_{-c_N}\Bigr)\otimes
T^{-s_N}_{-p_N}(x)
\end{array}
$$
So we have the relations
\begin{equation}
 \label{DL-8}
\begin{array}{l}
\wh{T}^i_k(x;\vec{u})\wh{\Omega}=
\wh{R}^{i,1}_{s_1,b_1}(x,u_1)\wh{R}^{s_1,1}_{s_2,b_2}(x,u_2)\ldots
\wh{R}^{s_{N-1},1}_{s_N,b_N}(x,u_N)\\[4pt]
\hskip30mm
R^{p_N,-c_N}_{p_{N-1},-1}(x,u_N)R^{p_{N-1},-c_{N-1}}_{p_{N-2},-1}(x,u_{N-1})\ldots
R^{p_1,-c_1}_{k,-1}(x,u_1)\\[4pt]
\hskip40mm
\Bigl(\MB{f}^{b_1}\otimes\ldots\otimes\MB{f}^{b_N}\Bigr)\otimes
\Bigl(\MB{e}_{-c_1}\otimes\ldots\otimes\MB{e}_{-c_N}\Bigr)\otimes
T^{s_N}_{p_N}(x)\omega\\[9pt]
\wh{T}^{-i}_{-k}(x;\vec{u})\wh{\Omega}=
\wh{R}^{-i,1}_{-s_1,b_1}(x,u_1)\wh{R}^{-s_1,1}_{-s_2,b_2}(x,u_2)\ldots
\wh{R}^{-s_{N-1},1}_{-s_N,b_N}(x,u_N)\\[4pt]
\hskip30mm
R^{-p_N,-c_N}_{-p_{N-1},-1}(x,u_N)R^{-p_{N-1},-c_{N-1}}_{-p_{N-2},-1}(x,u_{N-1})\ldots
R^{-p_1,-c_1}_{-k,-1}(x,u_1)\\[4pt]
\hskip40mm
\Bigl(\MB{f}^{b_1}\otimes\ldots\otimes\MB{f}^{b_N}\Bigr)\otimes
\Bigl(\MB{e}_{-c_1}\otimes\ldots\otimes\MB{e}_{-c_N}\Bigr)\otimes
T^{-s_N}_{-p_N}(x)\omega\,.
\end{array}
\end{equation}

Especially, we have the relation
$$
\begin{array}{l}
\wh{T}^1_2(x;\vec{u})\wh{\Omega}=
\wh{R}^{1,1}_{s_1,b_1}(x,u_1)\wh{R}^{s_1,1}_{s_2,b_2}(x,u_2)\ldots
\wh{R}^{s_{N-1},1}_{s_N,b_N}(x,u_N)\\[4pt]
\hskip30mm
R^{p_N,-c_N}_{p_{N-1},-1}(x,u_N)R^{p_{N-1},-c_{N-1}}_{p_{N-2},-1}(x,u_{N-1})\ldots
R^{p_1,-c_1}_{2,-1}(x,u_1)\\[4pt]
\hskip40mm
\Bigl(\MB{f}^{b_1}\otimes\ldots\otimes\MB{f}^{b_N}\Bigr)\otimes
\Bigl(\MB{e}_{-c_1}\otimes\ldots\otimes\MB{e}_{-c_N}\Bigr)\otimes
T^{s_N}_{p_N}(x)\omega\\[9pt]
\wh{T}^{-2}_{-1}(x;\vec{u})\wh{\Omega}=
\wh{R}^{-2,1}_{-s_1,b_1}(x,u_1)\wh{R}^{-s_1,1}_{-s_2,b_2}(x,u_2)\ldots
\wh{R}^{-s_{N-1},1}_{-s_N,b_N}(x,u_N)\\[4pt]
\hskip30mm
R^{-p_N,-c_N}_{-p_{N-1},-1}(x,u_N)R^{-p_{N-1},-c_{N-1}}_{-p_{N-2},-1}(x,u_{N-1})\ldots
R^{-p_1,-c_1}_{-1,-1}(x,u_1)\\[4pt]
\hskip40mm
\Bigl(\MB{f}^{b_1}\otimes\ldots\otimes\MB{f}^{b_N}\Bigr)\otimes
\Bigl(\MB{e}_{-c_1}\otimes\ldots\otimes\MB{e}_{-c_N}\Bigr)\otimes
T^{-s_N}_{-p_N}(x)\omega
\end{array}
$$
Since $R^{p,-c}_{2,-1}(x,u)=\delta^p_2\delta^c_1$ and
$\wh{R}^{-2,1}_{-s,b}=\delta^2_s\delta^1_b$, we obtain
$$
\begin{array}{l}
\wh{T}^1_2(x;\vec{u})\wh{\Omega}=
\wh{R}^{1,1}_{s_1,b_1}(x,u_1)\wh{R}^{s_1,1}_{s_2,b_2}(x,u_2)\ldots
\wh{R}^{s_{N-1},1}_{s_N,b_N}(x,u_N)\\[4pt]
\hskip30mm
\Bigl(\MB{f}^{b_1}\otimes\ldots\otimes\MB{f}^{b_N}\Bigr)\otimes
\Bigl(\MB{e}_{-1}\otimes\ldots\otimes\MB{e}_{-1}\Bigr)\otimes
T^{s_N}_2(x)\omega\\[9pt]
\wh{T}^{-2}_{-1}(x;\vec{u})\wh{\Omega}=
R^{-p_N,-c_N}_{-p_{N-1},-1}(x,u_N)R^{-p_{N-1},-c_{N-1}}_{-p_{N-2},-1}(x,u_{N-1})\ldots
R^{-p_1,-c_1}_{-1,-1}(x,u_1)\\[4pt]
\hskip40mm
\Bigl(\MB{f}^{1}\otimes\ldots\otimes\MB{f}^{1}\Bigr)\otimes
\Bigl(\MB{e}_{-c_1}\otimes\ldots\otimes\MB{e}_{-c_N}\Bigr)\otimes
T^{-2}_{-p_N}(x)\omega
\end{array}
$$
Further, the relations
$\wh{R}^{1,1}_{s,b}(x,u)=\delta^1_s\delta^1_b$ and
$R^{-p,-c}_{-1,-1}(x,u)=\delta^p_1\delta^c_1$, are true, and since
$\omega$ is the vacuum vector, we obtain
$$
\begin{array}{l}
\wh{T}^1_2(x;\vec{u})\wh{\Omega}=
\Bigl(\MB{f}^{1}\otimes\ldots\otimes\MB{f}^{1}\Bigr)\otimes
\Bigl(\MB{e}_{-1}\otimes\ldots\otimes\MB{e}_{-1}\Bigr)\otimes
T^1_2(x)\omega=0\\[6pt]
\wh{T}^{-2}_{-1}(x;\vec{u})\wh{\Omega}=
\Bigl(\MB{f}^{1}\otimes\ldots\otimes\MB{f}^{1}\Bigr)\otimes
\Bigl(\MB{e}_{-1}\otimes\ldots\otimes\MB{e}_{-1}\Bigr)\otimes
T^{-2}_{-1}(x)\omega=0\,.
\end{array}
$$

\bigskip

If we put $i=k=2$ in (\ref{DL-8}), we have
$$
\begin{array}{l}
\wh{T}^2_2(x;\vec{u})\wh{\Omega}=
\wh{R}^{2,1}_{s_1,b_1}(x,u_1)\wh{R}^{s_1,1}_{s_2,b_2}(x,u_2)\ldots
\wh{R}^{s_{N-1},1}_{s_N,b_N}(x,u_N)\\[4pt]
\hskip30mm
R^{p_N,-c_N}_{p_{N-1},-1}(x,u_N)R^{p_{N-1},-c_{N-1}}_{p_{N-2},-1}(x,u_{N-1})\ldots
R^{p_1,-c_1}_{2,-1}(x,u_1)\\[4pt]
\hskip40mm
\Bigl(\MB{f}^{b_1}\otimes\ldots\otimes\MB{f}^{b_N}\Bigr)\otimes
\Bigl(\MB{e}_{-c_1}\otimes\ldots\otimes\MB{e}_{-c_N}\Bigr)\otimes
T^{s_N}_{p_N}(x)\omega\\[9pt]
\wh{T}^{-2}_{-2}(x;\vec{u})\wh{\Omega}=
\wh{R}^{-2,1}_{-s_1,b_1}(x,u_1)\wh{R}^{-s_1,1}_{-s_2,b_2}(x,u_2)\ldots
\wh{R}^{-s_{N-1},1}_{-s_N,b_N}(x,u_N)\\[4pt]
\hskip30mm
R^{-p_N,-c_N}_{-p_{N-1},-1}(x,u_N)R^{-p_{N-1},-c_{N-1}}_{-p_{N-2},-1}(x,u_{N-1})\ldots
R^{-p_1,-c_1}_{-2,-1}(x,u_1)\\[4pt]
\hskip40mm
\Bigl(\MB{f}^{b_1}\otimes\ldots\otimes\MB{f}^{b_N}\Bigr)\otimes
\Bigl(\MB{e}_{-c_1}\otimes\ldots\otimes\MB{e}_{-c_N}\Bigr)\otimes
T^{-s_N}_{-p_N}(x)\omega
\end{array}
$$
From the equations $R^{p,-c}_{2,-1}(x,u)=\delta^p_2\delta^c_1$ and
$\wh{R}^{-2,1}_{-s,b}=\delta^2_s\delta^1_b$ we then obtain
$$
\begin{array}{l}
\wh{T}^2_2(x;\vec{u})\wh{\Omega}=
\wh{R}^{2,1}_{s_1,b_1}(x,u_1)\wh{R}^{s_1,1}_{s_2,b_2}(x,u_2)\ldots
\wh{R}^{s_{N-1},1}_{s_N,b_N}(x,u_N)\\[4pt]
\hskip30mm
\Bigl(\MB{f}^{b_1}\otimes\ldots\otimes\MB{f}^{b_N}\Bigr)\otimes
\Bigl(\MB{e}_{-1}\otimes\ldots\otimes\MB{e}_{-1}\Bigr)\otimes
T^{s_N}_{2}(x)\omega\\[9pt]
\wh{T}^{-2}_{-2}(x;\vec{u})\wh{\Omega}=
R^{-p_N,-c_N}_{-p_{N-1},-1}(x,u_N)R^{-p_{N-1},-c_{N-1}}_{-p_{N-2},-1}(x,u_{N-1})\ldots
R^{-p_1,-c_1}_{-2,-1}(x,u_1)\\[4pt]
\hskip30mm
\Bigl(\MB{f}^{1}\otimes\ldots\otimes\MB{f}^{1}\Bigr)\otimes
\Bigl(\MB{e}_{-c_1}\otimes\ldots\otimes\MB{e}_{-c_N}\Bigr)\otimes
T^{-2}_{-p_N}(x)\omega
\end{array}
$$
Since $\omega$ is the vacuum vector, we have
$$
T^{s_N}_{2}(x)\omega=\lambda_2(x)\delta^{s_N}_2\omega\,,\qquad
T^{-2}_{-p_N}(x)\omega=\lambda_{-2}(x)\delta^2_{p_N}\omega\,,
$$
and so
$$
\begin{array}{l}
\wh{T}^2_2(x;\vec{u})\wh{\Omega}= \lambda_2(x)
\wh{R}^{2,1}_{s_1,b_1}(x,u_1)\wh{R}^{s_1,1}_{s_2,b_2}(x,u_2)\ldots
\wh{R}^{s_{N-1},1}_{2,b_N}(x,u_N)\\[4pt]
\hskip30mm
\Bigl(\MB{f}^{b_1}\otimes\ldots\otimes\MB{f}^{b_N}\Bigr)\otimes
\Bigl(\MB{e}_{-1}\otimes\ldots\otimes\MB{e}_{-1}\Bigr)\otimes\omega\\[9pt]
\wh{T}^{-2}_{-2}(x;\vec{u})\wh{\Omega}= \lambda_{-2}(x)
R^{-2,-c_N}_{-p_{N-1},-1}(x,u_N)R^{-p_{N-1},-c_{N-1}}_{-p_{N-2},-1}(x,u_{N-1})\ldots
R^{-p_1,-c_1}_{-2,-1}(x,u_1)\\[4pt]
\hskip30mm
\Bigl(\MB{f}^{1}\otimes\ldots\otimes\MB{f}^{1}\Bigr)\otimes
\Bigl(\MB{e}_{-c_1}\otimes\ldots\otimes\MB{e}_{-c_N}\Bigr)\otimes\omega
\end{array}
$$
The relations
$$
\wh{R}^{s,1}_{2,b}(x,u)=
\frac1{f(u,x)}\,\delta^s_2\delta^1_b\,,\qquad
R^{-1,-c}_{-p,-1}(x,u)= \frac1{f(x,u)}\,\delta^1_p\delta^c_2\,,
$$
then lead to the equations
$$
\begin{array}{l}
\wh{T}^2_2(x;\vec{u})\wh{\Omega}=
\dfrac{\lambda_2(x)}{F(\ol{u},x)}
\Bigl(\MB{f}^{1}\otimes\ldots\otimes\MB{f}^{1}\Bigr)\otimes
\Bigl(\MB{e}_{-1}\otimes\ldots\otimes\MB{e}_{-1}\Bigr)\otimes\omega=
\dfrac{\lambda_2(x)}{F(\ol{u},x)}\,\wh{\Omega}\\[9pt]
\wh{T}^{-2}_{-2}(x;\vec{u})\wh{\Omega}=
\dfrac{\lambda_{-2}(x)}{F(x,\ol{u})}
\Bigl(\MB{f}^{1}\otimes\ldots\otimes\MB{f}^{1}\Bigr)\otimes
\Bigl(\MB{e}_{-c_1}\otimes\ldots\otimes\MB{e}_{-c_N}\Bigr)\otimes\omega=
\dfrac{\lambda_{-2}(x)}{F(x,\ol{u})}\,\wh{\Omega}
\end{array}
$$

\bigskip

For $i=k=1$ relation (\ref{DL-8}) gives
$$
\begin{array}{l}
\wh{T}^1_1(x;\vec{u})\wh{\Omega}=
\wh{R}^{1,1}_{s_1,b_1}(x,u_1)\wh{R}^{s_1,1}_{s_2,b_2}(x,u_2)\ldots
\wh{R}^{s_{N-1},1}_{s_N,b_N}(x,u_N)\\[4pt]
\hskip30mm
R^{p_N,-c_N}_{p_{N-1},-1}(x,u_N)R^{p_{N-1},-c_{N-1}}_{p_{N-2},-1}(x,u_{N-1})\ldots
R^{p_1,-c_1}_{1,-1}(x,u_1)\\[4pt]
\hskip40mm
\Bigl(\MB{f}^{b_1}\otimes\ldots\otimes\MB{f}^{b_N}\Bigr)\otimes
\Bigl(\MB{e}_{-c_1}\otimes\ldots\otimes\MB{e}_{-c_N}\Bigr)\otimes
T^{s_N}_{p_N}(x)\omega\\[9pt]
\wh{T}^{-1}_{-1}(x;\vec{u})\wh{\Omega}=
\wh{R}^{-1,1}_{-s_1,b_1}(x,u_1)\wh{R}^{-s_1,1}_{-s_2,b_2}(x,u_2)\ldots
\wh{R}^{-s_{N-1},1}_{-s_N,b_N}(x,u_N)\\[4pt]
\hskip30mm
R^{-p_N,-c_N}_{-p_{N-1},-1}(x,u_N)R^{-p_{N-1},-c_{N-1}}_{-p_{N-2},-1}(x,u_{N-1})\ldots
R^{-p_1,-c_1}_{-1,-1}(x,u_1)\\[4pt]
\hskip40mm
\Bigl(\MB{f}^{b_1}\otimes\ldots\otimes\MB{f}^{b_N}\Bigr)\otimes
\Bigl(\MB{e}_{-c_1}\otimes\ldots\otimes\MB{e}_{-c_N}\Bigr)\otimes
T^{-s_N}_{-p_N}(x)\omega
\end{array}
$$
Since the relations $\wh{R}^{1,1}_{s,b}(x,u)=\delta^1_s\delta^1_b$ and
$R^{-p,-c}_{-1,-1}(x,u)=\delta^p_1\delta^c_1$ are true, we obtain
$$
\begin{array}{l}
\wh{T}^1_1(x;\vec{u})\wh{\Omega}=
R^{p_N,-c_N}_{p_{N-1},-1}(x,u_N)R^{p_{N-1},-c_{N-1}}_{p_{N-2},-1}(x,u_{N-1})\ldots
R^{p_1,-c_1}_{1,-1}(x,u_1)\\[4pt]
\hskip30mm
\Bigl(\MB{f}^{1}\otimes\ldots\otimes\MB{f}^{1}\Bigr)\otimes
\Bigl(\MB{e}_{-c_1}\otimes\ldots\otimes\MB{e}_{-c_N}\Bigr)\otimes
T^{1}_{p_N}(x)\omega\\[9pt]
\wh{T}^{-1}_{-1}(x;\vec{u})\wh{\Omega}=
\wh{R}^{-1,1}_{-s_1,b_1}(x,u_1)\wh{R}^{-s_1,1}_{-s_2,b_2}(x,u_2)\ldots
\wh{R}^{-s_{N-1},1}_{-s_N,b_N}(x,u_N)\\[4pt]
\hskip30mm
\Bigl(\MB{f}^{b_1}\otimes\ldots\otimes\MB{f}^{b_N}\Bigr)\otimes
\Bigl(\MB{e}_{-1}\otimes\ldots\otimes\MB{e}_{-1}\Bigr)\otimes
T^{-s_N}_{-1}(x)\omega
\end{array}
$$
For the vacuum vector $\omega$ the relations
$$
T^{1}_{p_N}(x)\omega=\lambda_1(x)\delta^1_{p_N}\omega\,,\qquad
T^{-s_N}_{-1}(x)\omega=\lambda_{-1}(x)\delta^{s_N}_1\omega\,,
$$
are valid, and so
$$
\begin{array}{l}
\wh{T}^1_1(x;\vec{u})\wh{\Omega}= \lambda_1(x)
R^{1,-c_N}_{p_{N-1},-1}(x,u_N)R^{p_{N-1},-c_{N-1}}_{p_{N-2},-1}(x,u_{N-1})\ldots
R^{p_1,-c_1}_{1,-1}(x,u_1)\\[4pt]
\hskip30mm
\Bigl(\MB{f}^{1}\otimes\ldots\otimes\MB{f}^{1}\Bigr)\otimes
\Bigl(\MB{e}_{-c_1}\otimes\ldots\otimes\MB{e}_{-c_N}\Bigr)\otimes\omega\\[9pt]
\wh{T}^{-1}_{-1}(x;\vec{u})\wh{\Omega}= \lambda_{-1}(x)
\wh{R}^{-1,1}_{-s_1,b_1}(x,u_1)\wh{R}^{-s_1,1}_{-s_2,b_2}(x,u_2)\ldots
\wh{R}^{-s_{N-1},1}_{-1,b_N}(x,u_N)\\[4pt]
\hskip30mm
\Bigl(\MB{f}^{b_1}\otimes\ldots\otimes\MB{f}^{b_N}\Bigr)\otimes
\Bigl(\MB{e}_{-1}\otimes\ldots\otimes\MB{e}_{-1}\Bigr)\otimes\omega
\end{array}
$$
The relations
\begin{eqnarray*}
R^{1,-c}_{p,-1}(x,u)&=& \bigl(1-k(x,u)\bigr)\delta^1_p\delta^c_1=
\dfrac{u-x+2}{u-x+1}\delta^1_p\delta^c_1=
f(u,x-1)\delta^1_p\delta^c_1\\
\wh{R}^{-s,1}_{-1,b}(x,u)&=&
\bigl(1-k(u,x)\bigr)\delta^s_1\delta^1_b=
\dfrac{x-u+2}{x-u+1}\delta^s_1\delta^1_b=
f(x+1,u)\delta^s_1\delta^1_b
\end{eqnarray*}
then give
\begin{eqnarray*}
\wh{T}^1_1(x;\vec{u})\wh{\Omega}&=&
\lambda_1(x)F(\ol{u},x-1)
\Bigl(\MB{f}^{1}\otimes\ldots\otimes\MB{f}^{1}\Bigr)\otimes
\Bigl(\MB{e}_{-1}\otimes\ldots\otimes\MB{e}_{-1}\Bigr)\otimes\omega=\\
&=&
\lambda_1(x)F(\ol{u},x-1)\wh{\Omega}\\[9pt]
\wh{T}^{-1}_{-1}(x;\vec{u})\wh{\Omega}&=&
\lambda_{-1}(x)F(x+1,\ol{u})
\Bigl(\MB{f}^{1}\otimes\ldots\otimes\MB{f}^{1}\Bigr)\otimes
\Bigl(\MB{e}_{-1}\otimes\ldots\otimes\MB{e}_{-1}\Bigr)\otimes\omega=\\
&=&
\lambda_{-1}(x)F(x+1,\ol{u})\wh{\Omega}
\end{eqnarray*}
which proves the Lemma.
\qed

\subsection*{A.8. Proof of Lemma 9}

The first four relations can be proved in the same way as similar claims in Lemma 1.

We will prove the last four claims of lemmas by induction over the number of elements
$P$ and $Q$ of sets $\ol{u}$ and $\ol{v}$.
For $P$ and $Q$ equal to one, these are commutation relations in the
RTT--algebra $\tilde{\MC{A}}$.

If we assume that the statements hold for $P$ a $Q$, we get
$$
\begin{array}{l}
\tilde{T}^1_1(x)\tilde{T}^{-1}_{-2}(\{\ol{w},w_{Q+1}\})=
F(x-2,\{\ol{w},w_{Q+1}\})\tilde{T}^{-1}_{-2}(\{\ol{w},w_{Q+1}\})\tilde{T}^1_1(x)+\\[4pt]
\hskip15mm+
{\tsum_{w_s\in\ol{w}}}g(x-2,w_s)F(w_s,\{\ol{w}_s,w_{Q+1}\})
\tilde{T}^2_1(x)\tilde{T}^{-1}_{-2}(\{\ol{w}_s,w_{Q+1}\})\tilde{T}^{-2}_{-2}(w_s)+\\[4pt]
\hskip15mm+
\Bigl(g(x-2,w_{Q+1})F(x-2,\ol{w})-{\tsum_{w_s\in\ol{w}}}g(x-2,w_s)g(w_s,w_{Q+1})F(w_s,\ol{w}_s)\Bigr)\\[4pt]
\hskip100mm
\tilde{T}^2_1(x)\tilde{T}^{-1}_{-2}(\ol{w})\tilde{T}^{-2}_{-2}(w_{Q+1})\\[6pt]
\tilde{T}^2_2(x)\tilde{T}^{-1}_{-2}(\{\ol{w},w_{Q+1}\})=
F(\{\ol{w},w_{Q+1}\},x-2)\tilde{T}^{-1}_{-2}(\{\ol{w},w_{Q+1}\})\tilde{T}^2_2(x)+\\[4pt]
\hskip15mm+
{\tsum_{w_s\in\ol{w}}}g(w_s,x-2)F(\{\ol{w}_s,w_{Q+1}\},w_s)
\tilde{T}^2_1(x)\tilde{T}^{-1}_{-2}(\{\ol{w}_s,w_{Q+1}\})\tilde{T}^{-1}_{-1}(w_s)+\\[4pt]
\hskip15mm+
\Bigl(g(w_{Q+1},x-2)F(\ol{w},x-2)-{\tsum_{w_s\in\ol{w}}}g(w_{Q+1},w_s)g(w_s,x-2)F(\ol{w}_s,w_s)\Bigr)\\[4pt]
\hskip100mm
\tilde{T}^2_1(x)\tilde{T}^{-1}_{-2}(\ol{w})\tilde{T}^{-1}_{-1}(w_{Q+1})
\end{array}
$$
$$
\begin{array}{l}
\tilde{T}^{-1}_{-1}(x)\tilde{T}^2_1(\{\ol{v},v_{P+1}\})=
F(\{\ol{v},v_{P+1}\},x+2)\tilde{T}^2_1(\{\ol{v},v_{P+1}\})\tilde{T}^{-1}_{-1}(x)+\\[4pt]
\hskip15mm+
{\tsum_{v_r\in\ol{v}}}g(v_r,x+2)F(\{\ol{v}_r,v_{P+1}\},v_r)
\tilde{T}^2_1(\{\ol{v}_r,v_{P+1}\})\tilde{T}^{-1}_{-2}(x)\tilde{T}^2_2(v_r)+\\[4pt]
\hskip15mm+
\Bigl(g(v_{P+1},x+2)F(\ol{v},x+2)-{\tsum_{v_r\in\ol{v}}}g(v_{P+1},v_r)g(v_r,x+2)F(\ol{v}_r,v_r)\Bigr)\\[4pt]
\hskip100mm
\tilde{T}^2_1(\ol{v})\tilde{T}^{-1}_{-2}(x)\tilde{T}^2_2(v_{P+1})\\[6pt]
\tilde{T}^{-2}_{-2}(x)\tilde{T}^2_1(\{\ol{v},v_{P+1}\})=
F(x+2,\{\ol{v},v_{P+1}\})\tilde{T}^2_1(\{\ol{v},v_{P+1}\})\tilde{T}^{-2}_{-2}(x)+\\[4pt]
\hskip15mm+
{\tsum_{v_r\in\ol{v}}}g(x+2,v_r)F(v_r,\{\ol{v}_r,v_{P+1}\})
\tilde{T}^2_1(\{\ol{v}_r,v_{P+1}\})\tilde{T}^{-1}_{-2}(x)\tilde{T}^1_1(v_r)+\\[4pt]
\hskip15mm+
\Bigl(g(x+2,v_{P+1})F(x+2,\ol{v})-{\tsum_{v_r\in\ol{v}}}g(x+2,v_r)g(v_r,v_{P+1})F(v_r,\ol{v}_r)\Bigr)\\[4pt]
\hskip100mm
\tilde{T}^2_1(\ol{v})\tilde{T}^{-1}_{-2}(x)\tilde{T}^1_1(v_{P+1})
\end{array}
$$
So, it is enough to prove the relations
$$
\begin{array}{l}
g(x-2,w_{Q+1})F(x-2,\ol{w})-{\tsum_{w_s\in\ol{w}}}g(x-2,w_s)g(w_s,w_{Q+1})F(w_s,\ol{w}_s)=\\[4pt]
\hskip80mm=
g(x-2,w_{Q+1})F(w_{Q+1},\ol{w})\\[6pt]
g(w_{Q+1},x-2)F(\ol{w},x-2)-{\tsum_{w_s\in\ol{w}}}g(w_{Q+1},w_s)g(w_s,x-2)F(\ol{w}_s,w_s)=\\[4pt]
\hskip80mm=
g(w_{Q+1},x-2)F(\ol{w},w_{Q+1})\\[6pt]
g(v_{P+1},x+2)F(\ol{v},x+2)-{\tsum_{v_r\in\ol{v}}}g(v_{P+1},v_r)g(v_r,x+2)F(\ol{v}_r,v_r)=\\[4pt]
\hskip80mm=
g(v_{P+1},x+2)F(\ol{v},v_{P+1})\\[6pt]
g(x+2,v_{P+1})F(x+2,\ol{v})-{\tsum_{v_r\in\ol{v}}}g(x+2,v_r)g(v_r,v_{P+1})F(v_r,\ol{v}_r)=\\[4pt]
\hskip80mm=
g(x+2,v_{P+1})F(v_{P+1},\ol{v})\,.
\end{array}
$$
But these are consequences of Lemma 0.
\qed

\subsection*{A.9. Proof of Lemma 10}

According to Lemma 9, we have
$$
\begin{array}{l}
\tilde{T}^1_1(x)\tilde{T}^2_1(\ol{v})\tilde{T}^{-1}_{-2}(\ol{w})=
F(x,\ol{v})\tilde{T}^2_1(\ol{v})\tilde{T}^1_1(x)\tilde{T}^{-1}_{-2}(\ol{w})-\\[6pt]
\hskip20mm-
{\tsum_{r=1}^P}g(x,v_r)F(v_r,\ol{v}_r)\tilde{T}^2_1(\ol{v}_r,x)\tilde{T}^1_1(v_r)\tilde{T}^{-1}_{-2}(\ol{w})=\\[6pt]
\hskip10mm=
F(x,\ol{v})F(x-2,\ol{w})\tilde{T}^2_1(\ol{v})\tilde{T}^{-1}_{-2}(\ol{w})\tilde{T}^1_1(x)-\\[4pt]
\hskip15mm-
{\tsum_{r=1}^P}g(x,v_r)F(v_r-2,\ol{w})F(v_r,\ol{v}_r)\tilde{T}^2_1(\{\ol{v}_r,x\})
\tilde{T}^{-1}_{-2}(\ol{w})\tilde{T}^1_1(v_r)+\\[4pt]
\hskip15mm+
{\tsum_{s=1}^Q}F(w_s,\ol{w}_s)\Bigl(g(x-2,w_s)F(x,\ol{v})-
{\tsum_{r=1}^P}g(x,v_r)g(v_r-2,w_s)F(v_r,\ol{v}_r)\Bigr)\\[4pt]
\hskip90mm
\tilde{T}^2_1(\{\ol{v},x\})\tilde{T}^{-1}_{-2}(\ol{w}_s)\tilde{T}^{-2}_{-2}(w_s)
\end{array}
$$
Since the relation $g(x-2,w_s)=g(x,v_s+2)$ is true,
following Lemma 0 we have
$$
{\tsum_{v_r\in\ol{v}}}g(x,v_r)g(v_r,v_s+2)F(v_r,\ol{v}_r)=
g(x,v_s+2)\bigl(F(x,\ol{v})-F(v_s+2,\ol{v})\bigr)\,,
$$
and the relation
$$
g(x-2,v_s)F(x,\ol{v})-{\tsum_{v_r\in\ol{v}}}g(x,v_r)g(v_r-2,v_s)F(v_r,\ol{v}_r)=
g(x-2,v_s)F(v_s+2,\ol{v})\,,
$$
holds. So the relation
$$
\begin{array}{l}
\tilde{T}^1_1(x)\tilde{T}^2_1(\ol{v})\tilde{T}^{-1}_{-2}(\ol{w})=
F(x,\ol{v})F(x-2,\ol{w})\tilde{T}^2_1(\ol{v})\tilde{T}^{-1}_{-2}(\ol{w})\tilde{T}^1_1(x)-\\[4pt]
\hskip30mm-
{\tsum_{r=1}^P}g(x,v_r)F(v_r-2,\ol{w})F(v_r,\ol{v}_r)\tilde{T}^2_1(\{\ol{v}_r,x\})
\tilde{T}^{-1}_{-2}(\ol{w})\tilde{T}^1_1(v_r)+\\[4pt]
\hskip30mm+
{\tsum_{s=1}^Q}g(x-2,v_s)F(v_s+2,\ol{v})F(w_s,\ol{w}_s)
\tilde{T}^2_1(\{\ol{v},x\})\tilde{T}^{-1}_{-2}(\ol{w}_s)\tilde{T}^{-2}_{-2}(w_s)
\end{array}
$$
is valid.

The other relations can be proved in a similar way:
$$
\begin{array}{l}
\tilde{T}^2_2(x)\tilde{T}^2_1(\ol{v})\tilde{T}^{-1}_{-2}(\ol{w})=
F(\ol{v},x)F(\ol{w},x-2)\tilde{T}^2_1(\ol{v})\tilde{T}^{-1}_{-2}(\ol{w})\tilde{T}^2_2(x)-\\[4pt]
\hskip15mm-
{\tsum_{v_r\in\ol{v}}}g(v_r,x)F(\ol{v}_r,v_r)F(\ol{w},v_r-2)
\tilde{T}^2_1(\ol{v}_r,x)\tilde{T}^{-1}_{-2}(\ol{w})\tilde{T}^2_2(v_r)+\\[4pt]
\hskip15mm+
{\tsum_{w_s\in\ol{w}}}F(\ol{w}_s,w_s)\Bigl(g(w_s,x-2)F(\ol{v},x)-
{\tsum_{v_r\in\ol{v}}}g(w_s,v_r-2)g(v_r,x)F(\ol{v}_r,v_r)\Bigr)\\[4pt]
\hskip100mm
\tilde{T}^2_1(\{\ol{v},x\})\tilde{T}^{-1}_{-2}(\ol{w}_s)\tilde{T}^{-1}_{-1}(w_s)\,,\\[9pt]
\tilde{T}^{-1}_{-1}(x)\tilde{T}^2_1(\ol{v})\tilde{T}^{-1}_{-2}(\ol{w})=
F(\ol{v},x+2)F(\ol{w},x)\tilde{T}^2_1(\ol{v})\tilde{T}^{-1}_{-2}(\ol{w})\tilde{T}^{-1}_{-1}(x)+\\[4pt]
\hskip15mm+
{\tsum_{v_r\in\ol{v}}}g(v_r,x+2)F(\ol{w},v_r-2)F(\ol{v}_r,v_r)
\tilde{T}^2_1(\ol{v}_r)\tilde{T}^{-1}_{-2}(\{\ol{w},x\})\tilde{T}^2_2(v_r)-\\[4pt]
\hskip15mm-
{\tsum_{w_s\in\ol{w}}}F(\ol{w}_s,w_s)\Bigl(g(w_s,x)F(\ol{v},x+2)-
{\tsum_{v_r\in\ol{v}}}g(w_s,v_r-2)g(v_r,x+2)F(\ol{v}_r,v_r)\Bigr)\\[4pt]
\hskip100mm
\tilde{T}^2_1(\ol{v})\tilde{T}^{-1}_{-2}(\{\ol{w}_s,x\})\tilde{T}^{-1}_{-1}(w_s)\\[9pt]
\tilde{T}^{-2}_{-2}(x)\tilde{T}^2_1(\ol{v})\tilde{T}^{-1}_{-2}(\ol{w})=
F(x+2,\ol{v})F(x,\ol{w})\tilde{T}^2_1(\ol{v})\tilde{T}^{-1}_{-2}(\ol{w})\tilde{T}^{-2}_{-2}(x)+\\[4pt]
\hskip15mm+
{\tsum_{v_r\in\ol{v}}}g(x+2,v_r)F(v_r-2,\ol{w})F(v_r,\ol{v}_r)
\tilde{T}^2_1(\ol{v}_r)\tilde{T}^{-1}_{-2}(\{\ol{w},x\})\tilde{T}^1_1(v_r)-\\[4pt]
\hskip15mm-
{\tsum_{w_s\in\ol{w}}}F(w_s,\ol{w}_s)\Bigl(g(x,w_s)F(x+2,\ol{v})-
{\tsum_{v_r\in\ol{v}}}g(x+2,v_r)g(v_r-2,w_s)F(v_r,\ol{v}_r)\Bigr)\\[4pt]
\hskip100mm
\tilde{T}^2_1(\ol{v})\tilde{T}^{-1}_{-2}(\{\ol{w}_s,x\})\tilde{T}^{-2}_{-2}(w_s)
\end{array}
$$
The assertion is derived from the equations
$$
\begin{array}{l}
g(w_s,x-2)F(\ol{v},x)-
{\tsum_{v_r\in\ol{v}}}g(w_s,v_r-2)g(v_r,x)F(\ol{v}_r,v_r)=
g(w_s,x-2)F(\ol{v},w_s+2)\\[4pt]
g(w_s,x)F(\ol{v},x+2)-{\tsum_{v_r\in\ol{v}}}g(w_s,v_r-2)g(v_r,x+2)F(\ol{v}_r,v_r)=
g(w_s,x)F(\ol{v},w_s+2)\\[4pt]
g(x,w_s)F(x+2,\ol{v})-{\tsum_{v_r\in\ol{v}}}g(x+2,v_r)g(v_r-2,w_s)F(v_r,\ol{v}_r)=
g(x,w_s)F(w_s+2,\ol{v})\,,
\end{array}
$$
which follow from Lemma 0.
\qed

\subsection*{A.10. Proof of Theorem  3}

If we use the relation $g(x,y)=-g(y,x)$, we have from Lemma 10
$$
\begin{array}{l}
\tilde{H}^{(+)}(x)|\ol{v};\ol{w}\bigr>=
\bigl(\tilde{T}^1_1(x)+\tilde{T}^2_2(x)\bigr)\bigl|\ol{v};\ol{w}\bigr>=\\[6pt]
\hskip5mm=
\Bigl(\mu_1(x)F(x,\ol{v})F(x-2,\ol{w})+\mu_2(x)F(\ol{v},x)F(\ol{w},x-2)\Bigr)
\bigl|\ol{v};\ol{w}\bigr>+\\[6pt]
\hskip10mm+
{\tsum_{v_r\in\ol{v}}}g(v_r,x)
\Bigl(\mu_1(v_r)F(v_r,\ol{v}_r)F(v_r-2,\ol{w})-
\mu_2(v_r)F(\ol{v}_r,v_r)F(\ol{w},v_r-2)\Bigr)\\[4pt]
\hskip100mm
\bigl|\{\ol{v}_r,x\};\ol{w}\bigr>+\\[6pt]
\hskip10mm+
{\tsum_{w_s\in\ol{w}}}g(w_s,x-2)
\Bigl(\mu_{-1}(w_s)F(\ol{v},w_s+2)F(\ol{w}_s,w_s)-\mu_{-2}(w_s)F(w_s+2,\ol{v})F(w_s,\ol{w}_s)\Bigr)\\[4pt]
\hskip100mm
\bigl|\{\ol{v},x\};\ol{w}_s\bigr>\,,\\[9pt]
\end{array}
$$
%%%%%%%%
$$
\begin{array}{l}
\tilde{H}^{(-)}(x)|\ol{v};\ol{w}\bigr>=
\bigl(\tilde{T}^{-1}_{-1}(x)+\tilde{T}^{-2}_{-2}(x)\bigr)\bigl|\ol{v};\ol{w}\bigr>=\\[6pt]
\hskip5mm=
\Bigl(\mu_{-1}(x)F(\ol{v},x+2)F(\ol{w},x)+\mu_{-2}(x)F(x+2,\ol{v})F(x,\ol{w})\Bigr)
\bigl|\ol{v};\ol{w}\bigr>+\\[6pt]
\hskip10mm+
{\tsum_{v_r\in\ol{v}}}g(x+2,v_r)
\Bigl(\mu_1(v_r)F(v_r,\ol{v}_r)F(v_r-2,\ol{w})-\mu_2(v_r)F(\ol{v}_r,v_r)F(\ol{w},v_r-2)\Bigr)\\[4pt]
\hskip100mm
\bigl|\ol{v}_r;\{\ol{w},x\}\bigr>+\\[6pt]
\hskip10mm+
{\tsum_{w_s\in\ol{w}}}g(x,w_s)
\Bigl(\mu_{-1}(w_s)F(\ol{v},w_s+2)F(\ol{w}_s,w_s)-\mu_{-2}(w_s)F(w_s+2,\ol{v})F(w_s,\ol{w}_s)\Bigr)\\[4pt]
\hskip100mm
\bigl|\ol{v};\{\ol{w}_s,x\}\bigr>
\end{array}
$$
Thus, if conditions (\ref{BP-TA}) are met, the relations
$$
\begin{array}{l}
\tilde{H}^{(+)}(x)|\ol{v};\ol{w}\bigr>=
\Bigl(\mu_1(x)F(x,\ol{v})F(x-2,\ol{w})+\mu_2(x)F(\ol{v},x)F(\ol{w},x-2)\Bigr)
\bigl|\ol{v};\ol{w}\bigr>,\\[6pt]
\tilde{H}^{(-)}(x)|\ol{v};\ol{w}\bigr>=
\Bigl(\mu_{-1}(x)F(\ol{v},x+2)F(\ol{w},x)+\mu_{-2}(x)F(x+2,\ol{v})F(x,\ol{w})\Bigr)
\bigl|\ol{v};\ol{w}\bigr>
\end{array}
$$
are true.
\qed

\end{document}